\documentclass[12pt]{article}

\usepackage{amsmath,amsfonts,amssymb,amsthm}
\usepackage{algorithm}
\usepackage{algorithmicx}
\usepackage{algpseudocode}
\usepackage[algo2e]{algorithm2e} 
\algnewcommand\algorithmicforeach{\textbf{for each}}
\algdef{S}[FOR]{ForEach}[1]{\algorithmicforeach\ #1\ \algorithmicdo}
\usepackage{subcaption}

\SetKwInOut{Input}{Input}
\SetKwInOut{Output}{Output}

\makeatletter
\def\algbackskip{\hskip-\ALG@thistlm}
\makeatother
\usepackage{setspace}
\usepackage{bbding}
\usepackage{graphicx,psfrag,epsf}
\usepackage{enumitem}
\usepackage{bm}
\usepackage{bbm}
\usepackage{soul} %
\usepackage[framemethod=TikZ]{mdframed}
\usepackage[margin=1in]{geometry}

\usepackage{natbib}

\usepackage{xurl} %
\usepackage[colorlinks,allcolors=black]{hyperref}

\newcommand{\red}[1]{{ #1}}

\usepackage{comment}
\usepackage{verbatim}
\usepackage{upquote}

\usepackage{algpseudocode}



\usepackage[section]{placeins}
\usepackage{rotating}
\usepackage{relsize}
\usepackage{scalerel}

\usepackage{tikz}
\usetikzlibrary{tikzmark, arrows, arrows.meta, bending, calc, shapes, patterns, decorations, decorations.pathreplacing, decorations.markings, fit, positioning, matrix}

\usepackage{epigraph}
\newcommand{\blind}{0}

\newcommand{\br}{\bm{r}}

\newcommand{\bU}{\bm{U}}
\newcommand{\bu}{\bm{u}}

\newcommand{\E}{\mathbb{E}}

\newcommand{\bbone}{\mathbbm{1}}
\newcommand{\indep}{\perp \!\!\! \perp}

\newtheorem{theorem}{Theorem}
\newtheorem{proposition}{Proposition}
\newtheorem{lemma}{Lemma}
\newtheorem{corollary}{Corollary}%
\newtheorem{definition}{Definition}

\newcommand{\repeatedstatementheading}{}
\newtheorem*{repeatedstatement}{\repeatedstatementheading}

\usepackage{booktabs}
\usepackage{tabularx}
\begin{document}

\widowpenalty0
\clubpenalty0

\def\spacingset#1{\renewcommand{\baselinestretch}%
{#1}\small\normalsize} \spacingset{2}
\pagenumbering{gobble}

\usetikzlibrary{shapes,shapes.geometric, decorations,arrows,calc,arrows.meta,fit,positioning}
\tikzset{
 -Latex,auto,node distance=1 cm and 1 cm,semithick,
 state/.style={ellipse, draw, minimum width=0.7 cm},
 point/.style={circle, draw, inner sep=0.04cm,fill,node contents={}},
 bidirected/.style={Latex-Latex,dashed},
 el/.style={inner sep=2pt, align=left, sloped}
}

 \if0\blind
{
 \begin{center}
 {\Large\bfseries A General Approach for Assessing Sensitivity to Violations of Causal Assumptions\par}
 \vspace{1.25em}
 Guilherme Duarte\par
 \vspace{0.75em}
 \begin{minipage}{0.75\textwidth}
 \centering
 \small
 Harvard University, Cambridge, Massachusetts 02138, USA
 (\href{mailto:gjduarte@fas.harvard.edu}{gjduarte@fas.harvard.edu}).
 \end{minipage}
 \end{center}
} \fi

\if1\blind
{\begin{center}
 {\LARGE\bf \bf }
\end{center}
 \medskip
} \fi

\thispagestyle{empty}

\begin{abstract}
Although causal sensitivity analyses are widely 
used to assess the robustness of conclusions, a key limitation 
is that  
existing methods are typically developed for specific designs. 
This paper develops a general sensitivity analysis framework 
that applies to a wide range of discrete causal models and estimands 
by deriving sharp bounds 
on causal estimands as a function of 
the maximum proportion of the population for which the assumption
is relaxed.
The framework is 
built on
the generalized principal strata approach, 
which allows the translation of assumptions, estimands, 
and data into  
polynomial programs.
In this approach, three classes of assumptions can be relaxed:
a) functional restrictions, 
which limit the ways that one variable can respond to another, e.g. 
strong monotonicity; b) exclusion restrictions, which rule out 
direct causal effects; c) unconfoundedness,
which rules out unobserved common causes of variables. 
The approach produces both sharp sensitivity curves
and robustness values, 
i.e., the smallest sensitivity level at which the sign of a causal
effect stops being identified.
The applicability of the method is demonstrated 
by generalizing existing proportion-of-violations approaches,
deriving new sensitivity analysis methods for instrumental variable 
and selection-on-observables designs, and reanalyzing
empirical results. 
\end{abstract}

\noindent%
\vfill
\thispagestyle{empty}
\newpage
\pagenumbering{arabic}
 \setcounter{page}{1}

\vspace{-.6cm}
\section{Introduction}
\label{sec:intro} 
\vspace{-.4cm}

In recent decades, the quality of scientific results in the social sciences has
improved due to a shift in focus toward better research designs and identification
strategies \citep{angrist2010credibility}. This shift, often referred to as the
``credibility revolution,'' emphasizes transparency in causal assumptions and
the use of domain expertise to justify them. Yet, every causal research design
relies on assumptions that cannot be fully verified from observed data alone.
For example, to study a voting-induced habit, i.e., the effect of turning out to
vote in one election on the propensity to vote in upcoming elections, one can
use randomized messages encouraging voters to turn out in the first election
as an instrumental variable \citep{coppock2016voting}. However, the validity of
the analysis depends on assumptions. One assumption, called 
``monotonicity,'' rules out ``defiers'':
voters who would vote without the encouragement but would not vote if
encouraged, e.g. those who might get annoyed and decide not to vote
in response to the encouragement.  Another assumption, the
 exclusion restriction, 
requires that 
the
encouragement affect voting in the second election only through its effect on
voting in the first one. This restriction would fail, for instance, if some
individuals remember the encouragement and vote in the second election due to it, 
independently of whether they voted in the first.

For many assumptions in applied research, debates about plausibility involve a
false dichotomy: either an assumption is defensible or it is not. In the voting
setting, either monotonicity holds or it does not, either the exclusion
restriction holds or it does not. Sensitivity analysis provides a way to move
beyond this binary perspective by evaluating how causal conclusions change
under relaxations of these assumptions. 
For example, one might ask how many
defiers or exclusion violators would be needed before the instrumental variables 
analysis no longer
supports the original conclusion.
However, despite the fact that these
 sensitivity questions arise naturally in applied research,
 existing sensitivity analysis approaches 
 are often specific to particular assumptions and research designs. 
While a large variety of methods in the literature  have been developed
to assess sensitivity to violations of unconfoundedness, comparable 
approaches remain unavailable for many other settings.

To address this gap, I propose a widely applicable approach that accommodates these
assumptions under a single unified framework and allows researchers to relax a
broad range of additional restrictions beyond unconfoundedness. The method
allows researchers to answer questions derived from the research problem above:
how would conclusions change if up to a specified percentage of voters defy
encouragement, or what is the smallest allowed percentage of exclusion
violations at which the original conclusion is no longer supported? The same
approach addresses questions about the robustness of causal conclusions to
violations of assumptions across different designs. The approach is developed
for discrete causal models, although it can be extended to continuous outcomes
through discretization and to many continuous covariates through modeling.

The approach proceeds as follows. A researcher specifies an initial causal
model and a relaxed model in which the assumption under evaluation is not
imposed. The sensitivity parameter $\theta\in[0,1]$ bounds the proportion of the
population for which the assumption is relaxed. Setting $\theta=0$ recovers the
initial model, while $\theta=1$ recovers the fully relaxed model. The sensitivity
analysis shows how the range of causal effects consistent with the data and
assumptions changes as $\theta$ increases. This allows researchers to report
not only how causal conclusions change with relaxations of an assumption but
also the sensitivity level at which the sign of the effect is no longer
determined. For discrete causal models, the proportion of units violating many common
assumptions can be written as a polynomial in the model parameters. Bounding
this polynomial by a sensitivity parameter turns the relaxation into a partial
identification problem in the framework of \citet{duarte2021automated}.
When the resulting program is globally optimized, its extrema are sharp for
the specified sensitivity model.

More specifically, this paper makes two main contributions.
First, Section~\ref{subsec:natural-sensitivity-analysis} 
introduces the natural sensitivity analysis framework, 
which formulates sensitivity analyses through restrictions 
on generalized principal strata and their probability distributions.
In particular,
relaxations of three types of causal assumptions are 
constructed (Section~\ref{sec:assumptiontype}). The first 
type involves violations of functional restrictions, which limit the possible ways that one
variable can respond to another, as does the monotonicity assumption described 
above. Other examples include linearity and interaction restrictions.
The second type involves violations of exclusion restrictions, which rule out
direct causal effects. The third type involves violations of unconfoundedness assumptions,
allowing unobserved common causes that the original model rules out. 
Because many causal assumptions can be understood as belonging to one of
these three classes, the framework provides a common language for comparing sensitivity
analyses across a wide range of research designs.
Second, the paper provides an algorithmic procedure for transforming a
sensitivity analysis problem into a partial identification problem 
(Section~\ref{sec:sensitivity-formal}).  Users can, therefore, describe 
their causal inference problem and the assumptions they wish to relax, and the procedure will automatically 
compute
sensitivity bounds for the estimand of interest and robustness values.

To show the applicability of the framework and 
as major contributions of this paper, in Section~\ref{sec:particular}, I 
derive new sensitivity analyses for two canonical 
research designs. First, I propose a sensitivity analysis method 
in terms of sharp bounds on Local Average Treatment Effect (LATE)
under maximum proportions of defiers or violations of the exclusion restriction, 
extending the results of \citet{richardson2010analysis} (Section~\ref{subsec:ivtheory}). 
Second, for binary treatment and outcome, 
I derive sharp bounds on the average treatment effect 
under a confounding mixture model with a 
bounded unrestricted component. 
This generalizes the approach of 
\citet{bonvini2022sensitivity} by removing 
additional independence restrictions on component membership 
(Section~\ref{subsec:classicconfounding}).
  Finally, Section~\ref{sec:empirical} illustrates 
these results through  empirical applications 
to habit formation in voter 
turnout \citep{coppock2016voting} and the effect of education on 
Brexit voting \citep{hobolt2016brexit}.

\section{Related Literature}
\label{sec:related}

This paper contributes to several strands of work on sensitivity analysis and 
causal inference. 
The closest precursor is the literature on sensitivity analysis 
under selection on observables, where researchers study how omitted 
confounding would have to behave to overturn a substantive conclusion. 
In that setting, a wide variety of methods have been proposed to 
assess whether conclusions are robust to assumption 
violations \citep[e.g.][]{cornfield1959smoking,
rosenbaum1983assessing, imbens2003sensitivity,tan2006distributional,
 blackwell2014selection,
vanderweele2017sensitivity, masten2018identification,
zhao2019sensitivity,
cinelli2020making, chernozhukov2021long,dorn2023sharp}. 
The paper also relates to a broader literature that 
studies sensitivity analysis outside the classic confounding setting, 
such as the many methods developed 
for instrumental variable designs \citep{small2007sensitivity, cinelli2025omitted}, 
mediation \citep{imai2010identification},
interference \citep{vanderweele2015interference}, external validity 
\citep{huang2024sensitivity}, 
and many other specific designs, estimands, and causal models.
One goal of the present paper is to 
replace that piecemeal approach with a unified framework that 
handles multiple assumption types using a common language.

More specifically, the proposed method is closely connected to a class of
approaches that I call \emph{natural sensitivity analyses}: methods that
index departures from an assumption by the proportion of units for which it
may fail. In assessing sensitivity to unconfoundedness, an early example 
is \citet{rosenbaum1987sensitivity}, which 
allows a specified fraction of matched pairs to be confounded, 
although the method is developed specifically for exact matching.
Recently, 
\citet{bonvini2022sensitivity} use a mixture model with a binary selector 
where a confounded distribution 
is selected with probability $\epsilon$ and a distribution satisfying 
unconfoundedness is selected with probability $1 - \epsilon$. 
In their main analysis, they derive 
sharp bounds on the average treatment effect as a function of $\epsilon$, but, 
for computational tractability reasons, under the additional assumption 
that the selection process is, conditionally on the observed covariates, independent of both 
 treatment and outcome. This paper uses its general framework to derive 
 new sharp bounds on the ATE for binary treatment and outcome, 
 without this additional assumption.
For instrumental variable models, 
\citet{richardson2010analysis} derive bounds on average 
causal effects for specified proportions of defiers 
while maintaining the exclusion restriction. 
This paper recovers the corresponding monotonicity bounds 
within a general optimization framework and obtains 
sensitivity bounds by allowing the defier proportion 
to vary up to a specified maximum. Section~\ref{subsec:ivtheory} 
also derives bounds under restrictions on the proportion of 
units violating the exclusion restriction.
  More recently, \citet{noack2021sensitivity} explicitly developed a sensitivity analysis 
 for violations of monotonicity that requires an additional 
 parameter measuring the discrepancy between the potential-outcome 
 distributions of 
 compliers and defiers. 
 Nevertheless,
the sensitivity models underlying these approaches can be derived 
 as special cases of the general approach proposed here.

 Finally, this paper relates to automated causal 
 identification \citep{shpitser2006identification,shpitser2023proximal,
 lee2020general,lee2024general} 
 and partial identification \citep{sachs2023general}. 
 It builds directly on \citet{duarte2021automated} by 
 extending their framework to automate sensitivity analysis.

\section{Preliminaries}
\label{sec:preliminaries}

\subsection{Notation}

I use uppercase letters for random variables, 
lowercase letters for their realizations, and
 $\Pr(\cdot)$ for probability. 
The potential outcome $Y(a)$ denotes the outcome 
under an intervention fixing $(A=a)$, and more generally,
 $Y(a_S)$ denotes the potential outcome 
when a subset $S$ of variables is intervened 
upon and set to $a_S$.  
I adopt the standard ``stable unit treatment value assumption'' \citep{imbens2015causal}.
The framework applies to categorical variables, but 
continuous outcomes can be approximated  through discretization. For exposition, 
I use a binary instrumental variable setting,
 where $Z \in \{z_0, z_1\}$ is an instrument, $A \in \{a_0, a_1\}$ is a 
treatment, and $Y \in \{y_0, y_1\}$ is the outcome. 
From 
causal identification theory, reviewed in  Appendix~\ref{subsection-id}, 
a model $\mathcal{M}$ 
represents a set of possible data generating processes (DGPs) over the observed 
variables $V={V_1,\dots,V_I}$ and the relevant potential outcomes, 
together with a collection of 
assumed constraints $\mathcal{C}={C_\lambda:\lambda\in\Lambda}$ on these distributions, 
and an estimand $\mathcal{E}$ represents a quantity of interest as a function of the DGP.
The goal of the identification is  to characterize all possible values 
of  $\mathcal{E}$ among DGPs of $\mathcal{M}$ that are compatible 
with an observed data distribution.

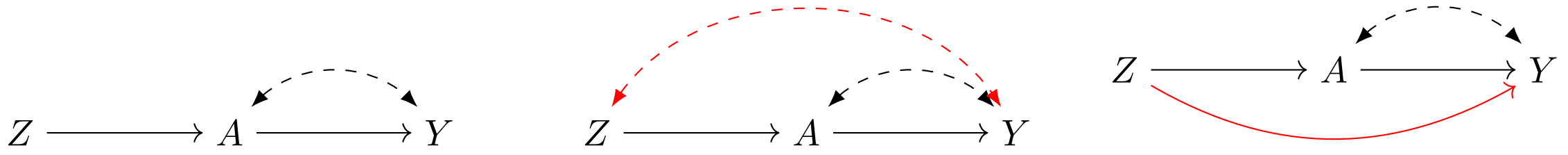
\begin{figure}[htp]
\footnotesize
\centering
\caption{\textbf{IV data generating process.}%
 \quad 
Panel (a) shows the standard IV graph. Panels (b) and (c)
add edges that permit violations of random assignment and
the exclusion restriction, respectively. Monotonicity is
imposed separately as a functional restriction.}
    \begin{subfigure}[b]{0.33\textwidth}
        \centering
 \begin{tikzpicture}
 \node (Z) at (0,0) {$Z$};
 \node (A) at (2,0) {$A$};
 \node (Y) at (4,0) {$Y$};
 \path (Z) edge [->] node [midway, below=-.25ex] {\scriptsize $ $} (A);
 \path (A) edge [->](Y);
 \path[bidirected] (A) edge[bend left=50] node[below] {$ $}  (Y);
 \end{tikzpicture}
 \subcaption{Standard IV}\label{fig:IV_DAG1}
 \end{subfigure}\hfill
    \begin{subfigure}[b]{0.33\textwidth}
        \centering \begin{tikzpicture}
 \node (Z) at (0,0) {$Z$};
 \node (A) at (2,0) {$A$};
 \node (Y) at (4,0) {$Y$};
 \path (Z) edge [->] node [midway, below=-.25ex] {\scriptsize $ $} (A);
 \path (A) edge [->](Y);
 \path[bidirected] (A) edge[bend left=50] node[below] {$ $}  (Y);
 \path[bidirected] (Z) edge[color=red,bend left=60] node[below] {$ $}  (Y);
 \end{tikzpicture}
 \subcaption{Randomization violation}\label{fig:IV_DAG2}
 \end{subfigure}\hfill
    \begin{subfigure}[b]{0.33\textwidth}
  \begin{tikzpicture}
 \node (Z) at (0,0) {$Z$};
 \node (A) at (2,0) {$A$};
 \node (Y) at (4,0) {$Y$};
 \path (Z) edge [->] node [midway, below=-.25ex] {\scriptsize $ $} (A);
 \path (A) edge [->](Y);
 \path[bidirected] (A) edge[bend left=50] node[below] {$ $}  (Y);
  \path (Z) edge[color=red, bend right=30] [->](Y);
 \end{tikzpicture}
 \subcaption{Exclusion restr. violation}\label{fig:IV_DAG3}
  \end{subfigure}\hfill
\label{fig:IV_DAG}
\end{figure}

Graphs are also used to encode sets of causal assumptions in our analysis. 
Specifically, I use acyclic directed mixed graphs (ADMGs), 
which consist of vertices $V$, directed edges ($E_d \subset V \times V$), and bidirected edges
($E_b \subset V \times V$), with no directed cycles \citep{richardson2017nested}.
Under a structural causal model associated with the graph, the potential outcome $Y(a_S)$ is 
the value of $Y$ induced after all arrows into the variables in $S \subset V$ 
are removed and those variables are externally set to $a_S$ \citep{pearl2009causality}.
Figure~\ref{fig:IV_DAG} illustrates an ADMG for the instrumental variable setting. 
The bidirected edge between $A$ and $Y$ indicates that both treatment and outcome are confounded by 
unobserved variables. Because $Z$ also has no parents, the absence of bidirected edges between  
$Z$ and $Y$ and between $Z$ and $A$ represents the assumption of random 
assignment, while the absence of a direct arrow from $Z$ to $Y$ represents the 
exclusion restriction. The typical ``no defier'' assumption  has no direct graphical representation and 
is separately notated as $\Pr(A(z_1) = a_0, A(z_0) = a_1) = 0$.

The sensitivity analysis theory developed here builds on
the generalized principal strata framework of
\citet{duarte2021automated}, which parameterizes the joint
distribution of potential outcomes while allowing for
unobserved confounding. The procedure called
``canonicalization'' constructs the generalized principal
strata for a model and provides the basis for the global
optimization problems used below. This procedure is reviewed
in Appendix~\ref{subsubsec:autobounds}.

\section{Theory for Natural Sensitivity Analyses}
\label{sec:sensitivity}

In this section, I show that sensitivity analysis of causal assumptions can be 
formulated as a causal identification problem. Specifically, I study how the 
identified set for a causal estimand $\mathcal{E}$ changes when some of the 
constraints in $\mathcal{C}$ are relaxed. A sensitivity analysis problem is 
therefore defined by a causal model, an observed data distribution, an estimand, 
and sensitivity parameters that govern the permitted relaxations of the original 
assumptions.

This formalization has two advantages. First, the same machinery developed for 
automated point and partial identification can be used to compute solutions to 
sensitivity analysis problems \citep{duarte2021automated, sachs2023general}.
I adopt this perspective below, showing how assumption relaxations can be displayed
using causal graphs and interpretable parameters and how the resulting problems can be
solved automatically. Second, the formalization provides a common language for comparing 
sensitivity analysis methods and clarifying how their assumptions and results differ.

\subsection{Natural Sensitivity Analysis Problem}
\label{subsec:natural-sensitivity-analysis}

The framework assesses sensitivity by changing the proportion of the
population for which an assumption is relaxed.
The main idea is to report how sharp bounds for an estimand
$\mathcal{E}$ vary as the permitted proportion increases. For instance, in the 
binary IV example, instead of imposing the ``no defiers'' assumption and 
obtaining a point identified LATE, one can ask what the identified set would be if 
defiers made up at most 1\%, 5\%, or 10\% of the population. 

A sensitivity analysis begins with a causal model $\mathcal{M}$, an observed data 
distribution $P_V$, and an estimand $\mathcal{E}$. Together, the triple 
$(\mathcal{M},P_V,\mathcal{E})$ defines the initial causal inference problem. The model 
$\mathcal{M}$ contains the DGPs over variables $V=\{V_1,\dots,V_I\}$ that satisfy the 
assumed constraints $\mathcal{C}=\{C_\lambda:\lambda\in\Lambda\}$. Compatibility with 
$P_V$ further restricts this model to DGPs that imply the observed data distribution. 
In the running binary IV example, the constraints include no defiers 
($C_{\mathrm{ND}}$), exclusion of a direct effect of the instrument on the outcome 
($C_{\mathrm{ER}}$), and random assignment of the instrument 
($C_{\mathrm{UC}}$), as described in Section~\ref{sec:preliminaries}.

The next step is to select a subset $\Lambda_R\subseteq\Lambda$ of assumptions to relax. 
For example, 
standard sensitivity methods may replace unconfoundedness with restrictions on the 
association between treatment $A$ and an unobserved confounder $U$, and between $U$ and 
the outcome, expressed using risk ratios \citep{vanderweele2017sensitivity} or partial 
$R^2$  \citep{cinelli2020making}. \red{The natural sensitivity analysis considered here instead requires,
for every $\lambda\in\Lambda_R$, an event $\mathcal V_\lambda$ whose
probability under DGP $P$ gives the proportion of the population
for which $C_\lambda$ is relaxed.} A sensitivity parameter $\theta_\lambda^{\max}$ then places the upper bound
$P(\mathcal V_\lambda)\leq\theta_\lambda^{\max}$.

Following the terminology introduced in
Section~\ref{sec:related}, I refer to this class of methods as 
\emph{natural sensitivity analysis}, because the 
sensitivity parameter has a direct population interpretation:
it bounds the proportion of the population for which the assumption
is relaxed.

\begin{definition}[Natural Sensitivity Problem]
\label{def:natural-sensitivity-problem}
Let $(\mathcal{M},P_V,\mathcal{E})$ be an initial causal inference problem with 
constraints $\mathcal{C}=\{C_\lambda:\lambda\in\Lambda\}$, and let 
$\Lambda_R\subseteq\Lambda$ index the constraints to be relaxed. Let 
$\mathcal{M}_R$ be the corresponding relaxed model, with all remaining
constraints retained. Let $\mathcal V_\lambda$ denote the event
identifying the part of the population for which $C_\lambda$ is relaxed.
For
$P\in\mathcal M_R$, its probability $P(\mathcal V_\lambda)\in[0,1]$
is the corresponding population proportion.
For a vector of upper bounds
$\Theta^{\max}=\{\theta_\lambda^{\max}:\lambda\in\Lambda_R\}$, define
$$
\mathcal{M}(\Theta^{\max})
=
\left\{
P\in\mathcal{M}_R:
P(\mathcal V_\lambda)\leq\theta_\lambda^{\max}
\text{ for every }\lambda\in\Lambda_R
\right\}.
$$
A natural sensitivity problem is the family of causal inference problems, 
$\bigl(\mathcal M(\Theta^{\max}),P_V,\mathcal E\bigr).$
\end{definition}

Notice that a natural sensitivity problem has 
 corresponding identified sets 
$\mathcal{I}_{\mathcal{E}}(\Theta^{\max})$ for each $\Theta^{\max} \in [0,1]^{|\Lambda_R|}$.
It characterizes how
$\mathcal{I}_{\mathcal{E}}(\Theta^{\max})$ changes as the allowed violation 
proportions in $\Theta^{\max}$ vary. 
The distinction between $P(\mathcal V_\lambda)$ and $\theta_\lambda^{\max}$ is relevant. 
The former is the exact proportion violated, while the 
latter is the largest proportion allowed by the sensitivity analysis. For example, 
$\theta_{\mathrm{ND}}^{\max}=0.05$ allows the proportion of defiers to take any value 
between 0\% and 5\%. It does not assert that exactly 5\% of the population consists of 
defiers. Varying $\Theta^{\max}$ traces how the identified set changes as 
larger violations are admitted. Later, I also show that two or 
more assumptions can be relaxed simultaneously without changing the method.

The natural sensitivity analysis problem also yields a quantity of particular interest: 
the \emph{robustness value}.
It is the sensitivity level at which the sign of the effect stops
being identified.

\begin{definition}[Robustness Value]
\label{def:robustness-value}
Let $\mathcal I_\lambda(\theta)$ be the identified set obtained
from the procedure in Definition~\ref{def:natural-sensitivity-problem}, and
assume $\mathcal I_\lambda(0)$ is nonempty.
The sign is robust at $\theta$ if the identified set is entirely positive
or entirely negative. The robustness value $\theta_\lambda^\star$ is the
\red{smallest sensitivity level} at which this sign conclusion
fails, taking the infimum if there is no smallest value. It is zero if
the sign is not identified at $\theta=0$. By convention, it is one
both when the first sign failure occurs at $\theta=1$ and when the sign
remains identified for every $\theta\in[0,1]$.
\end{definition}

\subsection{Representing and Interpreting Assumption Violations}
\label{sec:assumptiontype}

As discussed, natural sensitivity analysis uses parameters
that bound the proportion of units violating an assumption. I distinguish
three types of assumptions: functional assumptions, exclusion restrictions,
and unconfoundedness assumptions. Each has a transparent graphical display.
The sensitivity variables introduced below are additional binary nodes in the
augmented graph.

First, \emph{functional assumptions} correspond to constraints on how one variable
affects another.
If a model assumes that encouragement $Z$ causes 
treatment $A$ but restricts the 
form of this causal function, 
this constitutes a functional constraint. The ``no defier'' assumption discussed
above is a unit level functional constraint: among the possible responses of treatment
$A$ to instrument $Z$, it rules out defiers while permitting compliers, never takers,
and always takers. Functional constraints can also be stated at the population level.
For example, the additive causal mean restriction
$\E[Y(z,a)]=\alpha+\beta a+\gamma z$ rules out an interaction in the mean potential
outcome surface.
Both individual-level and population-level
functional restrictions are representable in the polynomial program.
Next, 
consider \emph{exclusion restriction} assumptions, which
rule out the presence of a direct causal influence 
from a variable $X$ to another variable $Y$ that is not
mediated by other variables. In graphical language, this 
is represented by the absence of an $X \to Y$ arrow. 
Finally, let $C$ collect the observed covariates on which the analysis
conditions, and consider unconfoundedness between treatment $A$ and its
vector of potential outcomes, $\mathbf Y_A=(Y(a):a\in\mathcal S(A))$.
Unconfoundedness means $A\indep\mathbf Y_A\mid C$.
Appendix~\ref{app:assumption-examples} gives examples of how these
restrictions and their relaxations enter the polynomial program.

To handle assumption relaxations, I introduce a new type of graphical 
representation, called a \emph{sensitivity graph},
which augments the original causal graph with additional nodes and edges 
that encode the relaxation of the assumptions and is formalized
 in Definition~\ref{def:sensitivity-graph} in
Appendix~\ref{app:sensitivity-graphs}. For each
relaxation, a sensitivity node pointing to the location of the
restriction is added, as illustrated in Table~\ref{fig:typesviolation}.

\begin{table}[htp]
\centering
\caption{{\textbf{Assumption relaxations in the IV model.}
The binary sensitivity nodes indicate where restrictions are relaxed:
monotonicity at $A$, exclusion at $Y$, and exogeneity through dependence
between $Z$ and the outcome response. Their probabilities are bounded
by the corresponding sensitivity parameters. The $+$ sign denotes
monotonicity in the original model.}}
\begin{tabular}{|c|c|>{\raggedright\arraybackslash}m{4cm}|}
    \hline
    \textbf{Assumption} & \textbf{Original Model} &
    \textbf{Sensitivity Graph} \\
    \hline
     Functional & 
    \begin{tikzpicture}[scale=0.75]
 \node (Z) at (0,0) {$Z$};
 \node (A) at (2,0) {$A$};
 \node (Y) at (4,0) {$Y$};
\path (Z) edge [->] node [midway, above=.25ex] {\scriptsize $ + $} (A);
 \path (A) edge[->] (Y);
 \path[bidirected] (A) edge[bend left=50] node[below] {$ $}  (Y);
 \end{tikzpicture} & 
     \begin{tikzpicture}[scale=0.75]
 \node (Z) at (0,0) {$Z$};
 \node (A) at (2,0) {$A$};
 \node (Y) at (4,0) {$Y$};
 \node[draw,circle,inner sep=1.5pt] (s1) at (2, -1) {$S_{\text{ND}}$};
\path (Z) edge [->] node [midway, below=-.25ex] {\scriptsize $ $} (A);
 \path (A) edge [->](Y);
 \path[bidirected] (A) edge[bend left=50] node[below] {$ $}  (Y);
  \path (s1) edge[->] (A);
 \end{tikzpicture}  \\
    \hline
     Exclusion Restriction & 
    \begin{tikzpicture}[scale=0.75]
 \node (Z) at (0,0) {$Z$};
 \node (A) at (2,0) {$A$};
 \node (Y) at (4,0) {$Y$};
\path (Z) edge [->] node [midway, below=-.25ex] {\scriptsize $ $} (A);
 \path (A) edge [->](Y);
 \path[bidirected] (A) edge[bend left=50] node[below] {$ $}  (Y);
 \end{tikzpicture} & 
     \begin{tikzpicture}[scale=0.75]
 \node (Z) at (0,0) {$Z$};
 \node (A) at (2,0) {$A$};
 \node (Y) at (4,0) {$Y$};
 \node[draw,circle,inner sep=1.5pt] (s2) at (2, -1.5) {$S_{\text{ER}}$};
\path (Z) edge [->] node [midway, below=-.25ex] {\scriptsize $ $} (A);
 \path (A) edge [->](Y);
 \path[bidirected] (A) edge[bend left=50] node[below] {$ $}  (Y);
  \path (Z) edge[bend right=30][->] (Y);
  \path (s2) edge[->] (Y);
 \end{tikzpicture}
     \\
    \hline
     Exogeneity &
      \begin{tikzpicture}[scale=0.75]
 \node (Z) at (0,0) {$Z$};
 \node (A) at (2,0) {$A$};
 \node (Y) at (4,0) {$Y$};
\path (Z) edge [->] node [midway, below=-.25ex] {\scriptsize $ $} (A);
 \path (A) edge [->](Y);
 \path[bidirected] (A) edge[bend left=50] node[below] {$ $}  (Y);
 \end{tikzpicture} & 
         \begin{tikzpicture}[scale=0.75]
 \node (Z) at (0,0) {$Z$};
 \node (A) at (2,0) {$A$};
 \node (Y) at (4,0) {$Y$};
 \node[draw,circle,inner sep=1.5pt] (s3) at (2, -1.5) {$S_{\text{UC}}$};
\path (Z) edge [->] node [midway, below=-.25ex] {\scriptsize $ $} (A);
 \path (A) edge [->](Y);
 \path[bidirected] (A) edge[bend left=50] node[below] {$ $}  (Y);
  \path[bidirected] (Z) edge[bend right=30] (Y);
  \path (s3) edge[->] (Z);
  \path (s3) edge[->] (Y);
 \end{tikzpicture}
   \\
    \hline
\end{tabular}
\label{fig:typesviolation}
\end{table}

\subsection{Formalizing the relaxations and the problem}
\label{sec:sensitivity-formal}

\subsubsection{Perturbation and sensitivity functions}

As we have defined, $\Lambda_R\subseteq\Lambda$ 
indexes the relaxed assumptions. Let
$K_R=|\Lambda_R|$, and for every $\lambda\in\Lambda_R$ let
$\mathcal V_\lambda$ be its specified violation event. Let
$\Theta=(\theta_\lambda:\lambda\in\Lambda_R)\in[0,1]^{K_R}$ be the vector of
realized violation proportions.  
The causal inference assumptions studied here
take the form $P(\mathcal V_\lambda)=0$. For example, the ``no defiers''
assumption is $P(A(z_0)=a_1,A(z_1)=a_0)=0$. Replacing 
$0$ with $\theta_{\mathrm{ND}}$, we have an exact 
perturbation, fixing the proportion of defiers. 
This construction can be repeated for each $\lambda\in\Lambda_R$.
In fact, even more complex constraints, such as 
exclusion restriction and unconfoundedness, can be represented in this way, as
will be shown.

For the vector $\Theta$, let $\mathcal M_{=}(\Theta;P_V)$ contain the DGPs that
remain compatible with the observed distribution and stated assumptions and satisfy
$P(\mathcal V_\lambda)=\theta_\lambda$ for every $\lambda \in \Lambda_R$.
 Solving the partial identification 
problem associated with this model gives the
range of estimand values under the violation proportions. This range defines a
perturbation function.

\begin{definition}[Perturbation function]
\label{def:perturbation}
    The perturbation function
    $f:[0,1]^{K_R}\to 2^{\mathrm{Codomain}(\mathcal E)}$ is
    \begin{align*}
    f(\Theta)=\{\mathcal E(P):P\in\mathcal M_{=}(\Theta;P_V)\}.
    \end{align*}
    For a scalar estimand, write $L(\Theta)=\inf f(\Theta)$ and
    $U(\Theta)=\sup f(\Theta)$ whenever $f(\Theta)$ is nonempty. If no compatible DGP
    has the exact violation vector $\Theta$, then $f(\Theta)=\varnothing$.
\end{definition}

The perturbation function evaluates conclusions at specified violation proportions. For
instance, suppose that the other IV assumptions are valid and exactly 20\% of units
are defiers. Then $\Theta=(0.2)$ and the modified constraint is
$P(A(z_0)=a_1,A(z_1)=a_0)=0.2$. The set $f(0.2)$ gives the admissible LATE values for
this exact perturbation.

One challenge with this approach is that researchers often lack precise information
about the exact proportion of violations. Consequently, the perturbation function may
be most useful for checking how solutions change as $\Theta$ varies. Researchers may be
more comfortable specifying plausible upper bounds. In the IV example, one can ask how
the results change when the proportion of defiers is at most $0.2$ rather than exactly
$0.2$. These cases yield sensitivity analysis functions.

\begin{definition}[Sensitivity analysis function]
\label{def:sensitivity}
    Let $\Theta^{\max}\in[0,1]^{K_R}$ be a vector of upper bounds for the
    $K_R$ violation proportions. The sensitivity analysis function
    $g:[0,1]^{K_R}\to 2^{\mathrm{Codomain}(\mathcal E)}$ is
    $$g(\Theta^{\max})
    =\bigcup_{\Theta\leq\Theta^{\max}}f(\Theta)
    =\mathcal I_{\mathcal E}(\Theta^{\max}),$$
    where the vector inequality is componentwise. For a scalar estimand, its sharp lower
    and upper bounds are
    $\inf_{\Theta\leq\Theta^{\max}}L(\Theta)$ and
    $\sup_{\Theta\leq\Theta^{\max}}U(\Theta)$ over feasible exact perturbations.
\end{definition}

\subsubsection{Individual-level relaxations}

I first consider individual-level constraints, i.e. 
those that eliminate a subset of generalized principal strata by 
setting their probability to $0$, as in Definition~\ref{def:individual-constraint}.

\begin{definition}[Individual-level constraint]
\label{def:individual-constraint}
A constraint $C_\lambda$ is \emph{individual-level} if it requires
$P(\mathcal V_\lambda)=0$, where $\mathcal V_\lambda$ is a fixed union
of generalized principal stratum events.
\end{definition}

For a natural sensitivity analysis, 
these constraints can be relaxed by allowing
$P(\mathcal V_\lambda)\leq\theta_\lambda^{\max}$.
The following theorem validates this procedure.
It shows that this analysis becomes a polynomial program 
by changing a constraint in the original causal problem. 
As the principal strata that violate an assumption are specified, 
their population proportion is obtained by summing their probabilities.
Setting this proportion equal to $\theta$ gives the 
perturbation problem, while bounding it by $\theta^{\max}$ 
gives the sensitivity problem. Solving the latter globally yields 
the sharp sensitivity curves.
The data constraints, 
the other assumptions, and the estimand remain unchanged. 
The proof is in Appendix~\ref{app:functional-autobounds}.

\begin{theorem}[Individual-level sensitivity]
\label{thm:functional-autobounds}
Let $C_\lambda$ be an individual-level constraint 
as in Definition~\ref{def:individual-constraint}.
 In the finite canonical model, define
$I_\lambda(\bu)=\bbone\{\{\bU=\bu\}\subseteq\mathcal V_\lambda\}$.
Let $\boldsymbol\eta$ be a vector of all the generalized principal strata 
probabilities
and any auxiliary variables. Let $\mathcal H_0(P_V)$ contain the
canonicalized parameter vectors
that reproduce $P_V$, satisfy the axioms of probability
and all retained restrictions in canonicalization,
including auxiliary-variable constraints but excluding $C_\lambda$,
and leave $\mathcal E(\boldsymbol\eta)$ defined. Then:
\begin{enumerate}[label=(\roman*)]
\item The event probability is the polynomial
\begin{equation}
\label{eq:functional-violation}
P_{\boldsymbol\eta}(\mathcal V_\lambda)
=\sum_{\bu\in\mathcal S(\bU)}I_\lambda(\bu)\prod_{k=1}^Kp_{k,u_k}.
\end{equation}
Here $P_{\boldsymbol\eta}$ denotes the distribution determined by $\boldsymbol\eta$.

\item For $\theta,\theta^{\max}\in[0,1]$, define
\begin{align*}
\mathcal H_{\lambda,=}(\theta;P_V)
&=\{\boldsymbol\eta\in\mathcal H_0(P_V):P_{\boldsymbol\eta}(\mathcal V_\lambda)=\theta\},\\
\mathcal H_{\lambda,\leq}(\theta^{\max};P_V)
&=\{\boldsymbol\eta\in\mathcal H_0(P_V):P_{\boldsymbol\eta}(\mathcal V_\lambda)\leq\theta^{\max}\}.
\end{align*}
These programs represent the exact perturbation problem and the
natural sensitivity problem of
Definition~\ref{def:natural-sensitivity-problem}, respectively.
Their images under $\mathcal E$ are $f(\theta)$ and
$g(\theta^{\max})$, with
\begin{align*}
g(\theta^{\max})=\bigcup_{0\leq\theta\leq\theta^{\max}}f(\theta).
\end{align*}

For each feasible problem, the infimum and supremum of the scalar estimand
over the corresponding polynomial program give its sharp bounds.

\end{enumerate}
\end{theorem}

The theorem has some immediate consequences. The first one  
is that allowing a larger maximum proportion of
violations cannot make the identified set smaller.

\begin{corollary}[Sensitivity Function Monotonicity]
\label{cor:sensitivity-monotonicity}
For any $0\leq\theta_1^{\max}\leq\theta_2^{\max}\leq1$,
$$g(\theta_1^{\max})\subseteq g(\theta_2^{\max}).$$
Consequently, wherever the sensitivity sets are nonempty, the sharp lower
endpoint is nonincreasing and the sharp upper endpoint is nondecreasing:
$$L^{\mathrm{sens}}(\theta_2^{\max})
\leq L^{\mathrm{sens}}(\theta_1^{\max}),
\qquad
U^{\mathrm{sens}}(\theta_2^{\max})
\geq U^{\mathrm{sens}}(\theta_1^{\max}).$$
More generally, for several sensitivity parameters,
$\Theta_1^{\max}\leq\Theta_2^{\max}$ componentwise implies
$g(\Theta_1^{\max})\subseteq g(\Theta_2^{\max})$. This monotonicity does not hold for the
exact perturbation function $f(\theta)$, because models with different exact
violation proportions are not in general nested.
\end{corollary}

Another relevant consequence is that this formulation admits 
a representation of the sensitivity problem as a local mixture model, 
in which the initial constrained model is chosen with probability 
$1 - \theta^{\max}$, while the relaxed model, in which the assumption is removed, 
with probability $\theta^{\max}$. This representation is especially 
relevant  because 
it suggests some mixture model sensitivity analyses, which 
were proposed in the 
literature, are a consequence of this construction.

\begin{corollary}[Mixture representation]
\label{cor:local-mixture}
Suppose $I_\lambda(\bu)$ depends only on one canonical block $U_j$. Let
$\mathcal S_j^{\mathrm{res}}$ contain the strata in which the restriction
holds, and suppose this set is nonempty. A principal strata distribution
satisfies $P_{\boldsymbol\eta}(\mathcal V_\lambda)\leq\theta^{\max}$ if and only if it can be written as
$$P_{U_j}=(1-\theta^{\max})P_{U_j}^{\mathrm{res}}
+\theta^{\max}P_{U_j}^{\mathrm{rel}},$$
where $P_{U_j}^{\mathrm{res}}$ is supported on
$\mathcal S_j^{\mathrm{res}}$ and $P_{U_j}^{\mathrm{rel}}$ is unrestricted on
$\mathcal S(U_j)$.
Substituting the combined principal strata distribution into every
original polynomial constraint gives an equivalent parameterization of the
same causal model.
\end{corollary}

The theorem applies only 
to assumptions expressed as individual-level constraints.  
This naturally includes 
 exclusion restrictions, as formalized in Lemma~\ref{cor:exclusion-autobounds}.
 The key idea is that 
 an exclusion restriction becomes a functional 
restriction once
the omitted arrow is added: the outcome principal strata must be invariant
to the added parent while its other inputs are held fixed. For example, 
in the binary IV example, the exclusion restriction can be 
represented either by the graph in Figure~\ref{fig:IV_DAG1}, which omits the arrow 
$Z \to Y$, or by 
the graph in Figure~\ref{fig:IV_DAG3}, which includes that arrow but 
imposes the individual-level assumption that 
$P(Y(z_1, a)=y , Y(z_0, a)=y') = 0$ for any $a$ and every pair $y \neq y'$.

\subsubsection{Population-level relaxations}
\label{sec:population-level}

In this subsection, I show how the results from the previous subsection
extend naturally to population-level assumptions, including unconfoundedness and
many functional constraints.

\begin{definition}[Population-level constraint]
\label{def:population-constraint}
A constraint $C_\lambda$ is \emph{population-level} if it requires
$h_\lambda(\boldsymbol\eta)=0$, $h_\lambda(\boldsymbol\eta){\color{black}\leq}0$ or $h_\lambda(\boldsymbol\eta){\color{black}\geq}0$, where $h_\lambda$ is a polynomial
function
depending on the strata probability  $\boldsymbol\eta$.
\end{definition}

By this definition, 
population-level constraints subsume individual-level ones, but do not need
to require the probability of a particular {\color{black} stratum} to be $0$.
For example,
population-level monotonicity, 
$\E[A(z_1)]\geq \E[A(z_0)]$, requires the proportion of compliers
{\color{black} to be at least as large as} that of defiers, while individual-level monotonicity excludes
defiers entirely. The population-level restriction can therefore hold
even when defiers are present, and its violation cannot in general be 
assigned to particular units. The current framework, therefore, does not allow 
a direct extension of the treatment of 
 individual-level to population-level constraints, but this can be  
circumvented through an additional construction.

To relax a particular population-level 
constraint, introduce a binary component indicator $H$.
Impose $C_\lambda$ on the response distribution conditional on $H=0$,
when that component has positive probability, and leave the $H=1$
component unrestricted by $C_\lambda$. The natural sensitivity parameter
bounds its weight, $P(H=1)\leq\theta_\lambda^{\max}$.
The data and all retained model restrictions apply to the recovered
population distribution. This extends the natural sensitivity framework
through an augmented model: $P(H=1)$ is a component proportion and need
not equal the probability of a fixed event violating $C_\lambda$.
Definition~\ref{def:augmented-model} and
Lemma~\ref{lem:unrestricted-augmentation} in
Appendix~\ref{app:augmented-graphs} give the formal construction and
establish invariance under unrestricted augmentation.

In this sense,
the enlargement is only a refinement of the generalized principal strata
parameterization, so that solving a problem without 
imposing any restriction on the
auxiliary variable yields the same result as solving the corresponding
problem without introducing that variable. Restrictions
on $H$, such as the componentwise independence 
condition imposed below, are
what can change the identified set. 
An important consequence of this augmentation is that
assumptions that were population-level under the original model admit 
individual-level representation under the augmented one. Therefore, 
all the results from Theorem~\ref{thm:functional-autobounds}, including 
its corollaries, apply 
directly to population-level constraints as well. This is stated in the following 
Lemma~\ref{lem:augmented-individual-representation}, and the formal proof
is  in Appendix~\ref{app:augmented-individual-representation}. 

\begin{lemma}[Individual-level representation under augmentation]
\label{lem:augmented-individual-representation}
Assume that $C_\lambda$ holds within the $H=0$
component of the augmented model.
Then imposing $P(H=1)=0$ recovers exactly the original model
that satisfies $C_\lambda$, by projection onto the original variables.
Therefore, relaxing this individual-level restriction to
$P(H=1)\leq\theta_\lambda^{\max}$ defines a sensitivity problem
to which Theorem~\ref{thm:functional-autobounds} and its corollaries
apply under their stated conditions.
\end{lemma}

Under the augmented model, unconfoundedness can therefore
 be represented in terms of a polynomial equality.
Let $A$ and $Y$ be finite-valued and write
$\mathbf Y_A=(Y(a):a\in\mathcal S(A))$. 
For the application below, take the selected district to contain $A$ and
$Y$. Each generalized principal stratum
$a_i\mathbf y_{\br}$ is thereby divided into two augmented strata,
$h_0a_i\mathbf y_{\br} \text{ and } h_1a_i\mathbf y_{\br}.$
The $H=0$ cells form the component in which unconfoundedness is imposed,
whereas the $H=1$ cells form an unrestricted component. Let
$q_{hi\br}=P(h_ha_i\mathbf y_{\br})$ denote 
 the probabilities of the augmented strata. The original
principal-stratum probabilities are recovered by marginalizing $H$:
$P(a_i\mathbf y_{\br})=q_{0i\br}+q_{1i\br}.$
The probability of the unrestricted component is then
$P(H=1)=\sum_{i,\br}q_{1i\br}.$ This assumption is studied  
in Appendix~\ref{app:augmented-graphs}, which includes its formalization in
Lemma~\ref{lem:unconfounded-mixture}.
For binary and confounded $A$ and $Y$, the augmented strata are
$h_sa_i y_{jk}$ for $s,i,j,k\in\{0,1\}$. Thus, for example, the original
stratum $a_0y_{01}$ is divided into $h_0a_0y_{01}$ and
$h_1a_0y_{01}$.

As a consequence of both Corollary~\ref{cor:local-mixture} and
Lemma~\ref{lem:augmented-individual-representation},
the representation of the sensitivity model as a mixture of two different
models where the first one that carries the constraint is chosen 
with probability $1 - \theta$ and the second one that relaxes that constraint with probability $\theta$
extends as well to population-level assumptions. 
Therefore, 
the mixture model used by
\citet{bonvini2022sensitivity}, discussed further in
Section~\ref{subsec:classicconfounding}, is a special case of this construction.

\subsection{Solving the sensitivity analysis problem}

To solve the optimization problems arising from natural sensitivity analysis, 
there are two main computational approaches.  The first one generates numerical solutions through 
branch-and-bound algorithms, employed in packages such as SCIP  \citep{hojny2025scip}.
The algorithm \texttt{autobounds} introduced by
\citet{duarte2021automated} can formulate these problems for numerical optimization and, in fact, 
 the exact perturbation function need not be derived first, as
the constraints $P(\mathcal V_\lambda)\leq\theta_\lambda^{\max}$ can be included directly in
the optimization problem defining $g(\Theta^{\max})$.
The numerical sensitivity curve is obtained by repeating this optimization over values
of $\Theta^{\max}$. 
The second approach involves obtaining closed-form solutions for some of those problems. 
One method uses packages for vertex enumeration, such as cddlib \citep{fukuda2021cddlib}, also used 
in  \texttt{autobounds} to solve causal problems that are linear programs. 
Finally, closed-form solutions can be derived analytically using standard mathematical techniques.

\subsection{Statistical Inference}

The estimated sensitivity bounds also carry sampling uncertainty.
In particular, the challenges related to doing inference on
bounds are related to the fact that those quantities are often
not Hadamard differentiable, so there are difficulties
in deriving their asymptotic distribution. To address these issues,
under the regularity conditions in
Appendix~\ref{app:statistical-inference}, recentered subsampling provides
separate pointwise one-sided confidence limits for the sensitivity bounds
and a one-sided lower confidence bound for the robustness value.
Details on potential issues and on the validity, consistency, and coverage of
the confidence sets as well as sufficient
regularity conditions are discussed in
Appendix~\ref{app:statistical-inference}.
Among the conditions, continuity and the existence
of a limiting probability distribution are required.

\section{Sensitivity Analysis in Standard Scenarios}
\label{sec:particular}

I next apply the approach to 
derive sensitivity curves for violations of exclusion  and ``no defiers'' in  instrumental variables and 
 unconfoundedness in selection-on-observables. 
Related results for other assumptions are developed
in \citet{duarte2024identification} and \citet{cooper2026learning}.

\subsection{Instrumental Variables}
\label{subsec:ivtheory}

Consider the binary instrumental variable setting shown in
Figure~\ref{fig:IV_DAG}. For clarity, I state the results without covariates.
As shown in Appendix~\ref{subsection-id},
standard assumptions of exogeneity, exclusion restriction, and monotonicity (``no defiers'') 
point-identify the LATE (Local Average Treatment Effect), which is denoted by 
$\E[Y(a_1) - Y(a_0) \mid A(z_1) > A(z_0)]$
\citep{imbens1994identification}. The robustness of the LATE to unit-level violation 
of  monotonicity and 
exclusion can be assessed by  the following method. 
First, perturbation functions are derived in the following two theorems,
whose proofs are in Appendix~\ref{app:late-monotonicity-exclusion}.

\begin{theorem}[Perturbation bounds on the LATE for monotonicity violations]
    \label{thm:late_as_func_defiers}
\singlespacing
Let $C=\{A(z_1)>A(z_0)\}$ denote the complier event.
Suppose exogeneity and exclusion hold, the observed distribution is
compatible with the model and a fixed specified $\theta_{\mathrm{ND}}$, and
$P(C)  = P(A(z_1)>A(z_0))>0$. Then the perturbation bounds for a particular $\theta_{\text{ND}}$
are {$l(\theta_{\text{ND}})/P(C) \leq LATE \leq u(\theta_{\text{ND}})/P(C)$},
where {$p_{ij}^{k}=P(Y=y_i,A=a_j\mid Z=z_k)$
for $i,j,k\in\{0,1\}$}.
{
The functions $l(\theta_{\mathrm{ND}})$ and $u(\theta_{\mathrm{ND}})$ are,
respectively,
\begin{equation*}
\max\begin{cases}
- p_{10}^{0}, \\
p_{11}^{1} - P(y_1|z_0), \\
p_{11}^{1} + p_{00}^{0} - 1 + \theta_{\text{ND}}, \\
-P(C), \\
p_{10}^{1} -  p_{10}^{0} - \theta_{\text{ND}},\\
p_{01}^{0} -  p_{01}^{1} - \theta_{\text{ND}},\\
P(y_0|z_0) -  P(y_0|z_1) - \theta_{\text{ND}},\\
- p_{01}^{1}, \\
p_{00}^{0} - P(y_0|z_1),
\end{cases},
\quad
\min\begin{cases}
1-p_{10}^{0}-p_{01}^{1}-\theta_{\text{ND}}, \\
p_{11}^{1}, \\
P(y_0|z_0) - p_{01}^{1}, \\
P(y_0|z_0) - P(y_0|z_1) + \theta_{\text{ND}}, \\
p_{00}^{0}, \\
P(C), \\
p_{00}^{0} - p_{00}^{1} + \theta_{\text{ND}}, \\
P(y_1|z_1) - p_{10}^{0}, \\
p_{11}^{1}-p_{11}^{0}+\theta_{\mathrm{ND}}, \\
\end{cases}
\end{equation*}
}
and {$P(C) = P(a_1|z_1) - P(a_1|z_0) + \theta_{\text{ND}}$}.
\end{theorem}

\doublespacing
Next, allow the direct $Z\to Y$ arrow shown in Figure~\ref{fig:IV_DAG3}. Let
$\theta_{\mathrm{ER}}$ denote the exact proportion of units for whom changing $Z$ changes
$Y(z,a)$ for at least one value of $a$. With binary variables, 
this is the total probability of strata satisfying
$Y(z_0,a)\neq Y(z_1,a)$ for at least one $a$.

\begin{theorem}[Perturbation bounds on the LATE under exclusion violations]
\label{thm:late_as_func_er}
\singlespacing
Here, with exclusion relaxed, write
$\operatorname{LATE}=\E[Y(Z,1)-Y(Z,0)\mid A(z_1)>A(z_0)]$,
which reduces to the usual LATE under exclusion.
Suppose exogeneity and monotonicity hold, the observed distribution is compatible with
the model and the specified $\theta_{\mathrm{ER}}$, and
$P(C)  = P(A(z_1)>A(z_0))>0$. { Then}
$\operatorname{LATE}$ satisfies
$l(\theta_{\mathrm{ER}})/P(C)\leq\operatorname{LATE}
\leq u(\theta_{\mathrm{ER}})/P(C)$,
{using $p_{ij}^{k}$ as above.}
{
\par\noindent
The functions $l(\theta_{\mathrm{ER}})$ and $u(\theta_{\mathrm{ER}})$ are,
respectively,
\medmuskip=2mu
\thickmuskip=3mu
\begin{equation*}
\sum_{k=0}^1 P(z_k)\max\begin{cases}
-P(C), \\
-p_{1-k,k}^{k}, \\
P(y_1|z_1)-P(y_1|z_0)-\theta_{\mathrm{ER}}, \\
p_{01}^{0}-p_{00}^{0}+p_{00}^{1}-p_{01}^{1}-\theta_{\mathrm{ER}}, \\
p_{10}^{1}-p_{11}^{1}-p_{10}^{0}+p_{11}^{0}-\theta_{\mathrm{ER}},
\end{cases},
\;
\sum_{k=0}^1 P(z_k)\min\begin{cases}
P(C), \\
p_{kk}^{k}, \\
P(y_1|z_1)-P(y_1|z_0)+\theta_{\mathrm{ER}}, \\
p_{00}^{0}-p_{01}^{0}-p_{00}^{1}+p_{01}^{1}+\theta_{\mathrm{ER}}, \\
p_{10}^{0}-p_{11}^{0}+p_{11}^{1}-p_{10}^{1}+\theta_{\mathrm{ER}}.
\end{cases}
\end{equation*}
}
with {$P(C) = P(a_1|z_1) - P(a_1|z_0)$}.

{
Let $p_{z1}=P(Z=1)$,
$\min_z=\min\{p_{z1},1-p_{z1}\}$,
and $\max_z=\max\{p_{z1},1-p_{z1}\}$. Define
\begin{align*}
a_{p_{z1}}&=\begin{cases}
p_{10}^{0},&p_{z1}\leq\tfrac12,\\
p_{01}^{1},&p_{z1}>\tfrac12,
\end{cases}
&
b_{p_{z1}}&=\begin{cases}
p_{00}^{0},&p_{z1}\leq\tfrac12,\\
p_{11}^{1},&p_{z1}>\tfrac12.
\end{cases}
\end{align*}
For $\theta=\theta_{\mathrm{ER}}$, the sharp perturbation bounds are
\begin{align*}
L_{\mathrm{ER}}^{\mathrm{pert}}(\theta)
&=\frac{1}{P(C)}\max\Big\{
 l(\theta),\,-\max_z P(C)+\min_z(\theta-1),\\
&\hspace{4em}-\min_z P(C)-|1-2p_{z1}|a_{p_{z1}}
 +\min_z(\theta-1)\Big\},\\
U_{\mathrm{ER}}^{\mathrm{pert}}(\theta)
&=\frac{1}{P(C)}\min\Big\{
 u(\theta),\,\max_z P(C)+\min_z(1-\theta),\\
&\hspace{4em}\min_z P(C)+|1-2p_{z1}|b_{p_{z1}}
 +\min_z(1-\theta)\Big\}.
\end{align*}
}
\end{theorem}

The sensitivity curves  can be calculated from the
perturbations.

\begin{theorem}[Natural sensitivity bounds for IV assumption violations]
\label{thm:iv-natural-sensitivity}
Consider either relaxation separately, under the maintained assumptions
of Theorems~\ref{thm:late_as_func_defiers} and~\ref{thm:late_as_func_er},
respectively, and suppose its sensitivity model is nonempty.
Let $q_a^z=P(A=a\mid Z=z)$ and $C$ denote the complier event.
\begin{enumerate}[label=(\roman*)]
\item For monotonicity sensitivity, let $\mathcal D$ be the set of feasible
values $d\leq\theta_{\mathrm{ND}}^{\max}$ with $P(C)=q_1^1-q_1^0+d>0$.
Let $\mathcal B$ consist of the endpoints of $\mathcal D$ and the values
of $d$ within $\mathcal D$ at which two nonparallel affine terms in
either $l(d)$ or $u(d)$ intersect.
The sharp LATE sensitivity endpoints are
$$
L_{\mathrm{ND}}(\theta_{\mathrm{ND}}^{\max})
=\min_{d\in\mathcal B}\frac{l(d)}{q_1^1-q_1^0+d},\qquad
U_{\mathrm{ND}}(\theta_{\mathrm{ND}}^{\max})
=\max_{d\in\mathcal B}\frac{u(d)}{q_1^1-q_1^0+d},
$$
using one-sided limits at excluded endpoints.
\item For exclusion sensitivity, $P(C)=q_1^1-q_1^0>0$ is fixed and the sharp sensitivity
 bounds for $\operatorname{LATE}$ are
$$
L_{\mathrm{ER}}(\theta_{\mathrm{ER}}^{\max})
=\frac{l(\theta_{\mathrm{ER}}^{\max})}{P(C)},
\qquad
U_{\mathrm{ER}}(\theta_{\mathrm{ER}}^{\max})
=\frac{u(\theta_{\mathrm{ER}}^{\max})}{P(C)}.
$$

\end{enumerate}
\end{theorem}

\subsection{Classic Confounding}
\label{subsec:classicconfounding}
Consider a binary treatment and outcome whose relationship may be confounded by
unobserved variables. I derive bounds that allow a limited proportion of
the population to belong to a component in which confounding is unrestricted.
Theorem~\ref{thm:confounding-bounds} extends the mixture approach of
\citet{bonvini2022sensitivity} by deriving closed-form bounds without
the additional restrictions discussed in
Section~\ref{sec:population-level}. A proof is given in Appendix~\ref{app:simpleconfounding}.

\begin{theorem}[Natural sensitivity bounds under classic confounding]
\label{thm:confounding-bounds}
\singlespacing
Let $\theta=\theta_{\mathrm{UC}}\in[0,1]$ be the nominal weight of the
component in which confounding is permitted, and suppose the mixture is
compatible with the observed distribution. Write
$p_{ay}=P(A=a,Y=y)$ and $t_{ha}=P(H=h,A=a)$, as in
Appendix~\ref{app:simpleconfounding}. The feasible range of $t_{0a}$ is
\begin{align*}
t_{0a}^{-}=\max\{0,P(A=a)-\theta\}
\leq t_{0a}\leq
\min\{1-\theta,P(A=a)\}=t_{0a}^{+}.
\end{align*}
At each fixed $t_{0a}>0$, define
\begin{align*}
L_{P(Y(A=a)=y)}(t_{0a})
&=p_{ay}+\max\left\{0,
\frac{(1-\theta-t_{0a})(t_{0a}-p_{a,1-y})}{t_{0a}}\right\},\\
U_{P(Y(A=a)=y)}(t_{0a})
&=\theta+t_{0a}-p_{a,1-y}
+\min\left\{1-\theta-t_{0a},
\frac{(1-\theta-t_{0a})p_{ay}}{t_{0a}}\right\}.
\end{align*}
At $t_{0a}=0$, these bounds are $p_{ay}$ and $1-p_{a,1-y}$,
respectively.
{
The sharp marginal bounds are the minimum of $L$ and maximum of $U$
over the two endpoints $t_{0a}^{-},t_{0a}^{+}$.

For the ATE, let $L_a(x)=L_{P(Y(A=a)=1)}(x)$ and
$U_a(x)=U_{P(Y(A=a)=1)}(x)$. Both potential outcomes must use the
same allocation $t=t_{01}$, with $t_{00}=1-\theta-t$. Define
\begin{align*}
L_{\mathrm{ATE}}(t)&=L_1(t)-U_0(1-\theta-t),&
U_{\mathrm{ATE}}(t)&=U_1(t)-L_0(1-\theta-t).
\end{align*}
Let $\mathcal T=[t_{01}^{-},t_{01}^{+}]$ and
\begin{align*}
\mathcal B_L&=\{t_{01}^{-},t_{01}^{+},p_{10},1-\theta-p_{01}\}
\cap\mathcal T,&
\mathcal B_U&=\{t_{01}^{-},t_{01}^{+},p_{11},1-\theta-p_{00}\}
\cap\mathcal T.
\end{align*}
The sharp ATE sensitivity bounds are
\begin{align*}
\mathcal L_{\mathrm{ATE}}(\theta)
&=\min_{t\in\mathcal B_L}L_{\mathrm{ATE}}(t),&
\mathcal U_{\mathrm{ATE}}(\theta)
&=\max_{t\in\mathcal B_U}U_{\mathrm{ATE}}(t).
\end{align*}
The boundary conventions above apply when $t=0$ or $1-\theta-t=0$.
}
The proof of Theorem~\ref{thm:confounding-bounds} is given in
Appendix~\ref{app:simpleconfounding}.
\end{theorem}

\doublespacing

\section{Empirical Results}
\label{sec:empirical}

\FloatBarrier

\subsection[Reanalysis of \citet{coppock2016voting}]{Reanalysis of \citet{coppock2016voting}}
\label{sec:results_gotv}

I reanalyze \citet{coppock2016voting}, who use the social pressure experiment of
\citet{Gerber:2008} to study whether voting is habit forming. In that experiment, voters
were randomly assigned to receive different turnout messages; the strongest treatment
informed recipients that their turnout would be monitored and reported to their
neighbors. The IV analysis treats the social pressure intervention as encouragement
$Z$, voting in the 2006 primary as treatment $A$, and voting in a later election as
outcome $Y$. The
question is whether voting in election $t$ increases the probability of voting in
election $t+1$.
{\color{black}
The replication data are from \citet{coppock2015votingdata}.
}

There are reasons to doubt that monotonicity holds in this setting: given the 
intrusive nature of the intervention, it is quite 
possible that social pressure caused some voters not to 
turn out in election $t$ who otherwise would have.  
In addition, the exclusion restriction implies that habit 
formation is the only mechanism through which turnout would 
increase in election $t+1$: that is, social pressure 
increases turnout only insofar as it affects 
turnout in election $t$, and there is no direct 
channel through which social pressure affects 
downstream turnout. However, it is plausible 
that individuals might have learned from 
the previous intervention, internalizing the 
behavior. If individuals have learned or simply 
remembered being encouraged to vote, then the exclusion restriction is violated.

\begin{figure}[htbp]
    \centering
    \caption{Sensitivity to monotonicity and exclusion violations.
    Panel (a) bounds LATE; panel (b) bounds the instrument-averaged
    complier effect. Solid curves are full-sample bounds; dashed curves
    give separate 95\% one-sided confidence limits for the lower and
    upper bounds. The black vertical line marks the estimated robustness
    value, ${\color{black} 0.01060}$; the blue dashed vertical line marks its 95\%
    lower confidence bound, ${\color{black} 0.0075}$, in both panels.
    Both panels use 2,000 subsamples of size {\color{black} 18,902}.}
    \label{fig:green_sensitivity}
    \begin{subfigure}[t]{0.49\linewidth}
        \centering
        \includegraphics[width=\linewidth]{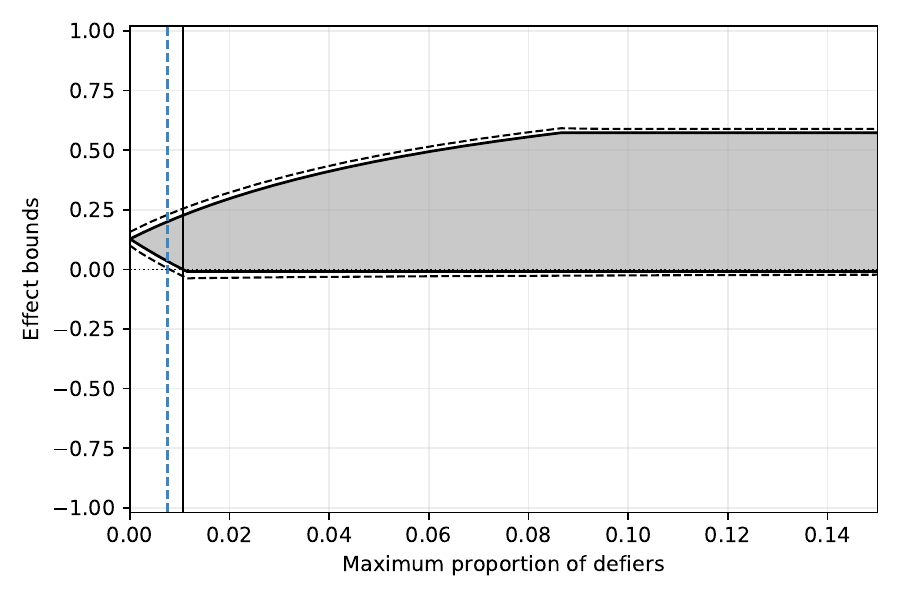}
        \caption{{Monotonicity: LATE.}}
        \label{fig:green_mono}
    \end{subfigure}\hfill
    \begin{subfigure}[t]{0.49\linewidth}
        \centering
        \includegraphics[width=\linewidth]{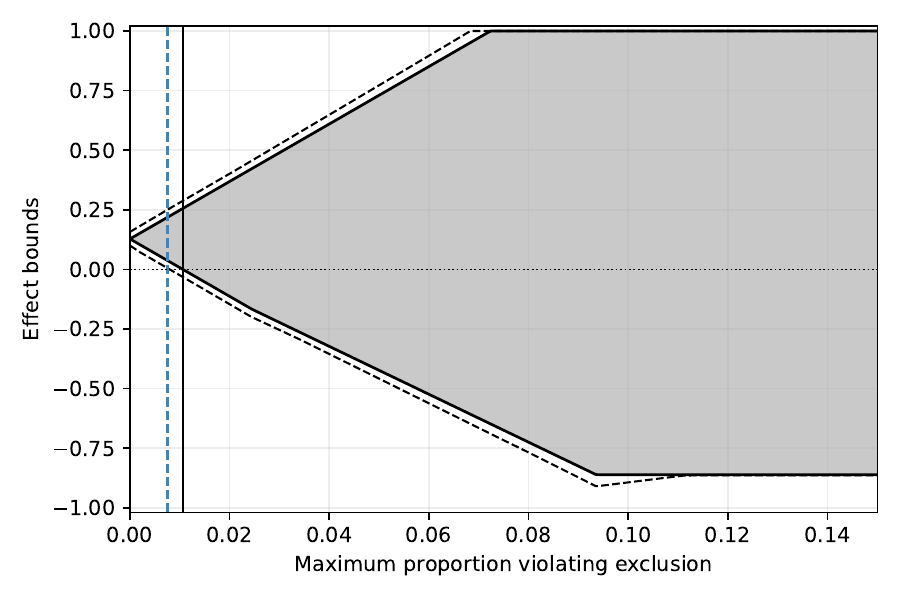}
        \caption{{Exclusion: instrument-averaged complier effect.}}
        \label{fig:green_er}
    \end{subfigure}
\end{figure}

Under monotonicity and exclusion, the estimated LATE is ${\color{black} 0.128}$,
with separate 95\% one-sided lower and upper confidence limits
of ${\color{black} 0.100}$ and ${\color{black} 0.157}$.
The analysis uses $n={\color{black} 221{,}642}$ observations and 2,000 subsamples of
size $m=\lfloor n^{0.8}\rfloor={\color{black} 18{,}902}$.
Figure~\ref{fig:green_mono} relaxes monotonicity while maintaining
exclusion and exogeneity. The estimated robustness value is ${\color{black} 0.01060}$:
the sharp lower bound reaches zero when approximately {\color{black} 1.06\%} of units
may be defiers. Its 95\% lower confidence bound is ${\color{black} 0.0075}$:
the analysis supports robustness to at least {\color{black} 0.75\%} defiers.
This bound addresses possible overstatement of robustness, without
seeking an upper confidence bound on the crossing location.

I next relax exclusion by allowing the direct $Z\to Y$ arrow in
Figure~\ref{fig:IV_DAG}(c). As explained in Section~\ref{subsec:ivtheory}, ordinary LATE
is not defined once exclusion fails, so Figure~\ref{fig:green_er} reports bounds for the
instrument averaged complier effect $\operatorname{LATE}$.
The lower bound reaches zero at ${\color{black} 0.01060}$, with a 95\% lower
confidence bound on robustness of ${\color{black} 0.0075}$. These are the sharp analytical
sensitivity bounds in Section~\ref{subsec:ivtheory}.
Both IV analyses use the one-sided procedure in
Appendix~\ref{sec:statisticalinference}, with grid spacing $0.0005$.
When the sensitivity parameter is $0$, {\color{black} 15.70\%} of the subsample
draws yielded infeasible problems, but none did at the grid point
nearest the estimated robustness value.
For infeasible subsamples, I use the bounds $[-1,1]$.
The inference results require the probability of infeasibility to
approach zero as the sample size increases, together with the other
conditions in Appendix~\ref{sec:statisticalinference}.
I do not relax exogeneity in this application
because the instrument was randomized by
design.

\subsection[Classic Confounding: Reanalysis of \citet{hobolt2016brexit}]{Classic Confounding: Reanalysis of \citet{hobolt2016brexit}}
\label{sec:results_soo}

To demonstrate how researchers might apply this method in a case of 
simple confounding, I replicate and 
extend \citet{hobolt2016brexit}, an investigation into 
the individual causes of support for ``Brexit,'' the 
landmark 2016 referendum in which British voters 
elected to leave the European Union. The study argues in the abstract
that ``the divide between winners and losers of globalization
was a key driver of the vote''. To test this claim,
it estimates several multivariate logistic regressions using 
survey data on support for Brexit. While the study makes 
numerous causal claims, for illustrative purposes I 
focus on the following statement: ``Going from A-level education 
to an undergraduate degree reduces the probability of voting 
Leave by about 10 percentage points, all other things 
being equal'' (1269). In other words, the study claims 
that voters' level of education was a causal 
factor influencing the Brexit outcome.

For this reanalysis, I take the estimand to be
$\operatorname{ATE}_{3\to4}=\E[Y(d=4)-Y(d=3)]$, the average treatment
effect of obtaining an undergraduate degree rather than an A-level
education.
The original analysis assumes a correctly specified logistic model
and conditional unconfoundedness given gender, age, income, and
perceived economic conditions (Table 2, Model 1).
With or without these controls, the estimated difference in support
is approximately $-10$ percentage points.
However, if the included control variables fail to 
account for all common causes of education 
and Brexit preference, then the regression contrast 
generally does not identify the ATE. This might be the 
case if a variable, such as region or personal connections, 
affects both obtaining an undergraduate degree and 
voting against Brexit. If confounding cannot be
limited in any way, the resulting bounds are 
often uninformative.

\begin{figure}[htbp]
    \centering
       \caption{{Contamination sensitivity analysis of the ATE.
       The horizontal axis gives the maximum weight of the component
       in which confounding is unrestricted, $\theta_{\mathrm{UC}}^{\max}$.
       Solid curves are full-sample bounds; dashed curves give separate
       95\% one-sided confidence limits from 2,000 recentered subsamples
       of size 1,982. The black vertical line marks the estimated
       robustness value, 6.52\%; the blue dashed vertical line marks
       its 95\% lower confidence bound, 5.75\%.}}
    \includegraphics[width=0.65\linewidth]{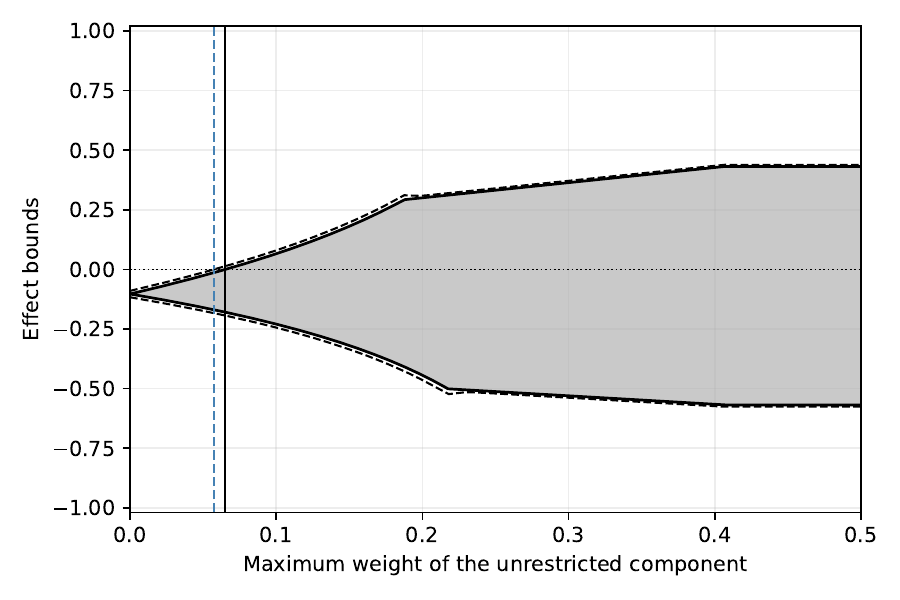}
    \label{fig:sensconf}
\end{figure}

Using the dataset with 13,229 eligible respondents,
the contrast under unconditional unconfoundedness is $-0.103$, with
separate 95\% one-sided lower and upper confidence limits of
$-0.117$ and $-0.090$.
The unrestricted ATE bounds are $[-0.568,0.432]$.
As emphasized in the introduction of this paper, there is no
need to restrict the alternative to unconfoundedness or not.
Using the approach advocated in this paper, I allow a component in which
education may be confounded to have weight at most
$\theta_{\mathrm{UC}}^{\max}$. In the remaining component, education is
unconfounded. This mixture is a formal decomposition of the population and
need not correspond to observed groups. Figure~\ref{fig:sensconf} plots the
resulting sensitivity curve, optimizing both potential outcomes
with a common mixture allocation as described in
Appendix~\ref{app:simpleconfounding}. The upper bound reaches zero
at $\theta_{\mathrm{UC}}^{\max}=0.06525$, approximately 6.52\%.
Its 95\% lower confidence bound on robustness is $0.0575$, supporting
the negative sign through at least 5.75\% contamination.
This uses 2,000 subsamples of size
$m=\lfloor n^{0.8}\rfloor=1{,}982$ and grid spacing $0.0005$.
Whether this represents substantial robustness depends on the plausible
weight of such a component.

\FloatBarrier

\section{Conclusion}
\label{sec:conclusion}

In this paper, I propose a natural sensitivity analysis approach
that applies to a broad class of causal models. The framework
represents relaxations of individual-level and population-level
constraints as polynomial optimization problems. These problems
can be solved numerically using \texttt{autobounds}, and closed-form
perturbation and sensitivity bounds can be obtained in some cases.
The applications examine habit formation in voting and the relationship
between education and Brexit preferences. In each case, the robustness
value measures the sensitivity level at which the sign of the effect
stops being identified. The substantive interpretation of this value
remains with the researcher. The resulting bounds are sharp when
the optimization reaches the global extrema of the stated model,
or when sharpness is established analytically.

\section*{Disclosure Statement}
The author reports there are no competing interests to declare.

\section*{Declaration of Generative AI Use}
The author used OpenAI GPT models, primarily GPT-6 Astra,
and Codex (GPT-6),
for assistance with generating and verifying code, checking mathematical
details and exploring counterexamples, reviewing and editing the manuscript,
generating suggestions, and writing. These tools were used to support
computational implementation, examine mathematical arguments, and improve
the clarity and presentation of the research. AI assistance was also used to
organize the submission materials and replication package.
The author retains responsibility for the accuracy, originality, and integrity
of the submitted work, including its mathematical arguments, computational
results, references, and final wording.

\bibliographystyle{chicago}

\bibliography{biblio}

\newpage

\appendix

\section{Identification Theory and Principal Strata methods}

\subsection{Identification}
\label{subsection-id}

Typical identification problems are formalized by first defining a causal model $\mathcal{M}$, 
which represents a set of possible data generating processes (DGPs) over the observed 
variables $V={V_1,\dots,V_I}$ and the relevant potential outcomes, 
together with a collection of 
assumed constraints $\mathcal{C}={C_\lambda:\lambda\in\Lambda}$ on these distributions. 
These constraints may restrict not only the observed data distribution $\Pr(V)$, but also the joint 
distribution of potential outcomes $\Pr(V(a_1),\dots,V(a_J))$ for possible treatment values $a_1,\dots,a_J$. 
For example, in the running binary IV setting, the assumed constraints include the no-defiers 
constraint $C_{\mathrm{ND}}:\Pr(A(z_1)=a_0,A(z_0)=a_1)=0$, the exclusion restriction
constraints
$C_{\mathrm{ER},a,z,z'}:\Pr\{Y(z,a)\neq Y(z',a)\}=0$ for every
$a,z,z'$, and the
unconfoundedness 
constraint $C_{\mathrm{UC}}:Z\indep\bigl(A(z_0),A(z_1),\{Y(z,a):a\in\{a_0,a_1\},z\in\{z_0,z_1\}\}\bigr)$.
This condition states that the instrument is jointly independent of all relevant potential treatments and potential outcomes.
The constraints here can also be separated by representation type in a pair $(\mathcal{G}, \mathcal{F})$, 
where $\mathcal{G}$ is a causal ADMG and $\mathcal{F}$ is a set of 
functional restrictions on the response types allowed by the graph.
The choice of using graphs is motivated by syntactic convenience, 
as they provide a compact representation of the constraints and 
facilitate computation. The constraints implied by a causal graph will be called
here \emph{structural constraints}, while the constraints implied by functional restrictions 
will be called \emph{functional constraints}. In subsection~\ref{sec:assumptiontype}, this difference will be 
further discussed in the context of sensitivity analysis.

Second, the identification problem requires a set of observed data, 
which is typically represented by the complete or partial 
    observational distribution $\Pr(V)$, as in the 
    standard identification case \citep{shpitser2006identification}, 
    or mixing observed distributions over potential outcomes as 
    in the causal data fusion case \citep{bareinboim2016causal}.
    Because the observed data imposes a 
     restriction on the full potential-outcome distribution, it can 
     be understood as a constraint on the model. This point of view 
     will be particularly useful in the context of the solution 
     for partial identification problems, which will require 
     a constrained optimization approach, as discussed below. It is also 
     worth noting that the observed data here is a population 
     distribution, which will be assumed to be known. In practice, the observed data is
     estimated from a finite sample, which introduces sampling uncertainty, and 
     will be discussed separately in Appendix~\ref{sec:statisticalinference}.

The third relevant part of the identification problem is 
the estimand $\mathcal{E}$, which represents the causal quantity of interest. 
This estimand is typically a computable function of
the distribution of potential outcomes $\Pr(V(a_1), \dots, V(a_J))$.
For example, in the running binary IV example, 
the estimand of interest is the local average 
treatment effect (LATE) among compliers, 
which can be expressed 
as $\mathcal{E} := \mathbb{E}(Y(a_1) - Y(a_0)\mid A(z_1) = a_1, A(z_0) = a_0)$. 
The identification problem is 
then to determine whether $\mathcal{E}$ can 
be uniquely computed from the model, given the observed data.

\begin{definition}[Point Identification]
Given a model $\mathcal{M}$, an observed data distribution, 
and an estimand $\mathcal{E}$, the problem of deciding whether 
every DGP in $\mathcal{M}$ that is consistent with the 
observed data distribution yields the same 
value of $\mathcal{E}$ is called the \emph{point identification problem}.
\end{definition}

A few points about point identification are worth noting. First, 
the point identification problem is a purely theoretical 
problem that involves population quantities, not statistical inference. 
Second, the point identification problem is a yes-or-no question:
either the estimand $\mathcal{E}$ can be uniquely determined from the
    model and observed data, or it cannot. For that reason, 
    proofs of non-identifiability often involve constructing 
    two DGPs that are consistent with the model and observed data, 
    but yield different values of $\mathcal{E}$.
Finally, causal inference practice focuses heavily on obtaining 
identification results, sometimes to the point of changing 
an estimand to make it identifiable. For instance, in the IV setting, 
the LATE is often used instead of the ATE because it is identifiable under 
the standard IV assumptions with monotonicity while the latter is not. 
However, the estimand should follow directly from the substantive theory and should not 
be chosen just for the sake of identifiability, a relevant point brought up by 
\citet{lundbergetal2021}.

Nonetheless, even if the identification problem 
yields a negative answer, useful results can still be obtained. In particular,
one can determine which values of the estimand $\mathcal{E}$ do not 
contradict the model and observed data. In other words, if a 
unique value for $\mathcal{E}$ cannot be determined, it is still possible to 
compute the range of values of $\mathcal{E}$ that satisfy the assumptions and 
constraints. This is the problem of partial identification, which 
generalizes point identification by allowing the identified set to contain 
more than one value.

\begin{definition}[Partial identification]
Given a model $\mathcal{M}$, an 
observed data distribution, and an estimand $\mathcal{E}$, 
the problem of characterizing the set of values of $\mathcal{E}$ 
yielded by all DGPs in $\mathcal{M}$ that are consistent 
with the observed data distribution is called 
the \emph{partial identification problem}. 
This set of values is called the \emph{identified set}.
\end{definition}

In simple terms, the partial identification problem asks 
which values of the estimand $\mathcal{E}$ are compatible with the observed data,
given the constraints $\mathcal{C}$ of the model $\mathcal{M}$.
If only one value is allowed, then the estimand is point identified;
if more than one value is allowed, then the estimand is partially identified. 
Thus, partial identification generalizes point identification.
As an example, in the running binary IV example,
the LATE is point identified under the standard IV assumptions including monotonicity
\citep{imbens1994identification}, 
but the ATE is only partially identified \citep{balke1997bounds}.  
Typically, the identified set is summarized by lower and upper bounds, which delimit the range 
of values in which the identified set must lie. When the identified set is a closed, bounded interval, it can be written as $[\mathcal{E}_{\min}, \mathcal{E}_{\max}]$. In particular,
a researcher would in general be interested in bounds that are \emph{sharp}, 
meaning that they are the tightest possible bounds for the estimand 
$\mathcal{E}$ given the model and observed data and are the most informative for the 
researcher. General algorithms for computing those bounds have been 
developed in \citet{sachs2023general} and \citet{duarte2021automated}.
It is worth noting that the identified set might not be represented 
by a closed interval, for instance, if the identified set is a union of disjoint intervals. 
For a bounded and closed identified set, the lower and upper bounds are respectively its minimum and maximum.
Those bounds are sharp as long as they cannot be improved, i.e. there are no 
other greater lower and smaller upper bounds that cover the identified set.  

In the following sections, I will formalize a general 
sensitivity analysis framework that builds on the above 
identification and partial identification problems. 
The key idea is to study how the identified 
set for $\mathcal{E}$ changes as I relax some 
of the constraints in $\mathcal{C}$, 
allowing for a proportion of units to violate 
those assumptions. This method allows the assessment of the robustness 
of causal conclusions to violations of key assumptions.

\subsection{A Brief Primer on Generalized Principal Stratification and Autobounds}
\label{subsubsec:autobounds}

Manually deriving sharp bounds on effects can be 
analytically complex and time-consuming. 
To simplify this process, \cite{duarte2021automated} offers a 
user-friendly algorithm that computes sharp bounds 
automatically based on the parameters of a causal question, called \emph{autobounds}. 
I will utilize this algorithm for the sensitivity 
analyses. For a comprehensive description of the algorithm, refer to
\cite{duarte2021automated}.
The algorithm takes four inputs: a causal theory, 
represented in an ADMG; any additional functional 
assumption not captured by the ADMG (e.g.\ ``no defiers'' or ``no interaction''); 
a causal estimand (e.g. the ATE); and empirical distributions 
over a set of discrete variables. The algorithm then constructs a polynomial
program whose feasible points represent causal models consistent with these
inputs. Global minimization and maximization yield the sharp endpoints. If
computation stops earlier, the algorithm's certified outer bounds remain
valid, although they need not yet be sharp.

Central to this procedure is the idea of 
principal stratification \citep{balke1997bounds, Frangakis2002}, 
which involves categorizing units in a study based on 
their potential responses under different conditions. 
For instance, in the classic IV framework, units are
divided into four principal strata with respect to 
the treatment variable: compliers, defiers, 
always-takers, and never-takers. The key idea is that 
 any discrete causal model can be 
represented in terms of principal strata based on how each 
 variable might respond to its causal past. 
As the complexity of causal systems increases, the 
number of principal strata grows rapidly.
However, as long 
as all variables are discrete, the number of strata 
remains finite, because there are only a 
limited number of ways discrete variables can interact. 
\citet{duarte2021automated} generalizes this procedure into 
\emph{generalized principal stratification} 
to account for 
 unobserved confounders that may be high-dimensional and continuous.
 When the observed variables are discrete, 
 the number of response types remains finite, 
while their probabilities, 
and hence the causal estimand, 
may vary continuously. This reduction of the causal graph is called 
\emph{canonicalization}. 

In formal terms, the method of generalized
principal stratification involves considering, 
for every variable $V_i$ in a graph, all possible 
functions that map all possible values of all 
proximal causes (parents in the graph) to 
possible values of $V_i$. If $V_i$ can take on $c$ values 
and its parents have $d$ possible unique 
combinations of values, then its principal strata space
 has cardinality $c^{d}$. 
For instance, in the instrumental 
variable scenario with binary variables, $Z$ has no observable parent, 
so its principal strata space is 
$\{z_0,z_1\}$, representing control and encouraged units.
$A$ has only 
one observable parent, $Z$, so the strata of $A$ must 
account for all possible functions from $Z$ to $A$, which 
can be compactly written as
$\{a_{00},a_{01},a_{10},a_{11}\}$,
representing respectively never-takers, compliers, defiers, 
and always-takers. In this notation, the strata are
indexed by binary variables, where the first digit 
indicates the value of $A$ if $Z$ were $0$, and 
the second digit, if $Z$ were $1$. 
Likewise, for $Y$, the strata are $\{y_{00},y_{01},y_{10},y_{11}\}$, as $Y$
has only $A$ as an observable parent.   
After the response types are enumerated, it is still necessary to account for
unobserved common causes. In the canonical parameterization of the standard
IV model, one disturbance $U_Z$ generates the response type of $Z$, and a
second disturbance $U_{AY}$ generates the combined response type of $A$ and
$Y$. A realization of $U_{AY}$ then defines one joint stratum:
$a_{rs}y_{tu}.$ In summary, the canonical parameterization of the standard IV model has 
produced two sets of principal strata: $\{z_0,z_1\}$ for $Z$ and
$\{a_{00}y_{00},\cdots,a_{11}y_{11}\}$ 
for the joint response type of $A$ and $Y$, which capture not only the potential outcomes of $A$ and $Y$, 
but also their dependence structure.
As $U_Z$ and $U_{AY}$ are independent, the joint distribution of the principal strata is
$P(z_i,a_{rs}y_{tu})=P(z_i)P(a_{rs}y_{tu}).$

The representation with two disturbances is exact for the standard IV model.
In more general graphs, canonicalization must retain the structure of the
independent disturbances.

Let $V=(V_1,\ldots,V_J)$ be the finite observed variables in a
canonical structural causal model and let $U=(U_1,\ldots,U_K)$ be its
exogenous disturbances. Canonical disturbances are roots and are mutually
independent. Moreover, no disturbance has a set of observed children
contained in the child set of another disturbance. The
canonicalization algorithm and the finite support construction are given
by \citet{duarte2021automated}.

Each observed variable obeys an equation
$$
V_j=f_j\{\operatorname{pa}_V(V_j),
          \operatorname{pa}_U(V_j)\}.
$$
Fixing the disturbances turns this equation into a deterministic
response function of the observed parents. 

By Proposition 1 of \citet{duarte2021automated}, a finite categorical
representation of the disturbances preserves the full data law.
Here, $\mathcal S(U_k)$ is the state space of the categorical $U_k$, and
an event $\{U_k=u_k\}$ corresponds to a generalized principal stratum.
A category may encode response functions for more than one observed variable.
Let
$$P_{U_k}=\{p_{k,u_k}=P(U_k=u_k):u_k\in\mathcal S(U_k)\},
\qquad
P_U=(P_{U_1},\ldots,P_{U_K}).$$
The entries of $P_U$ are the probabilities of the generalized principal strata and
are the variables used in the polynomial program. For each $k$,
$p_{k,u_k}\geq0$ and $\sum_{u_k\in\mathcal S(U_k)}p_{k,u_k}=1$.
For $\bu=(u_1,\ldots,u_K)$, independence gives
$$P(\bU=\bu)=\prod_{k=1}^Kp_{k,u_k}.$$
Fixing $\bu$ determines a vector of strata, denoted
$\rho(\bu)$. Hence, for any event $E$ defined by a finite conjunction or
union of counterfactual statements,
\begin{equation}
\label{eq:autobounds-event}
P(E)=
\sum_{\bu\in\mathcal S(U_1)\times\cdots\times\mathcal S(U_K)}
\bbone\{\rho(\bu)\in E\}
\prod_{k=1}^Kp_{k,u_k}.
\end{equation}
This is the polynomial representation established by Proposition 2 of
\citet{duarte2021automated}. Observed data restrictions, other assumptions,
and causal estimands can therefore be represented by polynomial equalities,
inequalities, and polynomial fractions in the canonical probabilities.
The exact construction retains all these restrictions; global optimization
of the complete program gives sharp bounds relative to the stated model.
\citet{duarte2021automated} provides the complete canonicalization algorithm.
A complete stratification of an instrumental variable model,
including the representations of estimands and observed data, can be 
found in Appendix~\ref{subsub:ivprograms}. The idea behind this 
procedure is that we can maximize or minimize the estimand 
given data and assumptions (constraints) over the causal model strata (parameters). 

Employing stratification, \texttt{autobounds} efficiently identifies 
all principal strata suggested by a causal model and its assumptions. 
It crucially discards strata that cannot exist given the 
specified assumptions and observed data, or those that 
provide redundant information. The program then represents the causal 
estimand as a polynomial expression based on these principal 
strata, which can be optimized computationally, subject to the 
constraints of the causal theory, assumptions, and data. In this form, 
finding sharp bounds becomes a computational problem. 
The main challenge is achieving solutions in  
reasonable time, especially as the number of principal strata grows 
quickly with the complexity of the variables.

\subsection{Examples of assumption restrictions}
\label{app:assumption-examples}

\paragraph{Mean additivity and individual-level no interaction.}
For binary $A,Z$, let $r$ index a generalized principal stratum with
probability $p_r$, and define
\begin{align*}
\Delta_r=Y_r(1,1)-Y_r(0,1)-Y_r(1,0)+Y_r(0,0).
\end{align*}
Mean additivity imposes $\sum_r p_r\Delta_r=0$, allowing cancellation
across generalized principal strata. Individual-level no interaction
instead imposes $p_r=0$ whenever $\Delta_r\neq0$.

\paragraph{Exclusion restrictions.}
To relax an exclusion restriction forbidding a direct $X\to Y$ effect,
allow this arrow and expand the generalized principal strata space
$\mathcal R_Y$ to include strata in which $Y$ changes with $X$ while
the other observed parents of $Y$ are held fixed.
For example, suppose $X$ and $Y$ are binary and $Y$ has no other
observed parents. Without the arrow, $Y$ has the two constant strata
$\mathcal R_Y=\{y_0,y_1\}$. Allowing the arrow produces
$\mathcal R_Y=\{y_{00},y_{01},y_{10},y_{11}\}$, where the indices record
$Y(x_0)$ and $Y(x_1)$. The exclusion restriction is violated in strata
$y_{01}$ and $y_{10}$. An exact perturbation therefore imposes
\begin{align*}
P(y_{01})+P(y_{10})=\theta_{\mathrm{ER}},
\end{align*}
whereas a natural sensitivity problem imposes
\begin{align*}
P(y_{01})+P(y_{10})\leq\theta_{\mathrm{ER}}^{\max}.
\end{align*}
At zero, with all other model restrictions retained, the expanded model
is observationally and causally equivalent to the model without the arrow.

\paragraph{Unconfoundedness.}
Let $C$ denote observed covariates and
$\mathbf Y_A=(Y(a):a\in\mathcal S(A))$ the potential-outcome vector.
For sensitivity analysis, introduce a component indicator $H$ and impose
$A\indep\mathbf Y_A\mid C,H=0$ wherever this conditional distribution
is defined. The $H=1$ component allows arbitrary dependence, and its
probability is bounded by the sensitivity parameter. The polynomial
factorization also covers a component of zero mass.
Lemma~\ref{lem:unconfounded-mixture} and its proof in
Appendix~\ref{app:unconfounded-mixture} establish the factorization and
mixture representation.

\section{General Sensitivity Theory}

\subsection{Proof for individual-level sensitivity analysis}
\label{app:functional-autobounds}

Recall Definition~\ref{def:individual-constraint}: a constraint
$C_\lambda$ is individual-level if it requires
$P(\mathcal V_\lambda)=0$, where $\mathcal V_\lambda$ is a fixed union
of generalized principal stratum events. These are the strata that
violate the restriction. Let $I_\lambda$ indicate membership in
$\mathcal V_\lambda$. Relaxing the restriction allows their total
probability to equal $\theta$ for perturbation analysis, or to be at
most $\theta^{\max}$ for sensitivity analysis.

\renewcommand{\repeatedstatementheading}{Theorem~\ref{thm:functional-autobounds}}
\begin{repeatedstatement}[Individual-level sensitivity]
Let $C_\lambda$ be an individual-level constraint
as in Definition~\ref{def:individual-constraint}.
 In the finite canonical model, define
$I_\lambda(\bu)=\bbone\{\{\bU=\bu\}\subseteq\mathcal V_\lambda\}$.
Let $\boldsymbol\eta$ be a vector of all the generalized principal strata
probabilities
and any auxiliary variables. Let $\mathcal H_0(P_V)$ contain the
canonicalized parameter vectors
that reproduce $P_V$, satisfy the axioms of probability
and all retained restrictions in canonicalization,
including auxiliary-variable constraints but excluding $C_\lambda$,
and leave $\mathcal E(\boldsymbol\eta)$ defined.

Then:
\begin{enumerate}[label=(\roman*)]
\item The event probability is the polynomial
\begin{equation*}
P_{\boldsymbol\eta}(\mathcal V_\lambda)
=\sum_{\bu\in\mathcal S(\bU)}I_\lambda(\bu)\prod_{k=1}^Kp_{k,u_k}.
\end{equation*}
Here $P_{\boldsymbol\eta}$ denotes the distribution determined by $\boldsymbol\eta$.

\item For $\theta,\theta^{\max}\in[0,1]$, define
\begin{align*}
\mathcal H_{\lambda,=}(\theta;P_V)
&=\{\boldsymbol\eta\in\mathcal H_0(P_V):P_{\boldsymbol\eta}(\mathcal V_\lambda)=\theta\},\\
\mathcal H_{\lambda,\leq}(\theta^{\max};P_V)
&=\{\boldsymbol\eta\in\mathcal H_0(P_V):P_{\boldsymbol\eta}(\mathcal V_\lambda)\leq\theta^{\max}\}.
\end{align*}
These programs represent the exact perturbation problem and the
natural sensitivity problem of
Definition~\ref{def:natural-sensitivity-problem}, respectively.
Their images under $\mathcal E$ are $f(\theta)$ and
$g(\theta^{\max})$, with
\begin{align*}
g(\theta^{\max})=\bigcup_{0\leq\theta\leq\theta^{\max}}f(\theta).
\end{align*}
For a scalar estimand, the sharp endpoints satisfy
\begin{align*}
L^{\mathrm{sens}}(\theta^{\max})=\inf_{0\leq\theta\leq\theta^{\max}}L(\theta),
\qquad
U^{\mathrm{sens}}(\theta^{\max})=\sup_{0\leq\theta\leq\theta^{\max}}U(\theta),
\end{align*}
omitting infeasible values of $\theta$.

For each feasible problem, the infimum and supremum of the scalar estimand
over the corresponding polynomial program give its sharp bounds.

\end{enumerate}
\end{repeatedstatement}

\begin{proof}[Proof of Theorem~\ref{thm:functional-autobounds}]
The event $\mathcal V_\lambda$ is the disjoint union of the canonical
cells $\{\bU=\bu\}$ for which $I_\lambda(\bu)=1$. The canonical blocks are
independent, so its probability is
\begin{align*}
\sum_{\bu\in\mathcal S(\bU)}
I_\lambda(\bu)
\prod_{k=1}^K p_{k,u_k}=P_{\boldsymbol\eta}(\mathcal V_\lambda).
\end{align*}
Each summand is a monomial in the principal stratum sizes, so
$P_{\boldsymbol\eta}(\mathcal V_\lambda)$ is a polynomial. This is also an immediate application of
Proposition 2 of \citet{duarte2021automated} to the event $\mathcal V_\lambda$.
Any auxiliary variables remain subject to their original defining
constraints and do not enter $P_{\boldsymbol\eta}(\mathcal V_\lambda)$.
This proves claim (i).

For any $\boldsymbol\eta\in\mathcal H_{\lambda,\leq}(\theta^{\max};P_V)$, set
$t=P_{\boldsymbol\eta}(\mathcal V_\lambda)$. Then $0\leq t\leq\theta^{\max}$ and
$\boldsymbol\eta\in\mathcal H_{\lambda,=}(t;P_V)$. This proves one inclusion in
\begin{align*}
\mathcal H_{\lambda,\leq}(\theta^{\max};P_V)
=\bigcup_{0\leq t\leq\theta^{\max}}
\mathcal H_{\lambda,=}(t;P_V).
\end{align*}
On the other hand, if $\boldsymbol\eta\in\mathcal H_{\lambda,=}(t;P_V)$ for some
$t\leq\theta^{\max}$, then
$P_{\boldsymbol\eta}(\mathcal V_\lambda)=t\leq\theta^{\max}$, proving the reverse inclusion. Applying $\mathcal E$ gives the union of identified sets.
Taking infima and suprema gives the envelope formulas.

The construction in terms of generalized principal strata 
represents the observed data,
counterfactual restrictions, and estimand by polynomials or polynomial
fractions in the generalized principal stratum sizes. The
displayed formula adds one polynomial equality or inequality. Theorem 1 of
\citet{duarte2021automated}, therefore, applies to this polynomial program,
so whenever its global extrema are obtained, they are the sharp extrema over
precisely the stated model. 
Proposition 1 of the same paper allows 
the model to be represented by finitely many 
generalized principal strata while 
preserving the joint distribution of observed variables and 
potential outcomes. 
When $\mathcal V_\lambda$ is defined by the potential outcomes,
preserving their joint distribution also preserves
$P_{\boldsymbol\eta}(\mathcal V_\lambda)$. This proves claim (ii).

\end{proof}

\begin{corollary}[Simultaneous individual-level relaxations]
\label{cor:simultaneous-individual-relaxations}
Under the conditions of Theorem~\ref{thm:functional-autobounds},
consider finitely many individual-level constraints on the same
canonical model. Retain all other model constraints and impose one
polynomial inequality per restriction:
\[
P_{\boldsymbol\eta}(\mathcal V_{\lambda_j})\leq\theta_j^{\max},
\qquad j=1,\ldots,J.
\]
The union and sharp endpoint formulas in that theorem hold with
componentwise bounds on $\Theta$. The violation events may overlap,
and independence among them is not required.
\end{corollary}

\begin{proof}
For a finite collection of assumptions, each
$P_{\boldsymbol\eta}(\mathcal V_{\lambda_j})$ has the same polynomial form.
If $\boldsymbol\eta$ satisfies
$P_{\boldsymbol\eta}(\mathcal V_{\lambda_j})\leq\theta_j^{\max}$ for every $j$,
set $\Theta=(P_{\boldsymbol\eta}(\mathcal V_{\lambda_j}))_{j=1}^J$. Then,
$0\leq\Theta\leq\Theta^{\max}$ and $\boldsymbol\eta$ belongs to the
corresponding exact vector set. If each violation proportion is at most its bound, all
the sensitivity constraints are satisfied by the model.
The strata that violate more than one restriction contribute
to the violation proportion of each.
Taking the image under the estimand and then its infimum and supremum
gives the union and sharp endpoint formulas.
\end{proof}

{
\renewcommand{\repeatedstatementheading}{Corollary~\ref{cor:sensitivity-monotonicity}}
\begin{repeatedstatement}[Sensitivity Function Monotonicity]
For any $0\leq\theta_1^{\max}\leq\theta_2^{\max}\leq1$,
$$g(\theta_1^{\max})\subseteq g(\theta_2^{\max}).$$
Consequently, wherever the sensitivity sets are nonempty, the sharp lower
endpoint is nonincreasing and the sharp upper endpoint is nondecreasing:
$$L^{\mathrm{sens}}(\theta_2^{\max})
\leq L^{\mathrm{sens}}(\theta_1^{\max}),
\qquad
U^{\mathrm{sens}}(\theta_2^{\max})
\geq U^{\mathrm{sens}}(\theta_1^{\max}).$$
More generally, for several sensitivity parameters,
$\Theta_1^{\max}\leq\Theta_2^{\max}$ componentwise implies
$g(\Theta_1^{\max})\subseteq g(\Theta_2^{\max})$. This monotonicity does not hold for the
exact perturbation function $f(\theta)$, because models with different exact
violation proportions are not in general nested.
\end{repeatedstatement}
}

\begin{proof}[Proof of Corollary~\ref{cor:sensitivity-monotonicity}]
If $\theta_1^{\max}\leq\theta_2^{\max}$, then every $\boldsymbol\eta$ satisfying
$P_{\boldsymbol\eta}(\mathcal V_\lambda)\leq\theta_1^{\max}$ also satisfies
$P_{\boldsymbol\eta}(\mathcal V_\lambda)\leq\theta_2^{\max}$. Therefore,
\begin{align*}
\mathcal H_{\lambda,\leq}(\theta_1^{\max};P_V)
\subseteq
\mathcal H_{\lambda,\leq}(\theta_2^{\max};P_V).
\end{align*}
Considering the compatible values of $\mathcal E$  under  each constrained problem proves
$g(\theta_1^{\max})\subseteq g(\theta_2^{\max})$. Therefore, its suprema and infima
satisfy the endpoint inequalities. The same argument applied to every component gives
the vector result. On the other hand, $P_{\boldsymbol\eta}(\mathcal V_\lambda)=\theta_1$ does not imply
$P_{\boldsymbol\eta}(\mathcal V_\lambda)=\theta_2$ for $\theta_1\neq\theta_2$, so the exact
perturbation sets need not be nested.
\end{proof}

{
\renewcommand{\repeatedstatementheading}{Corollary~\ref{cor:local-mixture}}
\begin{repeatedstatement}[Mixture representation]
Suppose $I_\lambda(\bu)$ depends only on one canonical block $U_j$. Let
$\mathcal S_j^{\mathrm{res}}$ contain the strata in which the restriction
holds, and suppose this set is nonempty. A principal strata distribution
satisfies $P_{\boldsymbol\eta}(\mathcal V_\lambda)\leq\theta^{\max}$ if and only if it can be written as
$$P_{U_j}=(1-\theta^{\max})P_{U_j}^{\mathrm{res}}
+\theta^{\max}P_{U_j}^{\mathrm{rel}},$$
where $P_{U_j}^{\mathrm{res}}$ is supported on
$\mathcal S_j^{\mathrm{res}}$ and $P_{U_j}^{\mathrm{rel}}$ is unrestricted on
$\mathcal S(U_j)$.
Substituting the combined principal strata distribution into every
original polynomial constraint gives an equivalent parameterization of the
same causal model.
\end{repeatedstatement}
}

\begin{proof}[Proof of Corollary~\ref{cor:local-mixture}]
Let $t=P_{\boldsymbol\eta}(\mathcal V_\lambda)\leq\theta^{\max}$. Since the violation depends only
on $U_j$, the probability outside $\mathcal S_j^{\mathrm{res}}$ is $t$ and
the probability inside it is $1-t$. For
$0<\theta^{\max}<1$, define
\begin{align*}
P_{U_j}^{\mathrm{res}}(u_j)=
\begin{cases}
p_{j,u_j}/(1-t), &u_j\in\mathcal S_j^{\mathrm{res}},\\
0, &u_j\notin\mathcal S_j^{\mathrm{res}},
\end{cases}
\end{align*}
and
\begin{align*}
P_{U_j}^{\mathrm{rel}}(u_j)=
\begin{cases}
\dfrac{\theta^{\max}-t}{\theta^{\max}(1-t)}p_{j,u_j},
    &u_j\in\mathcal S_j^{\mathrm{res}},\\[6pt]
\dfrac{p_{j,u_j}}{\theta^{\max}},
    &u_j\notin\mathcal S_j^{\mathrm{res}}.
\end{cases}
\end{align*}
Both are probability distributions, $P_{U_j}^{\mathrm{res}}$ is supported
on the restricted strata, and direct substitution gives
\begin{align*}
P_{U_j}=(1-\theta^{\max})P_{U_j}^{\mathrm{res}}
+\theta^{\max}P_{U_j}^{\mathrm{rel}}.
\end{align*}
If $\theta^{\max}=0$, take $P_{U_j}^{\mathrm{res}}=P_{U_j}$ and choose
$P_{U_j}^{\mathrm{rel}}$ arbitrarily. If $\theta^{\max}=1$, take
$P_{U_j}^{\mathrm{rel}}=P_{U_j}$ and choose
$P_{U_j}^{\mathrm{res}}$ arbitrarily on $\mathcal S_j^{\mathrm{res}}$.

Conversely, suppose the displayed mixture holds. Since
$P_{U_j}^{\mathrm{res}}$ assigns no probability to a violation,
\begin{align*}
P_{\boldsymbol\eta}(\mathcal V_\lambda)
=\theta^{\max}
\sum_{u_j\notin\mathcal S_j^{\mathrm{res}}}
P_{U_j}^{\mathrm{rel}}(u_j)
\leq\theta^{\max}.
\end{align*}

Only one generalized principal strata distribution has been
reparameterized. The other independent blocks remain unchanged. 
Substituting $P_{U_j}$ into every original expression
reproduces the same $\prod_kp_{k,u_k}$. Therefore,
the two parameterizations
 have the same feasible full data laws and the same estimand
values.
\end{proof}

\begin{lemma}[Exclusion restriction as individual-level constraint]
\label{cor:exclusion-autobounds}
Suppose the original graph excludes the arrow $X\to Y$, and form the
relaxed graph by adding it. Let $W$ collect the other observed parents of
$Y$ in the relaxed graph. Within a generalized principal stratum, the
exclusion restriction is
\begin{align*}
Y(x,w)=Y(x',w)
\quad\text{for every }x,x',w.
\end{align*}
Thus, the exclusion restriction is an individual-level
 restriction in the relaxed
graph. Taking $\lambda$ in Theorem~\ref{thm:functional-autobounds} to be this
restriction, $I_\lambda$ identifies the principal strata for which
$Y(x,w)\neq Y(x',w)$ for some $x,x',w$, and
$P_{\boldsymbol\eta}(\mathcal V_\lambda)$ is their population proportion. Consequently, all the
conclusions of Theorem~\ref{thm:functional-autobounds} apply.
\end{lemma}

\subsection{Sensitivity Graphs}
\label{app:sensitivity-graphs}

A sensitivity graph locates the assumptions being relaxed, 
as defined below.
The graph and the stated constraints determine
the sensitivity model.
{
A sensitivity model consists of a sensitivity graph together with the
support restrictions and any additional constraints on the component
and recovered probability distributions.
}

\begin{definition}[Sensitivity graph]
\label{def:sensitivity-graph}
Let $(\mathcal G_R,\mathcal F_R)$ represent the relaxed model on the
original variables $V$, and let $\Lambda_R$ index its designated
relaxations. For each $\lambda\in\Lambda_R$, specify a binary sensitivity
indicator $S_\lambda$, a bound
$P(S_\lambda=1)\leq\theta_\lambda^{\max}$, and a location
$\chi_\lambda\subseteq V$. The location contains the variables whose
response functions are constrained by a functional restriction, the
child $Y$ for exclusion of $X\to Y$, or the treatment and outcome
variables involved in the unconfoundedness restriction. The sensitivity
graph adds the vertices $S_{\Lambda_R}$ to $\mathcal G_R$ and the edges
$$
E_{\mathrm{loc}}=
\{S_\lambda\to V_j:V_j\in\chi_\lambda,\ \lambda\in\Lambda_R\}.
$$

Let $I_\lambda$ be the indicator
of the strata that violate an individual-level restriction.
 Suppose the selected response system cannot violate
the restriction when $S_\lambda=0$. For a population-level restriction,
$S_\lambda$ is an augmented component indicator and the restriction is
imposed on the conditional response distribution given $S_\lambda=0$,
when that component has positive probability mass. The population 
distribution must satisfy all retained restrictions and observed data.
\end{definition}

For the particular case of 
individual-level restrictions, $I_\lambda$ indicates the
generalized principal strata that violate the restriction, so
$P(I_\lambda=1)$ is their exact population proportion.
The sensitivity indicator $S_\lambda$ identifies the unrestricted
component, whose proportion is $P(S_\lambda=1)$.
Since all violating strata must belong to this component,
but nonviolating strata may also belong to it,
$P(I_\lambda=1)\leq P(S_\lambda=1)\leq\theta_\lambda^{\max}$.

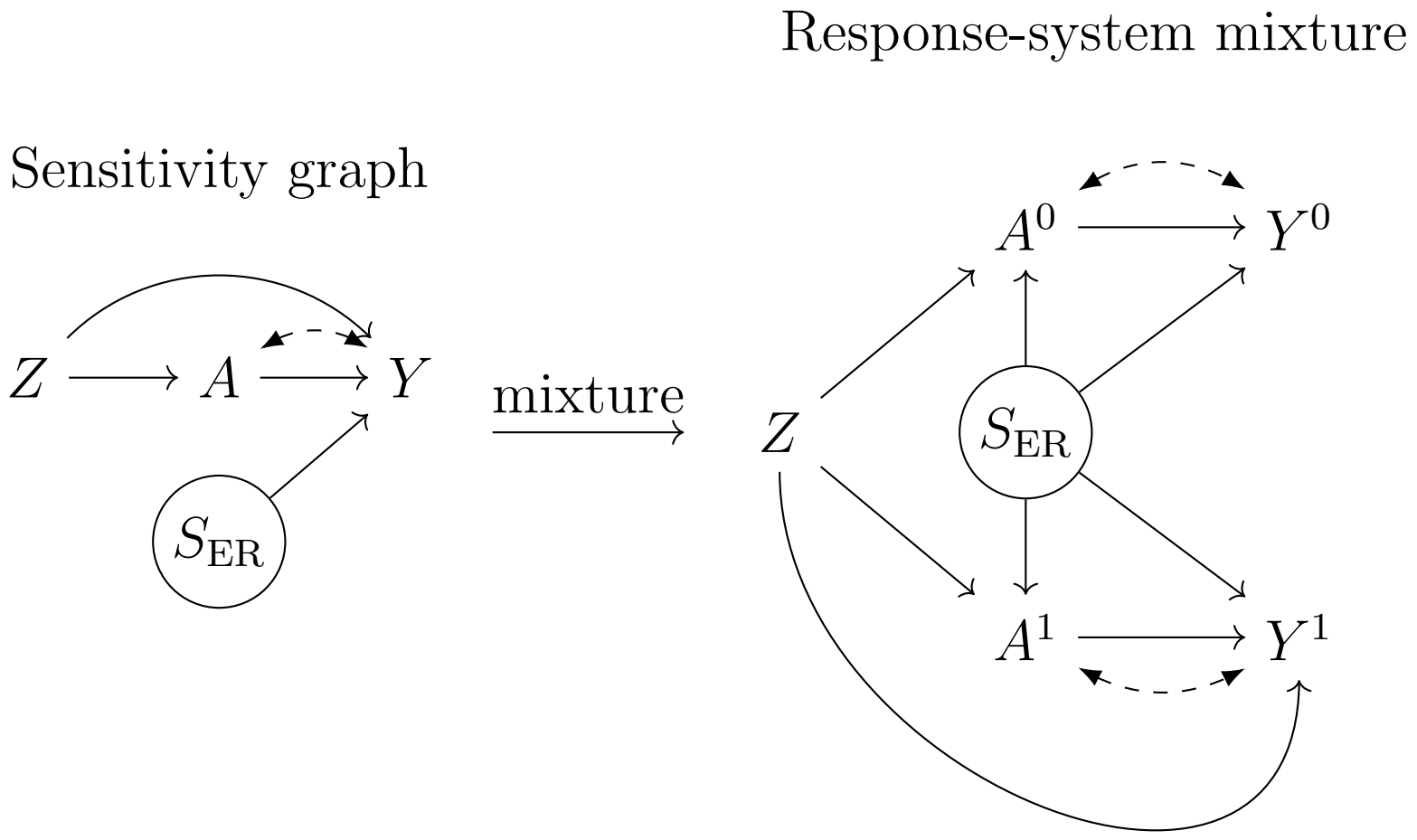
\begin{figure}[htp]
\centering
\caption{\textbf{Sensitivity graph and mixture representation for IV exclusion.}
The left panel locates the restriction at $Y$. The right panel shows 
the doubled representation with
restricted and unrestricted response parts sharing the independent
instrument input $Z$. Directed arrows from $S_{\mathrm{ER}}$ indicate
selection of the relaxed or original response system. This representation
preserves all the retained constraints from the original model.}
\label{fig:sensitivity-projected-admgs}
\begin{tikzpicture}
\node (z) at (0,0) {$Z$};
\node (a) at (1.4,0) {$A$};
\node (y) at (2.8,0) {$Y$};
\node[draw,circle,inner sep=2pt] (s) at (1.4,-1.2) {$S_{\mathrm{ER}}$};
\path (z) edge[->] (a);
\path (a) edge[->] (y);
\path (z) edge[->,bend left=45] (y);
\path[bidirected] (a) edge[bend left=35] (y);
\path (s) edge[->] (y);
\node at (1.4,1.5) {Sensitivity graph};
\draw[->] (3.4,-.4) -- node[above,align=center] {mixture} (4.8,-.4);
\node (zi) at (5.5,-.4) {$Z$};
\node (a0) at (7.3,1.1) {$A^0$};
\node (y0) at (9.3,1.1) {$Y^0$};
\node (a1) at (7.3,-1.9) {$A^1$};
\node (y1) at (9.3,-1.9) {$Y^1$};
\node[draw,circle,inner sep=2pt] (ss) at (7.3,-.4) {$S_{\mathrm{ER}}$};
\node at (7.8,2.5) {Response-system mixture};
\path (zi) edge[->] (a0);
\path (zi) edge[->] (a1);
\path (a0) edge[->] (y0);
\path (a1) edge[->] (y1);
\path (zi) edge[->,out=-90,in=-90,looseness=1.4] (y1);
\path[bidirected] (a0) edge[bend left=35] (y0);
\path[bidirected] (a1) edge[bend right=30] (y1);
\path (ss) edge[->] (a0);
\path (ss) edge[->] (y0);
\path (ss) edge[->] (a1);
\path (ss) edge[->] (y1);
\end{tikzpicture}
\end{figure}

Figure~\ref{fig:sensitivity-projected-admgs} illustrates a sensitivity 
graph for an instrumental variable model with a relaxed exclusion restriction.
The violation indicator is determined by the outcome strata.
Because treatment and outcome belong to the same district,
their strata may be dependent, so the indicator may also be
dependent on the treatment strata. The exogeneity assumption
 makes $Z$ independent of the joint treatment and outcome 
 strata. Conditioning the response law
on this violation, then, preserves instrument independence.
The doubled representation, defined below,
 draws the two branch systems independently
of the selector and evaluates the selected system using an independent
instrument with the observed distribution. 
The selector chooses the entire $\{A,Y\}$ response system,
although the restriction is located at $Y$ in the sensitivity graph.
The restricted branch, because it imposes exclusion restriction, 
omits the arrow $Z\to Y$. The unrestricted branch, on the other hand,
allows it.
The selector weight and the branch support restrictions cannot be 
derived from the ordinary edges alone and must be specified separately.

\FloatBarrier

\subsection{The enlarged doubled representation}
\label{app:sensitivity-selector}

The doubled representation makes the selection rule explicit. Let
$\mathbf R^0$ and $\mathbf R^1$ be complete generalized principal strata in the
restricted and unrestricted branches. 
The equivalence of the program derived by canonicalization 
of this ADMG with the sensitivity problem that relaxes 
an individual-level restriction is proved in Appendix~\ref{app:sensitivity-selector}.
This is a mixture construction. 
 Appendix~\ref{app:multiple-model-selectors} extends
this construction to many selectors.

The doubled representation is sound in the sense that every recovered 
model satisfies the original sensitivity constraints and the other 
 stated restrictions,  including
$P(I_{\lambda} = 1) \leq P(S_\lambda = 1) \leq \theta_{\lambda}^{\max}$.

\begin{proposition}[Soundness of the doubled representation for individual-level restrictions]
\label{prop:doubled-soundness}
Let $\mathbf R^0$ and $\mathbf R^1$ be the complete generalized principal strata
in the restricted and relaxed copies, and let $I_\lambda$ indicate a
violation of the functional restriction. Suppose
\begin{align*}
I_\lambda(\mathbf R^0)=0\quad\text{almost surely},
\qquad P(S=1)\leq\theta_\lambda^{\max}.
\end{align*}
Recover the original strata by $\mathbf R=\mathbf R^S$,
using the same $S$ and impose the other original
assumptions and observed data constraints on the recovered model.
Then the set of recovered models is a subset of the original
sensitivity model class.
\end{proposition}

\begin{proof}
The restricted copy contains no violations, so
\begin{align*}
P\{I_\lambda(\mathbf R)=1\}
&=P\{S=0,I_\lambda(\mathbf R^0)=1\}
  +P\{S=1,I_\lambda(\mathbf R^1)=1\}\\
&=P\{S=1,I_\lambda(\mathbf R^1)=1\}\\
&\leq P(S=1)\leq\theta_\lambda^{\max}.
\end{align*}

As all the retained assumptions are being imposed,
this is sufficient to prove soundness.
\end{proof}

It remains to be proved that the doubled representation is also complete, i.e. 
if the original sensitivity model is a subset of the recovered models. 
As that is the case, the doubled representation is equivalent 
to the original sensitivity model for individual-level restrictions.

\begin{proposition}[Equivalence of the enlarged doubled program]
\label{prop:enlarged-double}
Let $r$ denote a generalized principal stratum, and let
$I_\lambda(r)$ equal 1 if that stratum violates the restriction
and 0 otherwise. Let $p_r(\boldsymbol\eta)$ be the probability 
of the strata under the original model. Assume
that the set $\{r:I_\lambda(r)=0\}$ is nonempty.
Keep $\boldsymbol\eta\in\mathcal H_0(P_V)$, which imposes the
observed data constraints and all original assumptions except
the restriction being relaxed.
Replace the original violation bound 
$P_{\boldsymbol\eta}(\mathcal V_\lambda)\leq\theta_\lambda^{\max}$ 
by the following constraints:
\begin{align*}
p_r(\boldsymbol\eta)&=(1-t)q_r^0+tq_r^1 &&\text{for every }r,\\
q_r^s&\geq0,\qquad \sum_rq_r^s=1 &&s=0,1,\\
q_r^0&=0 &&\text{if }I_\lambda(r)=1,\\
0&\leq t\leq\theta_\lambda^{\max}.&&
\end{align*}
The original independence restrictions are imposed on the combined
distribution, not separately on $q^0$ and $q^1$.
The values of $\boldsymbol\eta$ compatible with these constraints
are exactly
\begin{align*}
\{\boldsymbol\eta\in\mathcal H_0(P_V):
  \sum_r I_\lambda(r)p_r(\boldsymbol\eta)\leq\theta_\lambda^{\max}\}.
\end{align*}
Therefore, solving the global optimization 
problem with the original estimand $\mathcal E(\boldsymbol\eta)$ as 
the objective function gives
the same bounds in the two programs.
\end{proposition}

\begin{proof}
Soundness follows from Proposition~\ref{prop:doubled-soundness}.
For equivalence, then, it remains to prove completeness.

Take any feasible original $\boldsymbol\eta$ and set
$t=\sum_r I_\lambda(r)p_r(\boldsymbol\eta)$.
If $0<t<1$, choose $q^0$ and $q^1$ to be the distributions
over strata conditional on $I_\lambda=0$ and $I_\lambda=1$,
respectively. This implies
\begin{align*}
q_r^s&\geq0,\qquad \sum_rq_r^s=1 &&s=0,1,\\
q_r^0&=0 &&\text{if }I_\lambda(r)=1.
\end{align*}
For $0 < t < 1$, the law of total probability guarantees
\begin{align*}
p_r(\boldsymbol\eta)&=(1-t)q_r^0+tq_r^1
&&\text{for every }r.
\end{align*}
For $t = 0$ , conditioning on $I_{\lambda} = 1$ is undefined, so 
 set $q_r^0=p_r(\boldsymbol\eta)$ and choose any
probability distribution for $q^1$.
Analogously, for $t = 1$, 
conditioning on $I_{\lambda} = 0$ is undefined,
so set $q_r^1=p_r(\boldsymbol\eta)$ and choose any
probability distribution for $q^0$ that 
assigns probability $0$ to all violating strata.

Finally, 
draw the strata in copies $0$ and $1$ independently with probabilities
$q_r^0$ and $q_r^1$, and independently draw $S$ with $P(S=1)=t$.
Select the stratum from copy $S$, keeping this choice fixed across
interventions. The probability of selecting stratum $r$ is
$(1-t)q_r^0+tq_r^1=p_r(\boldsymbol\eta)$, so this construction
reproduces the original strata probabilities. That proves completeness 
and equivalence.
\end{proof}

\subsection{Augmented graphs}
\label{app:augmented-graphs}

\begin{definition}[Augmented  model]
\label{def:augmented-model}
Let $\mathcal M_R$ denote a  relaxed causal model, and
let $C_\lambda$  be a population-level constraint for a
natural sensitivity analysis problem.
An \emph{augmented model} $M_A$ is a causal model constructed
by appending a binary auxiliary variable $H$ to the original model $\mathcal M_R$,
where the distribution conditional on $H=0$ satisfies
the population-level constraint $C_\lambda$, whereas no such
requirement is imposed on the distribution conditional on $H=1$.
\end{definition}

\begin{lemma}[Invariance under unrestricted augmentation]
\label{lem:unrestricted-augmentation}
Fix one district of a finite causal model and form an augmented enlargement
with indicator $H$. If no observed data restriction, structural assumption, or
estimand involves $H$, then projecting the augmented model onto the original
variables gives exactly the original feasible set of distributions. Therefore,
the identified set, and the respective sharp bounds, for any estimand defined on
the original variables remain invariant.
\end{lemma}

\begin{proof}[Proof of Lemma~\ref{lem:unrestricted-augmentation}]
\label{app:unrestricted-augmentation}
Firstly, take any data probability distribution in the augmented model and 
marginalize over
$H$. As $H$ has no children, the structural equations and principal
strata of the original variables are the same. 
At last, as there is no constraint involving $H$, the resulting 
marginal distribution belongs to the original model.

Any feasible distribution in the original model can be extended
by setting $H=0$, without changing the original variables.
As the estimand does not involve $H$, the two models have
the same feasible estimand values and sharp bounds.
\end{proof}

\renewcommand{\repeatedstatementheading}{Lemma~\ref{lem:augmented-individual-representation}}
\begin{repeatedstatement}[Individual-level representation under augmentation]
Assume that $C_\lambda$ holds within the $H=0$
component of the augmented model.
Then imposing $P(H=1)=0$ recovers exactly the original model that satisfies $C_\lambda$,
by projection onto the
original variables.
Therefore, relaxing this individual-level restriction to
$P(H=1)\leq\theta_\lambda^{\max}$ defines a sensitivity problem
to which Theorem~\ref{thm:functional-autobounds} and its corollaries
apply under their stated conditions.
\end{repeatedstatement}

\begin{proof}[Proof of Lemma~\ref{lem:augmented-individual-representation}]
\label{app:augmented-individual-representation}
Assume $P(H=1)=0$. Then the total population belongs to the $H=0$
component, so its probability distribution satisfies $C_\lambda$.
Projection does not change the original data constraints or the
retained model restrictions.
Any probability distribution in the original model satisfying
$C_\lambda$ can also be augmented by setting $H=0$ almost surely.
This does not change those constraints and satisfies $C_\lambda$
within $H=0$, proving equality of the projected feasible sets.

The restriction $P(H=1)=0$ excludes all augmented strata with $H=1$
and is therefore individual-level.
Keeping the requirement that $C_\lambda$ hold within $H=0$ among
the constraints in $\mathcal H_0(P_V)$, apply
Theorem~\ref{thm:functional-autobounds} with violation event $\{H=1\}$.
Finally, the corollaries of the theorem also apply under their stated conditions.
\end{proof}

\begin{lemma}[From an augmented indicator to an independent selector]
\label{lem:augmented-independent-selector}
Let $R$ denote the complete generalized principal stratum and let $H$
be a binary augmented indicator, with any joint distribution $P(H,R)$.
There exist two branch strata $R^0,R^1$ and a binary selector $S$,
independent of $(R^0,R^1)$, such that:
\begin{align*}
P(S=s,R^S=r)=P(H=s,R=r)
\qquad\text{for every }s,r.
\end{align*}
So, the doubled representation does not change the dependence between the
indicator and the selected strata, all constraints on their joint
probabilities, and any population-level restriction conditional on
$H=0$, which becomes conditional on $S=0$ when this event has positive probability.
\end{lemma}

\begin{proof}
Let $P(S=s)=P(H=s)$. If $P(H=s)=0$, choose any distribution
over strata for $R^s$.
For each $s$ with $P(H=s)>0$, let $R^s$ have
probabilities $P(R^s=r)=P(R=r\mid H=s)$.
Draw the strata for each branch independently of each other and of $S$,
and select $R^S$. 

Now, if $P(H=s)=0$,$P(S=s,R^S=r) = P(H=s,R=r) = 0$.
And for each $s$ with $P(H=s)>0$,
\begin{align*}
P(S=s,R^S=r)
&=P(S=s,R^s=r)\\
&=P(S=s)P(R^s=r)\\
&=P(H=s)P(R=r\mid H=s)\\
&=P(H=s,R=r).
\end{align*}

Therefore, $P(S=s,R^S=r)=P(H=s,R=r)$ for every $s,r$.
Then, replacing $(H,R)$ by $(S,R^S)$ does not change any
constraint on these probabilities or any estimand calculated from them.
Any doubled distribution also defines an augmented distribution
by setting $P(H=s,R=r)=P(S=s,R^S=r)$.
And the two representations give the same feasible distributions and 
estimand values, proving the statement.
\end{proof}

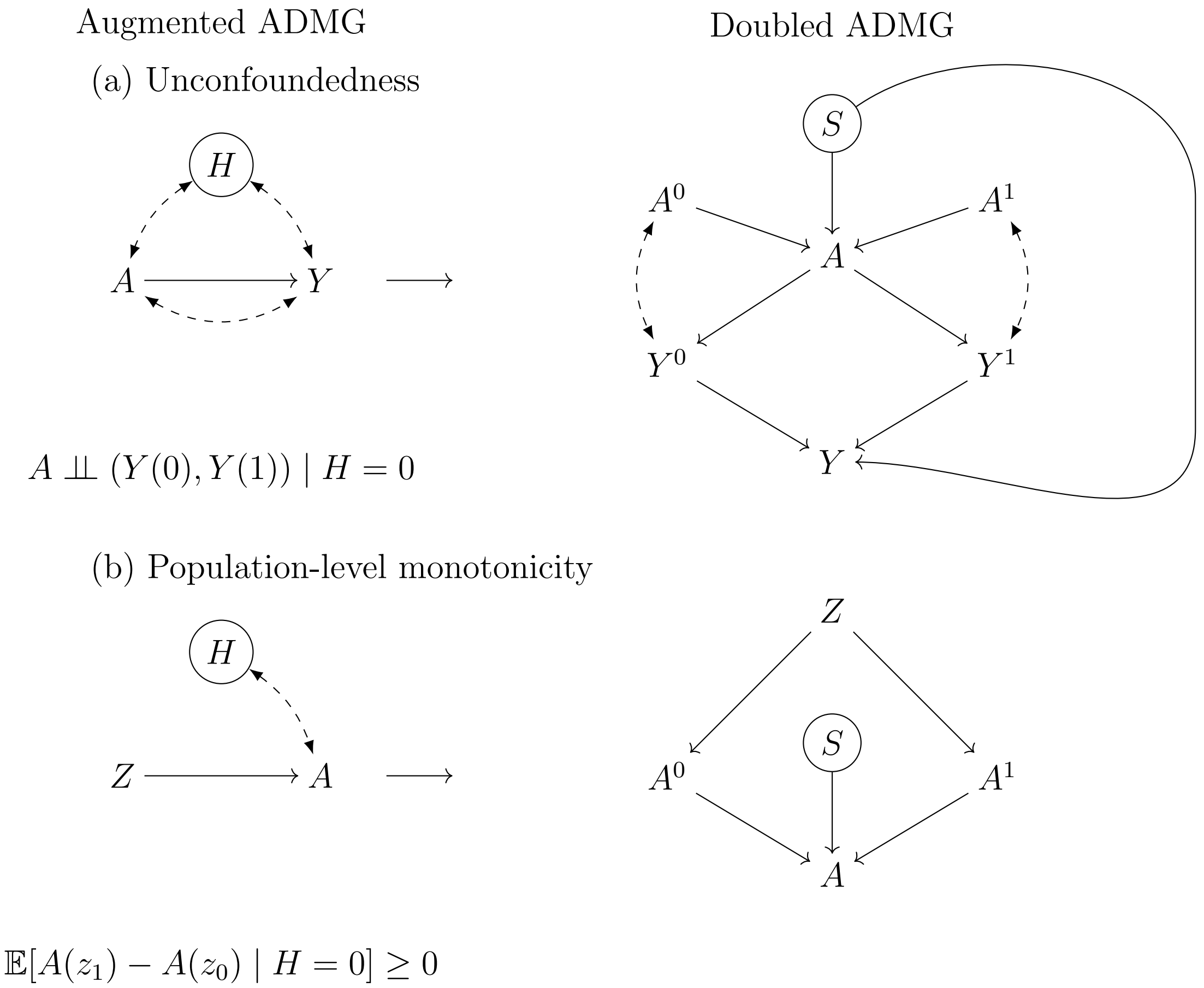
\begin{figure}[p]
\centering
\begin{tikzpicture}[every node/.style={inner sep=3pt}]
\node at (1.2,1) {Augmented ADMG};
\node at (8.6,1) {Doubled ADMG};
\node[anchor=west] at (-0.5,0.3) {(a) Unconfoundedness};
\node (a) at (0,-2.1) {$A$};
\node (y) at (2.4,-2.1) {$Y$};
\node[draw,circle] (h) at (1.2,-0.7) {$H$};
\path[bidirected] (h) edge[bend right=20] (a) edge[bend left=20] (y);
\path[bidirected] (a) edge[bend right=35] (y);
\path (a) edge[->] (y);
\draw[->] (3.2,-2.1) -- (4,-2.1);

\node (a0) at (6.6,-1.1) {$A^0$};
\node (a1) at (10.6,-1.1) {$A^1$};
\node (aa) at (8.6,-1.8) {$A$};
\node (y0) at (6.6,-3.1) {$Y^0$};
\node (y1) at (10.6,-3.1) {$Y^1$};
\node (yy) at (8.6,-4.3) {$Y$};
\node[draw,circle] (ss) at (8.6,-0.2) {$S$};
\path[bidirected] (a0) edge[bend right=30] (y0);
\path[bidirected] (a1) edge[bend left=30] (y1);
\path (a0) edge[->] (aa);
\path (a1) edge[->] (aa);
\path (ss) edge[->] (aa);
\draw[->] (ss) to[out=35,in=90] (13,-1.1) -- (13,-3.9)
  to[out=-90,in=0] (yy);
\path (aa) edge[->] (y0) edge[->] (y1);
\path (y0) edge[->] (yy);
\path (y1) edge[->] (yy);
\node at (1.2,-4.4) {$A\indep(Y(0),Y(1))\mid H=0$};

\node[anchor=west] at (-0.5,-5.6) {(b) Population-level monotonicity};
\node (zm) at (0,-8.1) {$Z$};
\node (dm) at (2.4,-8.1) {${\color{black}A}$};
\node[draw,circle] (hm) at (1.2,-6.6) {$H$};
\path[bidirected] (hm) edge[bend left=20] (dm);
\path (zm) edge[->] (dm);
\draw[->] (3.2,-8.1) -- (4,-8.1);
\node (zz) at (8.6,-6.1) {$Z$};
\node (d0) at (6.6,-8.1) {${\color{black}A^0}$};
\node (d1) at (10.6,-8.1) {${\color{black}A^1}$};
\node (ds) at (8.6,-9.3) {${\color{black}A}$};
\node[draw,circle] (sm) at (8.6,-7.7) {$S$};
\path (zz) edge[->] (d0) edge[->] (d1);
\path (d0) edge[->] (ds);
\path (d1) edge[->] (ds);
\path (sm) edge[->] (ds);
\node at (1.2,-10.4) {${\color{black}\E[A(z_1)-A(z_0)\mid H=0]\geq0}$};
\end{tikzpicture}
\caption{\textbf{Augmented and doubled ADMGs for population-level restrictions.}
Circles denote the latent component indicators $H$ and $S$.
Bidirected edges represent dependence through unobserved common causes.
On the left, $H$ may depend on the original strata. On the right,
branch $s$ has the original strata distribution conditional on $H=s$.
The branch strata and $S$ are drawn independently, with
$P(S=1)=P(H=1)\leq\theta^{\max}$.
In (a), $A=A^S$, each $Y^s$ is evaluated at this selected treatment,
and $Y=Y^S$. So, branch $0$ satisfies unconfoundedness as a restriction
on its strata probabilities. In (b), ${\color{black}A^s}$ is evaluated at the common
independent instrument $Z$, ${\color{black}A=A^S}$, and branch $0$ satisfies
${\color{black}\E[A^0(z_1)-A^0(z_0)]\geq0}$.
These branch restrictions and the deterministic selection equations
are imposed in addition to the ADMGs. All retained constraints apply
to the selected population distribution.}
\label{fig:population-doubled}
\end{figure}

Figure~\ref{fig:population-doubled} illustrates the augmented 
construction for
unconfoundedness and population-level monotonicity. In both cases, the
selector is independent of the two branches, but may be dependent on
the selected strata.

\FloatBarrier

\begin{lemma}[Unconfoundedness as a population-level constraint]
\label{lem:unconfounded-mixture}
For finite-valued $A$ and $Y$, unconfoundedness
$C_{\mathrm{UC}}:A\indep\mathbf Y_A$ is a population-level constraint
in Definition~\ref{def:population-constraint}.
In the augmented representation, when $q_{0++}>0$,
$A\indep\mathbf Y_A\mid H=0$ is equivalent to
\begin{equation}
\label{eq:unconfounded-strata-factorization}
q_{0i\br}q_{0++}=q_{0i+}q_{0+\br}
\quad\text{for every }a_i,\mathbf y_{\br},
\end{equation}
where a $+$ denotes summation over the corresponding coordinate.
When $q_{0++}=0$, these equations are vacuous.
Its component relaxation is therefore represented by these polynomial
equalities together with $\sum_{i,\br}q_{1i\br}=P(H=1)\leq\theta_{\mathrm{UC}}^{\max},$
with no unconfoundedness restriction on the $H=1$ component.
\end{lemma}

\begin{proof}[Proof of Lemma~\ref{lem:unconfounded-mixture}]
\label{app:unconfounded-mixture}
Write $p_{i\br}=P(a_i\mathbf y_{\br})$, with a $+$ denoting summation
over the corresponding coordinate. Unconfoundedness is equivalent to
$p_{i\br}=p_{i+}p_{+\br}$ for every $i,\br$, or to the single polynomial
constraint
\begin{align*}
h_{\mathrm{UC}}(\mathbf p)
=\sum_{i,\br}(p_{i\br}-p_{i+}p_{+\br})^2=0.
\end{align*}
Thus it is population-level under
Definition~\ref{def:population-constraint}.

If $q_{0++}>0$, divide
Equation~\eqref{eq:unconfounded-strata-factorization} by $q_{0++}^2$ to
obtain
\begin{align*}
P(a_i\mathbf y_{\br}\mid H=0)
=P(a_i\mid H=0)
P(\mathbf y_{\br}\mid H=0).
\end{align*}
Thus the polynomial equalities imply
$A\indep\mathbf Y_A\mid H=0$. Multiplying this conditional factorization
by $q_{0++}^2$ proves the converse part. If $q_{0++}=0$, the equalities are
vacuous and the conditional distribution on the null component can be arbitrarily chosen.

Marginalizing $H$ gives
$P(a_i\mathbf y_{\br})=q_{0i\br}+q_{1i\br}$.
Finally, summing the $H=1$ cells gives
$P(H=1)=\sum_{i,\br}q_{1i\br}$, which proves the sensitivity statement.

The mixture representation follows directly. Write $t=P(H=1)$.
For $0<t<1$, define the probability distributions
\begin{align*}
\alpha_i=\frac{q_{0i+}}{1-t},\qquad
\beta_{\br}=\frac{q_{0+\br}}{1-t},\qquad
\rho_{i\br}=\frac{q_{1i\br}}{t}.
\end{align*}
The factorization gives $q_{0i\br}=(1-t)\alpha_i\beta_{\br}$,
so marginalizing $H$ yields
\begin{align*}
P(a_i\mathbf y_{\br})
=(1-t)\alpha_i\beta_{\br}+t\rho_{i\br}.
\end{align*}
On the other hand, any such mixture defines augmented cells
$q_{0i\br}=(1-t)\alpha_i\beta_{\br}$ and
$q_{1i\br}=t\rho_{i\br}$, which satisfy
Equation~\eqref{eq:unconfounded-strata-factorization}.
At $t=0$ or $t=1$, choose the unused component distributions arbitrarily.
Therefore imposing $t\leq\theta_{\mathrm{UC}}^{\max}$ gives precisely the
contamination mixture with the stated weight bound.

\end{proof}

\subsection{Multiple sensitivity nodes and overlapping relaxations}
\label{app:multiple-model-selectors}

It is possible to relax many different assumptions by extending
the doubled representation with many binary selectors $S_1,\dots, S_J$.
In particular, the model implies  
up to $2^J$
branches, with branch $s$ retaining restriction $j$ when $s_j=0$ and
allowing its violation when $s_j=1$. Let
\begin{align*}
S=(S_1,\ldots,S_J),\qquad
\pi_s=P(S=s),\qquad s\in\{0,1\}^J.
\end{align*}
The joint distribution allows for the possibility 
of overlap among the relaxations, which typically happens when strata 
are constrained by multiple selectors.

I first prove that multiple relaxations of individual-level assumptions
have an equivalent mixture representation. I then show that, under the
stated assumptions on sensitivity graphs, recursively doubling the graphs
produces an ADMG with up to $2^J$ copies. Finally, I extend these results
to relaxations of population-level assumptions.

\begin{proposition}[Mixture representation of multiple relaxations]
\label{prop:multiple-model-selectors}
Let $I_1(r),\ldots,I_J(r)$ be individual-level violation indicators
on the complete states of a common finite canonical model.
Retain all observed-data and other model constraints on the recovered
distribution $p_r(\boldsymbol\eta)$.
For each $s\in\{0,1\}^J$, introduce a component distribution $q^s$
supported on states satisfying $I_j(r)=0$ whenever $s_j=0$.
Empty branches receive zero weight and are omitted.
Then the program
\begin{align*}
p_r(\boldsymbol\eta)
&=\sum_{s\in\{0,1\}^J}\pi_s q_r^s,
&&\text{for every }r,\\
\pi_s&\geq0,\qquad \sum_s\pi_s=1,\\
\sum_{s:s_j=1}\pi_s&\leq\theta_j^{\max},
&&j=1,\ldots,J,
\end{align*}
with each nonempty component normalized, has exactly the same feasible
recovered distributions as the simultaneous sensitivity model of
\red{Corollary~\ref{cor:simultaneous-individual-relaxations}}.

\end{proposition}

\begin{proof}
For any feasible mixture, the support restrictions imply
\begin{align*}
\sum_r I_j(r)p_r(\boldsymbol\eta)
=\sum_s\pi_s\sum_r I_j(r)q_r^s
\leq\sum_{s:s_j=1}\pi_s
\leq\theta_j^{\max}.
\end{align*}
The remaining model constraints hold by construction.
This suffices to prove that the feasible mixture program  produces a distribution satisfying the sensitivity bounds.

To prove the converse implication, take a feasible original distribution and let $\pi_s$
be the probability that its vector of violation indicators equals $s$.
For each branch with positive probability, choose $q^s$ as the conditional
probability distribution given that vector. Then the mixture equations follow from
the law of total probability, and the total probability of branches with $s_j = 1$ is equal to the 
corresponding violation proportion. A nonempty branch with zero probability
can use any distribution on its allowed support.
\end{proof}

\begin{proposition}[Recursive graphical representation]
For each selector configuration $s \in \{0,1\}^J$, let $\mathcal G_s$ be 
a corresponding branch ADMG. 
Suppose the directed edges of all the branch ADMGs have an acyclic union. 
{
Assume that each component distribution $q^s$ admits a causal
representation that is compatible with its branch ADMG $\mathcal G_s$.
}

{
An expanded ADMG can be constructed from disjoint copies of the branch
ADMGs, with one variable copy $V^s$ for each branch and selector arrows
entering the affected variables. Proposition~\ref{prop:enlarged-double}
establishes the corresponding equivalence with a mixture of two components.
}
Dependence among selectors might be represented by a common latent parent.

From the expanded ADMG, one can construct a sensitivity graph by merging
 variable copies and keeping all branch edges and selector arrows. 
The component probability distributions must allow any additional dependence 
induced by conditioning on selectors.
All the other constraints on the recovered distribution are kept.
\end{proposition}
\begin{proof}

To construct the expanded ADMG, first for each branch $s$, 
take a separate copy of its causal representation over a variable 
set $V^s$. Let the branch states be independent across copies,
and add the elements of the selector vector with joint probability 
$\pi$, independently of all branch states. 
The original variables can be recovered by 
selecting the same branch $s$ for all potential outcomes of a 
particular unit. So the recovered probability distribution is 
$\sum_s \pi_s q^s$.

Now construct the expanded graph from the disjoint ADMGs and add 
the selector arrows into the affected variables in each copy. 
It remains to show that this graph is acyclic.
First, each branch copy has an acyclic directed part, and there are no 
directed edges between the distinct copies. Second,  
arrows from selectors point to the copies, but no directed arrow from a copy 
points to a selector. At last, there are no 
directed edges among selectors. Dependence 
among selectors can be represented by a common latent parent whose only children 
are the selectors. Its latent projection introduces bidirected edges among selectors 
without changing the directed part of the graph. Therefore, the expanded graph is acyclic.

Now to construct the sensitivity graph, merge corresponding variable copies
in the expanded ADMG, keeping all branch edges and selector arrows. 
Then, the resulting graph will have no directed cycles because the union of branch 
directed edges is acyclic and no directed arrow from a branch copy points to a selector.

Finally, adding the retained constraints on the recovered 
probability distribution guarantees that the recovered
distribution satisfies the observed data and other model requirements.

\end{proof}

\begin{corollary}[Mixture representation of population-level relaxations]
Suppose the population-level constraints are jointly represented
in a common finite augmented canonical model, as in
Lemma~\ref{lem:augmented-individual-representation}.
Then, Proposition~\ref{prop:multiple-model-selectors} gives an equivalent
mixture representation of their simultaneous relaxations.
All retained constraints, including the population-level ones
in the augmented components, are imposed on the recovered
augmented distribution, rather than separately in every
selector branch.
\end{corollary}

\begin{proof}
Let $H_j$ be the augmented indicator for restriction $j$.
On each complete augmented state $r$, define
$I_j(r)=\mathbf{1}\{H_j(r)=1\}$.
The relaxation bounds are then bounds on individual-level
violation indicators. Applying
Proposition~\ref{prop:multiple-model-selectors} to these indicators and 
 retaining all other constraints on the recovered augmented
distribution suffices to  prove the result.
\end{proof}

Different assumptions may be relaxed for the same units. 
The joint probability distribution of the selectors allows this overlap without 
requiring the selectors to be independent or mutually exclusive.
Consider two relaxations, 
and let 
$\theta_j=P(S_j=1)$ and
$\omega=P(S_1=1,S_2=1)$. Their overlap satisfies the Frechet bounds:
\begin{align*}
\max\{0,\theta_1+\theta_2-1\}
\leq\omega\leq\min\{\theta_1,\theta_2\}.
\end{align*}
Any value in those inequalities is feasible for the overlap of the two relaxations.
Assume first the selectors are mutually independent, then 
their overlap is $\omega=\theta_1\theta_2$.
In fact, imposing it can restrict the sensitivity model.
Mutual exclusivity, on the other hand, is different. 
It requires $\pi_s=0$ whenever
more than one coordinate of $s$ is one, a support restriction that
an ordinary ADMG alone does not encode.

Furthermore, the mixture representation 
is invariant to the order in which restrictions are
listed, if the branch definitions and all compatibility constraints
are fixed on the same expanded model. Reordering merely permutes the
coordinates of $s$ and the corresponding proportions and probability distributions 
but does not change $\sum_s\pi_s q_r^s$. 
For example, jointly relaxing no defiers and exclusion in the IV model
gives four branches: a) both restrictions retained, b) only no defiers relaxed,
c) only exclusion relaxed, d) and both relaxed.

Alternatively, a single selector can distinguish a component
satisfying all restrictions from one in which all are relaxed. This gives
a one-dimensional analysis, with the proportion of the latter component
bounded by a single sensitivity parameter. For individual-level
restrictions, the corresponding compound violation event is their union.
In the instrumental variable case, that would mean for the mixture representation 
a branch between a model satisfying at the same time
 exclusion restriction and monotonicity and another branch, relaxing both.
For two restrictions, the probability is
\begin{align*}
P(I_1=1\text{ or }I_2=1)
=P(I_1=1)+P(I_2=1)-P(I_1=1,I_2=1).
\end{align*}
Therefore, this  analysis does not in general coincide with assigning
the same marginal bound to each separate selector.

\section{Results for Standard Sensitivity Scenarios}
\label{app:scenario-results}

\subsection{Symbolic sensitivity function for simple confounding}
\label{app:simpleconfounding}

{
\renewcommand{\repeatedstatementheading}{Theorem~\ref{thm:confounding-bounds}}
\begin{repeatedstatement}[Natural sensitivity bounds under classic confounding]
\singlespacing
Let $\theta=\theta_{\mathrm{UC}}\in[0,1]$ be the nominal weight of the
component in which confounding is permitted, and suppose the mixture is
compatible with the observed distribution. Write
$p_{ay}=P(A=a,Y=y)$ and $t_{ha}=P(H=h,A=a)$, as in
Appendix~\ref{app:simpleconfounding}. The feasible range of $t_{0a}$ is
\begin{align*}
t_{0a}^{-}=\max\{0,P(A=a)-\theta\}
\leq t_{0a}\leq
\min\{1-\theta,P(A=a)\}=t_{0a}^{+}.
\end{align*}
At each fixed $t_{0a}>0$, define
\begin{align*}
L_{P(Y(A=a)=y)}(t_{0a})
&=p_{ay}+\max\left\{0,
\frac{(1-\theta-t_{0a})(t_{0a}-p_{a,1-y})}{t_{0a}}\right\},\\
U_{P(Y(A=a)=y)}(t_{0a})
&=\theta+t_{0a}-p_{a,1-y}
+\min\left\{1-\theta-t_{0a},
\frac{(1-\theta-t_{0a})p_{ay}}{t_{0a}}\right\}.
\end{align*}
At $t_{0a}=0$, these bounds are $p_{ay}$ and $1-p_{a,1-y}$,
respectively.
{
The sharp marginal bounds are the minimum of $L$ and maximum of $U$
over the two endpoints $t_{0a}^{-},t_{0a}^{+}$.

For the ATE, let $L_a(x)=L_{P(Y(A=a)=1)}(x)$ and
$U_a(x)=U_{P(Y(A=a)=1)}(x)$. Both potential outcomes must use the
same allocation $t=t_{01}$, with $t_{00}=1-\theta-t$. Define
\begin{align*}
L_{\mathrm{ATE}}(t)&=L_1(t)-U_0(1-\theta-t),&
U_{\mathrm{ATE}}(t)&=U_1(t)-L_0(1-\theta-t).
\end{align*}
Let $\mathcal T=[t_{01}^{-},t_{01}^{+}]$ and
\begin{align*}
\mathcal B_L&=\{t_{01}^{-},t_{01}^{+},p_{10},1-\theta-p_{01}\}
\cap\mathcal T,&
\mathcal B_U&=\{t_{01}^{-},t_{01}^{+},p_{11},1-\theta-p_{00}\}
\cap\mathcal T.
\end{align*}
The sharp ATE sensitivity bounds are
\begin{align*}
\mathcal L_{\mathrm{ATE}}(\theta)
&=\min_{t\in\mathcal B_L}L_{\mathrm{ATE}}(t),&
\mathcal U_{\mathrm{ATE}}(\theta)
&=\max_{t\in\mathcal B_U}U_{\mathrm{ATE}}(t).
\end{align*}
The boundary conventions above apply when $t=0$ or $1-\theta-t=0$.
}
\end{repeatedstatement}
}

\begin{proof}
{
The bounds over the full feasible interval and their sharpness follow from
Theorems~\ref{eq:proof_sharpness_conf},
\ref{thm:confounding-marginal-endpoints}, and
\ref{thm:confounding-ate-candidates}.
Theorem~\ref{thm:confounding-marginal-endpoints} reduces the marginal
optimization to the feasible endpoints, with the other treatment and
outcome values obtained by relabeling.
Theorem~\ref{thm:confounding-ate-candidates} reduces the ATE optimization
to the candidate sets $\mathcal B_L$ and $\mathcal B_U$.
}
\end{proof}

If $P(A=a)>0$, $\theta=0$ gives the point-identified value
$P(Y=y\mid A=a)$, while $\theta=1$ gives the usual
no assumption bounds $[p_{ay},1-p_{a,1-y}]$.
If $P(A=a)=0$ and $\theta=0$, the target instead remains unrestricted in
$[0,1]$.

\subsubsection{Setting the optimization problem}
For binary treatment $A$ and outcome $Y$, write $h_ha_i y_{rs}$
for the augmented stratum satisfying $H=h, A=i, (Y(0),Y(1))=(r,s),$
where all indices are binary. The $h_0$ strata form the unconfounded
component, and the $h_1$ strata form the unrestricted component.
Write $\theta=\theta_{\mathrm{UC}}=P(H=1)$.
By consistency, the observed cells are
\begin{align*}
P(A=1,Y=1)
&=P(h_0a_1y_{01})+P(h_0a_1y_{11})+P(h_1a_1y_{01})+P(h_1a_1y_{11}),\\
P(A=1,Y=0)
&=P(h_0a_1y_{00})+P(h_0a_1y_{10})
+P(h_1a_1y_{00})+P(h_1a_1y_{10}),\\
P(A=0,Y=1)
&=P(h_0a_0y_{10})+P(h_0a_0y_{11})
+P(h_1a_0y_{10})+P(h_1a_0y_{11}),\\
P(A=0,Y=0)
&=P(h_0a_0y_{00})+P(h_0a_0y_{01})
+P(h_1a_0y_{00})+P(h_1a_0y_{01})
\end{align*}
The stratum probabilities are nonnegative and sum to one, with
$\sum_{i,r,s}P(h_1a_i y_{rs})=\theta.$
Unconfoundedness within $H=0$ imposes
$$
(1-\theta_{\mathrm{UC}})P(h_0a_i y_{rs})
=\left\{\sum_{r',s'}P(h_0a_i y_{r's'})\right\}
 \left\{\sum_{i'}P(h_0a_{i'}y_{rs})\right\}
\quad\text{for every }i,r,s
$$

The first objective function is $P(Y(A=1)=1)$. 
The translation of this estimand into strata 
shows that it is composed of a point identified part,
$P(A=1,Y=1)$, and a partially identified part,
$P(H=0,A=0,Y(A=1)=1)$ and $P(H=1,A=0,Y(A=1)=1)$, 
so bounds on the latter are sufficient. In particular, 
the partially identified part has
quantities that belong respectively to the 
unconfounded and the confounded arms. 
\begin{align*}
P(Y(A=1)=1)
=\sum_{k=0}^1\sum_{i=0}^1
\big[P(h_ka_i y_{01})+P(h_k a_i y_{11})\big] = \\
\begin{aligned}
&\underbrace{P(h_0a_1y_{01})+P(h_0a_1y_{11})
+P(h_1a_1y_{01})+P(h_1a_1y_{11})}_{P(A=1,Y=1)}\\
&+\underbrace{P(h_0a_0y_{01})+P(h_0a_0y_{11})}_{P(H=0,A=0,Y(A=1)=1)}
+\underbrace{P(h_1a_0y_{01})+P(h_1a_0y_{11})}_{P(H=1,A=0,Y(A=1)=1)}.
\end{aligned}
\end{align*}
We bound both contributions at a fixed 
treatment allocation and then optimize their sum over all feasible allocations.

For the ATE, the strata with $Y(0)=Y(1)$ cancel, leaving
\begin{align*}
\operatorname{ATE}
&=P(Y(1)=1)-P(Y(0)=1)\\
&=\sum_{k=0}^1\sum_{i=0}^1
\big[P(h_ka_i y_{01})-P(h_ka_i y_{10})\big]\\
&=P(h_0a_0y_{01})+P(h_0a_1y_{01})
-P(h_0a_0y_{10})-P(h_0a_1y_{10})\\
&\quad+P(h_1a_0y_{01})+P(h_1a_1y_{01})
-P(h_1a_0y_{10})-P(h_1a_1y_{10}).
\end{align*}
There is still an identified contribution from the observed outcomes.
By consistency, the same objective can be written as
\begin{align*}
\operatorname{ATE}
&=\underbrace{P(A=1,Y=1)-P(A=0,Y=1)}_{\text{identified contribution}}\\
&\quad+\sum_{k=0}^1\big[P(h_ka_0y_{01})+P(h_ka_0y_{11})\big]\\
&\quad-\sum_{k=0}^1\big[P(h_ka_1y_{10})+P(h_ka_1y_{11})\big].
\end{align*}

The global solutions for these optimization problems
are at the same time the perturbation and the sensitivity functions, because the 
former is suitably monotonic, as shown in Theorem~\ref{thm:monotonic-conf-bounds}.

\subsubsection{Bounds for $P(Y(A=1)=1)$ and the ATE}
\label{app:bounds_clas_confound}

The two nonidentified elements can be optimized. 
Call $p_u = P(H=0,A=0,Y(A=1)=1)$ and $p_c = P(H=1,A=0,Y(A=1)=1)$.

The idea is to characterize the feasible set in parts.
Denote $t_{ha} = P(H=h,A=a)$. Then let $t_{01} = P(H=0,A=1) = \sum_{r,s} P(h_0a_1y_{r,s})$. 
The feasible values for $t$ are upper-bounded by $P(H=0) = 1 - \theta$ and $P(A=1)$ 
by containment. Analogously, $P(H=0,A=1) = P(A=1) - P(H=1,A=1) \geq P(A=1) - \theta$,
so:
\begin{align*}
   \max\{0,  P(A=1) - \theta \} \leq t_{01} \leq \min\{1 - \theta, P(A=1)\}
\end{align*} 

Notice that by fixing $t_{01}$, then the other t-quantities are immediately identifiable:
\begin{align*}
    t_{11} = P(A=1) - t_{01}, \quad t_{00} = 1 - \theta - t_{01}, \quad t_{10} = \theta + t_{01} - P(A=1)
\end{align*}

Before targeting $p_u$,
consider $P(H=0, A=1, Y(A=1)=1)$, which is equal to 
$P(H=0, A=1, Y=1)$, by consistency. 
Denote $P(H=h,A=a,Y=y)$ by $q_{hay}$, so $q_{011} = P(H=0, A=1, Y(A=1)=1)$.
By containment, 
$q_{011}$ is upper-bounded by $t_{01}$ and $P(A=1,Y=1)$. 
Analogously,  
 $q_{011} = P(A=1,Y=1) - q_{111} \geq P(A=1,Y=1) - \min\{P(A=1,Y=1), t_{11}\}$.
As $t_{11}$ is a function of $t_{01}$:
\begin{equation}
    \label{eq:bounds_simple_3}
    \max\{0,  t_{01} - P(A=1,Y=0)\} \leq q_{011} \leq \min\{t_{01}, P(A=1,Y=1)\}
\end{equation}

With fixed $q_{011}, t_{00}, t_{01}, t_{10}, t_{11}$, 
it is possible to consider the bounds for $p_u$. 
Unconfoundedness implies the identity
$$
p_u t_{01}=q_{011}t_{00}.
$$
When both treatment-group masses are positive, this is equivalent to
$P(Y(A=1)=1\mid H=0,A=1)=P(Y(A=1)=1\mid H=0,A=0)$.
Assume first $t_{01}$ is positive. The cases where $t_{01} = 0$ will
be handled separately.

For further simplification, denote $p_{ay} = P(A=a,Y=y)$.
Substituting into Equation~\ref{eq:bounds_simple_3}:
\begin{align*}
    \frac{t_{00}}{t_{01}}\cdot\max\{0, t_{01} - p_{10} \} 
    \leq p_u \leq 
    \frac{t_{00}}{t_{01}}\cdot\min\{t_{01}, p_{11}\} 
\end{align*}
Substituting $t_{00}=1-\theta-t_{01}$ and rearranging gives:
\begin{equation}
    \label{eq:bounds_pu}
    \max\Big\{0, \frac{(1 - \theta - t_{01}) (t_{01} - p_{10})}{t_{01}} \Big\} 
    \leq p_u \leq 
    \min\Big\{(1 - \theta - t_{01}), \frac{(1 - \theta - t_{01})p_{11}}{t_{01}}\Big\}
\end{equation}

Now consider $p_c$. By containment,   $0 \leq p_c \leq t_{10}$.
As $t_{10}$ is a function of $t_{01}$:
\begin{equation}
    \label{eq:bounds_pc}
    0 \leq p_c \leq \theta + t_{01}- P(A=1)
\end{equation}

That is sufficient to show that $p_u$ and $p_c$  can be fully bounded 
considering the feasible values of $t_{01}$. 
Then the bounds for $P(Y(A=1)=1)$ can be computed by 
summing the bounds of $p_u$, $p_c$, and the identified part, which 
is equal to $p_{11}$.

For a particular $t_{01}$, the bounds $L_{P(Y(A=1)=1)}(t_{01})$ and $U_{P(Y(A=1)=1)}(t_{01})$ are:
\begin{align*}
L_{P(Y(A=1)=1)}(t_{01}) &= p_{11} + \max\Big\{0,
\frac{(1 - \theta - t_{01}) (t_{01} - p_{10})}{t_{01}} \Big\} \\
U_{P(Y(A=1)=1)}(t_{01}) &= \theta + t_{01} - p_{10} +     
\min\Big\{1 - \theta - t_{01}, \frac{(1 - \theta - t_{01})p_{11}}{t_{01}}\Big\} 
\end{align*}
For the case $t_{01} = 0$, then
$L_{P(Y(A=1)=1)}(0) = p_{11}$ and $U_{P(Y(A=1)=1)}(0) = 1 - p_{10}$.

Notice that by changing the value of $A$ from $1$ to $0$, 
the bounds for $P(Y(A=0)=1)$, when $t_{00}>0$, are:
\begin{align*}
L_{P(Y(A=0)=1)}(t_{00}) &= p_{01} + \max\Big\{0,
\frac{(1 - \theta - t_{00}) (t_{00} - p_{00})}{t_{00}} \Big\} \\
U_{P(Y(A=0)=1)}(t_{00}) &= \theta + t_{00} - p_{00} +     
\min\Big\{1 - \theta - t_{00}, \frac{(1 - \theta - t_{00})p_{01}}{t_{00}}\Big\} 
\end{align*}
At $t_{00}=0$, the corresponding bounds are
$L_{P(Y(A=0)=1)}(0)=p_{01}$ and
$U_{P(Y(A=0)=1)}(0)=1-p_{00}$.

To calculate the bounds for the ATE, first reconstruct the 
bounds for $P(Y(A=0)=1)$ as a function of $t_{01}$, using the fact 
that  $t_{00} = 1 - \theta - t_{01}$:
\begin{align*}
L_{P(Y(A=0)=1)}(t_{01}) &= p_{01} + \max\Big\{0,
\frac{(t_{01}) (1 - \theta - t_{01} - p_{00})}{1 - \theta - t_{01}} \Big\} \\
U_{P(Y(A=0)=1)}(t_{01}) &= 1 - t_{01} - p_{00} +     
\min\Big\{t_{01}, \frac{(t_{01})p_{01}}{1 - \theta - t_{01}}\Big\} 
\end{align*}

At last, for a particular $t_{01}$, valid bounds for the ATE can be calculated as:
\begin{align*}
L_{\text{ATE}}(t_{01})
&=L_{P(Y(A=1)=1)}(t_{01})-U_{P(Y(A=0)=1)}(t_{01})\\
&=p_{11}+p_{00}+t_{01}-1
+\max\Big\{0,\frac{(1-\theta-t_{01})(t_{01}-p_{10})}{t_{01}}\Big\}\\
&\quad-\min\Big\{t_{01},\frac{t_{01}p_{01}}{1-\theta-t_{01}}\Big\},\\[6pt]
U_{\text{ATE}}(t_{01})
&=U_{P(Y(A=1)=1)}(t_{01})-L_{P(Y(A=0)=1)}(t_{01})\\
&=\theta+t_{01}-p_{10}-p_{01}
+\min\Big\{1-\theta-t_{01},
\frac{(1-\theta-t_{01})p_{11}}{t_{01}}\Big\}\\
&\quad-\max\Big\{0,
\frac{t_{01}(1-\theta-t_{01}-p_{00})}{1-\theta-t_{01}}\Big\}.
\end{align*}
The fractional expressions apply when $0<t_{01}<1-\theta$.
At $t_{01}=0$, use the bounds $[p_{11},1-p_{10}]$ for
$P(Y(1)=1)$, and, at $t_{00}=0$, use $[p_{01},1-p_{00}]$ for
$P(Y(0)=1)$. 

Let the feasible set of $t_{01}$ be $\mathcal{T}_{01} = [  \max\{0,  P(A=1) - \theta \} , \min\{1 - \theta, P(A=1)\}]$.
Therefore, the general lower and upper bounds for both considered estimands, $(\mathcal{L}_{P(Y(A=1)=1)}, \mathcal{U}_{P(Y(A=1)=1)})$ 
and $(\mathcal{L}_{ATE}, \mathcal{U}_{ATE})$ are:
\begin{equation} 
    \label{eq:general_bound_on_ya1_cond_t01}
    (\mathcal{L}_{P(Y(A=1)=1)},\quad \mathcal{U}_{P(Y(A=1)=1)}) = (\min_{t_{01} \in \mathcal{T}_{01} } L_{P(Y(A=1)=1)}(t_{01}),\quad \max_{t_{01} \in \mathcal{T}_{01} } U_{P(Y(A=1)=1)}(t_{01}))
\end{equation}
and 
\begin{equation} 
        \label{eq:general_bound_on_ate_cond_t01}
    (\mathcal{L}_{ATE},\quad \mathcal{U}_{ATE}) = (\min_{t_{01} \in \mathcal{T}_{01} } L_{ATE}(t_{01}),\quad \max_{t_{01} \in \mathcal{T}_{01} } U_{ATE}(t_{01}) )
\end{equation}

\subsubsection{Sharpness for $P(Y(A=1)=1)$ and ATE bounds}

\begin{theorem}
    \label{eq:proof_sharpness_conf}
    The bounds calculated in Appendix~\ref{app:bounds_clas_confound} 
    are valid and sharp for $P(Y(A=1)=1)$ and the ATE.
\end{theorem}
\begin{proof}
Validity follows from Appendix~\ref{app:bounds_clas_confound}. For
sharpness, fix any feasible $t_{01}$ and the resulting treatment masses
$t_{ha}$. Independently for $a=0,1$, choose
$$
q_{0a1}\in[\max\{0,t_{0a}-p_{a0}\},\min\{t_{0a},p_{a1}\}],
\qquad q_{0a0}=t_{0a}-q_{0a1},
$$
and set $q_{1ay}=p_{ay}-q_{0ay}$. Both intervals for $q_{001}$ and $q_{011}$ 
contain feasible values that determine 
probabilities that satisfy the observed data and sum to the fixed $t_{ha}$.

When $t_{00},t_{01}>0$, construct the unconfounded strata by
$$P(h_0a_i y_{rs})
=t_{0i}\frac{q_{00r}}{t_{00}}\frac{q_{01s}}{t_{01}},
\qquad i,r,s\in\{0,1\}.
$$
Those probabilities reproduce every $q_{0ay}$. Both treatment and control groups
have the same joint potential outcome distribution in $H=0$, so all the unconfoundedness
equations hold. Then, every pair $(q_{001}, q_{011})$ can be attained jointly, 
including opposite endpoints of their intervals.
For $H=1$, each probability $q_{1ay}$ can be split arbitrarily between the two
strata with $Y(a)=y$, because the other potential outcome is unobserved.

Without loss of generality, consider $P(Y(1)=1)$, as
 the marginal bounds
for $P(Y(0)=1)$ follow by relabeling treatment. For positive
$t_{00},t_{01}$, the construction gives
\begin{align*}
P(Y(1)=1)
=p_{11}+\frac{t_{00}}{t_{01}}q_{011}
  +P(h_1a_0y_{01})+P(h_1a_0y_{11}),
\end{align*}
where $t_{00}=1-\theta-t_{01}$,
$t_{10}=\theta+t_{01}-P(A=1)$, and $t_{11}=P(A=1)-t_{01}$.
For the lower bound on $P(Y(1)=1)$, choose
$q_{011}=\max\{0,t_{01}-p_{10}\}$ and put the untreated $H=1$
masses in $h_1a_0y_{00}$ and $h_1a_0y_{10}$, making their
contribution zero. For the upper bound, choose
$q_{011}=\min\{t_{01},p_{11}\}$ and put those masses in
$h_1a_0y_{01}$ and $h_1a_0y_{11}$, making their contribution
$t_{10}$. Thus the attained values are
\begin{align*}
L_{P(Y(A=1)=1)}(t_{01})
&=p_{11}+\frac{t_{00}}{t_{01}}\max\{0,t_{01}-p_{10}\}\\
&=p_{11}+\max\Big\{0,
\frac{(1-\theta-t_{01})(t_{01}-p_{10})}{t_{01}}\Big\},\\
U_{P(Y(A=1)=1)}(t_{01})
&=p_{11}+\frac{t_{00}}{t_{01}}\min\{t_{01},p_{11}\}+t_{10}\\
&=\theta+t_{01}-p_{10}+\min\Big\{1-\theta-t_{01},
\frac{(1-\theta-t_{01})p_{11}}{t_{01}}\Big\}.
\end{align*}
To complete the strata distribution, choose any 
feasible $q_{001}$, which will not affect $P(Y(A=1)=1)$.

For the ATE, we also use
$$P(Y(0)=1)
=p_{01}+\frac{t_{01}}{t_{00}}q_{001}
  +P(h_1a_1y_{10})+P(h_1a_1y_{11}).$$

For the lower ATE bound, choose
$q_{011}=\max\{0,t_{01}-p_{10}\}$ and
$q_{001}=\min\{t_{00},p_{01}\}$. Within $H=1$, assign $Y(1)=0$
to untreated units and $Y(0)=1$ to treated units, preserving their
observed outcomes. Their missing-outcome contributions are then
$0$ and $t_{11}$, respectively. Subtracting the two expressions above gives
\begin{align*}
L_{\mathrm{ATE}}(t_{01})
&=p_{11}-p_{01}
  +\frac{t_{00}}{t_{01}}\max\{0,t_{01}-p_{10}\}\\
&\quad-\frac{t_{01}}{t_{00}}\min\{t_{00},p_{01}\}-t_{11}\\
&=L_{P(Y(A=1)=1)}(t_{01})-U_{P(Y(A=0)=1)}(t_{01}).
\end{align*}
For the upper ATE bound, choose
$q_{011}=\min\{t_{01},p_{11}\}$ and
$q_{001}=\max\{0,t_{00}-p_{00}\}$. Assign $Y(1)=1$ to untreated
$H=1$ units and $Y(0)=0$ to treated $H=1$ units, giving contributions
$t_{10}$ and $0$, respectively. Consequently,
\begin{align*}
U_{\mathrm{ATE}}(t_{01})
&=p_{11}-p_{01}
  +\frac{t_{00}}{t_{01}}\min\{t_{01},p_{11}\}+t_{10}\\
&\quad-\frac{t_{01}}{t_{00}}\max\{0,t_{00}-p_{00}\}\\
&=U_{P(Y(A=1)=1)}(t_{01})-L_{P(Y(A=0)=1)}(t_{01}).
\end{align*}

By this construction, both lower and upper bounds 
for the ATE are attained by distributions of strata 
that satisfy all the constraints, so proving sharpness.

It is relevant to consider the case 
where $t_{0a}=0$. Without loss of generality, 
consider $t_{01}=0$. 
As $P(Y(A=1)=1) = P(A=1,Y=1) + P(A=0,Y(A=1)=1)$, 
if the unconfounded component 
contains no treated units, setting $Y(A=1)=0$ 
for all untreated units implies $P(A=0, Y(A=1)=1) = 0$ and
 $P(Y(A=1)=1) = P(A=1,Y=1)$, attaining the lower bound.  
 On the other hand, setting $Y(A=1)=1$ for all the untreated units 
 implies $P(A=0, Y(A=1)=1)=P(A=0)$, so 
 $P(Y(A=1)=1) = P(A=1,Y=1) + P(A=0)$, attaining the upper bound. 
 Each probability assignment satisfies the  observed data
  by consistency and 
 satisfies unconfoundedness within $H=0$, as treatment is constant in
 that component. For ATE sharpness, at $t_{01}=0$, 
 setting $Y(A=1)=0$ for all untreated units and $Y(A=0)=1$
for all the treated units and, conversely,
setting $Y(A=1)=1$ for all untreated units and $Y(A=0)=0$
for all the treated units will give respectively, 
$L_{ATE} = -p_{10} - p_{01}$ and $U_{ATE} = p_{11} + p_{00}$. 
This preserves observed data, and the ATE bounds are also sharp.

\end{proof}

\subsubsection{Optimizing over $t_{01}$}

The bounds calculated in Equations~\ref{eq:general_bound_on_ya1_cond_t01}
and~\ref{eq:general_bound_on_ate_cond_t01} require searching
over the feasible set of $t_{01}$ of infinite cardinality. 
In fact, it is possible to reduce this search to finite sets.
This is because possible maxima and minima can be found either 
in points of degeneracy or among the extrema of feasible values of $t_{01}$. 
This can be shown for both $P(Y(A=1)=1)$ and for the ATE.

\begin{theorem}[Evaluation of bounds for $P(Y(A=1)=1)$ at finite points of $\mathcal{T}_{01}$]
\label{thm:confounding-marginal-endpoints}
Fix $\theta$ and use the fixed-allocation bounds derived above.
The sharp bounds for $P(Y(A=1)=1)$ are
\begin{align*}
\mathcal L_{P(Y(A=1)=1)}(\theta)
&=\min\Big\{
L_{P(Y(A=1)=1)}\big(\max\{0,P(A=1)-\theta\}\big),\\
&\hspace{2em}L_{P(Y(A=1)=1)}\big(\min\{1-\theta,P(A=1)\}\big)
\Big\},\\
\mathcal U_{P(Y(A=1)=1)}(\theta)
&=\max\Big\{
U_{P(Y(A=1)=1)}\big(\max\{0,P(A=1)-\theta\}\big),\\
&\hspace{2em}U_{P(Y(A=1)=1)}\big(\min\{1-\theta,P(A=1)\}\big)
\Big\}.
\end{align*}
Similarly, the sharp bounds for $P(Y(A=0)=1)$ are
\begin{align*}
\mathcal L_{P(Y(A=0)=1)}(\theta)
&=\min\Big\{
L_{P(Y(A=0)=1)}\big(\max\{0,P(A=1)-\theta\}\big),\\
&\hspace{2em}L_{P(Y(A=0)=1)}\big(\min\{1-\theta,P(A=1)\}\big)
\Big\},\\
\mathcal U_{P(Y(A=0)=1)}(\theta)
&=\max\Big\{
U_{P(Y(A=0)=1)}\big(\max\{0,P(A=1)-\theta\}\big),\\
&\hspace{2em}U_{P(Y(A=0)=1)}\big(\min\{1-\theta,P(A=1)\}\big)
\Big\}.
\end{align*}
All four fixed-allocation bounds use the same argument $t_{01}$,
with $t_{00}=1-\theta-t_{01}$. At any feasible allocation with
$t_{01}=0$ or $t_{00}=0$, use
\begin{align*}
L_{P(Y(A=a)=1)}(t_{01})=p_{a1},\qquad
U_{P(Y(A=a)=1)}(t_{01})=1-p_{a0},\quad a\in\{0,1\}.
\end{align*}
For $0<t_{01}<1-\theta$, use the fractional formulas derived above.
Their branches change at $p_{10}$ and $p_{11}$ for $P(Y(A=1)=1)$,
and at $1-\theta-p_{00}$ and $1-\theta-p_{01}$ for $P(Y(A=0)=1)$.
Only points within the feasible interval are considered.

\end{theorem}

\begin{proof}
    The key idea is to
optimize
the bounds derived above over the feasible interval
$\mathcal{T}_{01} = [t_{01}^{-},t_{01}^{+}]$, where
$$t_{01}^{-}=\max\{0,P(A=1)-\theta\},\qquad
t_{01}^{+}=\min\{1-\theta,P(A=1)\}.
$$

First, if $0 \in \mathcal{T}_{01}$, then $L_{P(Y(A=1)=1)}(0)=p_{11}$,
as no other bound can be smaller, as this is exactly the natural 
bound for confounded models. 
Second, if $0 \not\in \mathcal{T}_{01}$, and if any feasible 
$t_{01} \leq p_{10}$, that will also include the minimum of the feasible
set, and the lower bound is again $p_{11}$, which is the lowest possible value.
At last, if $0 \not\in \mathcal{T}_{01}$ and 
every feasible $t_{01} > p_{10}$, the second derivative of the lower bound 
with respect to $t_{01}$ is nonpositive. Therefore, the function is concave, 
and the lower bound must be found at an endpoint of the feasible interval.

\begin{align*}
L_{P(Y(A=1)=1)}(t_{01})
&=1-\theta+P(A=1)-t_{01}
-\frac{(1-\theta)p_{10}}{t_{01}},\\
\frac{d^2}{dt_{01}^2}L_{P(Y(A=1)=1)}(t_{01})
&=-\frac{2(1-\theta)p_{10}}{t_{01}^3}\leq0.
\end{align*}

Notice that this result works for computing lower bounds 
for $P(Y(A=1)=0), P(Y(A=0)=1), P(Y(A=0)=0)$, by changing 
the polarity of each binary $A$ or $Y$.

The upper bound of $P(Y(A=1)=1)$  can be computed by 
maximizing $1 - P(Y(A=1)=0)$, which is equivalent to minimizing 
$P(Y(A=1)=0)$, so necessarily it requires  checking the endpoints of $\mathcal{T}_{01}$.

\end{proof}

Both $\mathcal{L}_{\text{ATE}}$ and $\mathcal{U}_{\text{ATE}}$ are constructed 
over differences of bounds of $P(Y(A=1)=1)$ and $P(Y(A=0)=1)$ evaluated 
for the whole feasible space of $t_{01}$.  
It remains to prove that the ATE bounds are attained at an endpoint 
of the feasible interval or at a feasible point where the formulas change.

\begin{theorem}[Evaluation of bounds for the ATE at finite points of $\mathcal{T}_{01}$]
\label{thm:confounding-ate-candidates}
Fix $\theta$ and combine the marginal bounds at the same allocation:
\begin{align*}
L_{\mathrm{ATE}}(t_{01})
&=L_{P(Y(A=1)=1)}(t_{01})-U_{P(Y(A=0)=1)}(t_{01}),\\
U_{\mathrm{ATE}}(t_{01})
&=U_{P(Y(A=1)=1)}(t_{01})-L_{P(Y(A=0)=1)}(t_{01}).
\end{align*}
The sharp ATE bounds are
\begin{align*}
\mathcal L_{\mathrm{ATE}}(\theta)
&=\min\Big\{
L_{\mathrm{ATE}}\big(\max\{0,P(A=1)-\theta\}\big),\\
&\hspace{2em}L_{\mathrm{ATE}}\big(\min\{1-\theta,P(A=1)\}\big),\\
&\hspace{2em}L_{\mathrm{ATE}}(p_{10}),\;
L_{\mathrm{ATE}}(1-\theta-p_{01})\Big\},\\
\mathcal U_{\mathrm{ATE}}(\theta)
&=\max\Big\{
U_{\mathrm{ATE}}\big(\max\{0,P(A=1)-\theta\}\big),\\
&\hspace{2em}U_{\mathrm{ATE}}\big(\min\{1-\theta,P(A=1)\}\big),\\
&\hspace{2em}U_{\mathrm{ATE}}(p_{11}),\;
U_{\mathrm{ATE}}(1-\theta-p_{00})\Big\}.
\end{align*}
In each ATE formula, include either of the last two evaluations only
if its argument belongs to
$[\max\{0,P(A=1)-\theta\},\min\{1-\theta,P(A=1)\}]$.
At a feasible zero-mass allocation, the marginal conventions give
$L_{\mathrm{ATE}}(t_{01})=-p_{10}-p_{01}$ and
$U_{\mathrm{ATE}}(t_{01})=p_{11}+p_{00}$.
\end{theorem}
\begin{proof}

First, notice that for feasible $t_{01} = 0$ or $t_{01} = 1 - \theta$, 
then the results in Theorem~\ref{thm:confounding-marginal-endpoints}
are sufficient.

Let $0<t_{01}<1-\theta$. If $t_{01} \leq p_{10}$, then 
the max inside the lower bound formula is $0$. Also, 
for the min formula, if $t_{01} \leq \frac{p_{01} t_{01}}{1 - \theta - t_{01}}$, 
which is equivalent to $t_{01} \geq 1 - \theta - p_{01}$, then 
the min formula is equal to $t_{01}$. When those two conditions hold,
 the lower bound is constant for any $t_{01}$.
For the other cases, it is sufficient to check the second 
derivative of the other 
nonaffine terms, which are equal to:
$$-\frac{2(1-\theta)p_{10}}{t_{01}^3},\qquad
-\frac{2(1-\theta)p_{01}}{(1-\theta-t_{01})^3}.
$$
As both are nonpositive, the lower bound is concave on each piece.
Its minimum therefore occurs at a feasible interval endpoint or at
one of these two points, when feasible. 

For the upper ATE bound, the formulas change
at $t_{01}=p_{11}$ and $t_{01}=1-\theta-p_{00}$.
The corresponding second derivatives are
$$
\frac{2(1-\theta)p_{11}}{t_{01}^3},\qquad
\frac{2(1-\theta)p_{00}}{(1-\theta-t_{01})^3}.
$$
Thus the upper bound is convex on each piece, and its maximum occurs
at an interval endpoint or at one of these two points.
That completes the proof.

\end{proof}

\subsubsection{Monotonicity of the bounds}

\begin{theorem}
\label{thm:monotonic-conf-bounds}
For both $P(Y(1)=1)$ and the ATE, the sharp lower bound is
nonincreasing in $\theta$, and the sharp upper bound is nondecreasing.
\end{theorem}

\begin{proof}

Let $0\leq\theta<\theta'\leq 1$ and any feasible strata distribution
$P$ at $\theta$. Construct a distribution at $\theta'$ by moving the
same fraction of every $H=0$ stratum into its corresponding $H=1$
stratum:
\begin{align*}
P'(h_0a_i y_{rs})
&=\frac{1-\theta'}{1-\theta}P(h_0a_i y_{rs}),\\
P'(h_1a_i y_{rs})
&=P(h_1a_i y_{rs})
 +\frac{\theta'-\theta}{1-\theta}P(h_0a_i y_{rs}).
\end{align*}
These probability masses are nonnegative, sum to one, and satisfy $P'(H=1)=\theta'$.
For $\theta'<1$, the conditional on $H=0$ strata distribution, i.e. the strata divided 
by $1- \theta'$, remains unchanged,   so the independence between treatment and potential outcomes is preserved.
On the other hand, 
 for $\theta'=1$, the conditional on $H=0$ strata distribution is undefined, because no unconfounded component exists anymore.
 Moreover, $\sum_{h=0}^1 P'(h_ha_i y_{rs})
=\sum_{h=0}^1 P(h_ha_i y_{rs}).$ 
Therefore, both quantities $P(A=a,Y(A=0)=y, Y(A=1)=y')$ and, when $\theta' < 1$,
$P(A=a,Y(A=0)=y, Y(A=1)=y'| H = 0)$ are invariant.

The key idea is to show that allowing more confounding cannot rule out 
an effect value that was previously possible. So, if 
both optimization problems impose the same constraints with the exception of 
the probability mass of the confounded components, which increases from $\theta$ to $\theta'$, 
that suffices to prove the claim, provided that the construction preserves the effect value.

 Let $P$ be a
strata distribution attaining the sharp lower bound $L(\theta)$. 
Then let the distributions  be preserved as above. By increasing $\theta$ to $\theta'$, 
then the constructed distribution is feasible at $\theta'$ and still attains $L(\theta)$, so the sharp bound $L(\theta') \leq L(\theta)$.
Similarly, applying the construction to a distribution attaining the sharp $U(\theta)$  will imply $U(\theta') \geq U(\theta)$.

\end{proof}

\subsection{Symbolic sensitivity functions in instrumental variable models}
\label{app:iv}

\subsubsection{Polynomial Program for ``no defiers'' and exclusion restriction violations}
\label{subsub:ivprograms}

Let $C=\{A(z_1)>A(z_0)\}$ denote the complier event.
First, consider violations of ``no defiers'', retaining exogeneity
and exclusion. Write $a_{rs}y_{tu}$ for the stratum satisfying
\begin{align*}
(A(z_0),A(z_1))=(r,s),\qquad (Y(a_0),Y(a_1))=(t,u),
\end{align*}
where all indices are binary. By consistency and exogeneity, the
observed cells are
  \begin{align*}
    \Pr(y_0, a_0|z_0) &= P(a_{00}y_{00})
      + P(a_{01}y_{00})
      + P(a_{00}y_{01})
      + P(a_{01}y_{01}), \\
    \Pr(y_0, a_0|z_1) &= P(a_{00}y_{00})
      + P(a_{10}y_{00})
      + P(a_{00}y_{01})
      + P(a_{10}y_{01}), \\
    \Pr(y_0, a_1|z_0) &= P(a_{11}y_{00})
      + P(a_{10}y_{00})
      + P(a_{11}y_{10})
      + P(a_{10}y_{10}), \\
    \Pr(y_0, a_1|z_1) &= P(a_{11}y_{00})
      + P(a_{01}y_{00})
      + P(a_{11}y_{10})
      + P(a_{01}y_{10}), \\
    \Pr(y_1, a_0|z_0) &= P(a_{00}y_{11})
      + P(a_{01}y_{11})
      + P(a_{00}y_{10})
      + P(a_{01}y_{10}), \\
    \Pr(y_1, a_0|z_1) &= P(a_{00}y_{11})
      + P(a_{10}y_{11})
      + P(a_{00}y_{10})
      + P(a_{10}y_{10}), \\
    \Pr(y_1, a_1|z_0) &= P(a_{11}y_{11})
      + P(a_{10}y_{11})
      + P(a_{11}y_{01})
      + P(a_{10}y_{01}), \\
    \Pr(y_1, a_1|z_1) &= P(a_{11}y_{11})
      + P(a_{01}y_{11})
      + P(a_{11}y_{01})
      + P(a_{01}y_{01}), \\
\end{align*}
The stratum probabilities satisfy
\begin{align*}
P(a_{rs}y_{tu})\geq0,\qquad
\sum_{r,s,t,u}P(a_{rs}y_{tu})=1,\qquad
\sum_{t,u}P(a_{10}y_{tu})=\theta_{\mathrm{ND}}.
\end{align*}
The last equality fixes the proportion of defiers. The objective is
\begin{equation}
\label{eq:late-nd-program}
\operatorname{LATE}
=\frac{P(a_{01}y_{01})-P(a_{01}y_{10})}
{\sum_{t,u}P(a_{01}y_{tu})}.
\end{equation}
Minimizing and maximizing this expression over the stratum probabilities,
subject to the observed-cell equations and the three restrictions above,
defines the perturbation problem. Its denominator is identified as
\begin{align*}
P(C)=\sum_{t,u}P(a_{01}y_{tu})
=P(A=1\mid Z=1)-P(A=1\mid Z=0)+\theta_{\mathrm{ND}}>0.
\end{align*}
Thus we can minimize and maximize the linear numerator and divide its
bounds by $P(C)$. For the ATE, the same constraints apply to the
objective $\sum_{r,s}[P(a_{rs}y_{01})-P(a_{rs}y_{10})]$.
Replacing the fixed-defier equality by
$\sum_{t,u}P(a_{10}y_{tu})\leq\theta_{\mathrm{ND}}^{\max}$ gives the
sensitivity problem; its LATE denominator then varies with the defier mass.

Next, consider violations of exclusion, retaining exogeneity and
monotonicity. Write $a_{ij}y_{rstu}$ for the stratum satisfying
\begin{align*}
(A(z_0),A(z_1))=(i,j),\qquad
\big(Y(z_0,a_0),Y(z_0,a_1),Y(z_1,a_0),Y(z_1,a_1)\big)=(r,s,t,u).
\end{align*}
All indices are binary. Monotonicity excludes the $a_{10}$ strata.
By consistency and exogeneity, the observed cells are, for $y=0,1$,
\begin{align*}
P(Y=y,A=0\mid Z=0)
&=\sum_{s,t,u}\big[P(a_{00}y_{ystu})+P(a_{01}y_{ystu})\big],\\
P(Y=y,A=1\mid Z=0)
&=\sum_{r,t,u}P(a_{11}y_{rytu}),\\
P(Y=y,A=0\mid Z=1)
&=\sum_{r,s,u}P(a_{00}y_{rsyu}),\\
P(Y=y,A=1\mid Z=1)
&=\sum_{r,s,t}\big[P(a_{01}y_{rsty})+P(a_{11}y_{rsty})\big].
\end{align*}
The stratum probabilities satisfy
\begin{align*}
P(a_{ij}y_{rstu})\geq0,\qquad
\sum_{i,j,r,s,t,u}P(a_{ij}y_{rstu})=1,\qquad
P(a_{10}y_{rstu})=0\quad\text{for every }r,s,t,u.
\end{align*}
Exclusion fails when $r\neq t$ or $s\neq u$, so its exact violation
proportion imposes the additional constraint
\begin{align*}
\sum_{i,j}\sum_{\substack{r,s,t,u:\\r\neq t\ \text{or}\ s\neq u}}
P(a_{ij}y_{rstu})=\theta_{\mathrm{ER}}.
\end{align*}
The two complier numerators are
\begin{align*}
N_0=\sum_{r,s,t,u}(s-r)P(a_{01}y_{rstu}),\qquad
N_1=\sum_{r,s,t,u}(u-t)P(a_{01}y_{rstu}).
\end{align*}
The objective function, after 
averaging over the observed distribution over $Z$, is
\begin{equation}
\label{eq:late-er-program}
\operatorname{LATE}
=\frac{P(z_0)N_0+P(z_1)N_1}
{\sum_{r,s,t,u}P(a_{01}y_{rstu})}.
\end{equation}
Minimizing and maximizing this expression over the stratum probabilities,
subject to the observed-cell equations, probability constraints,
monotonicity, and the specified proportion of exclusion violations, defines the
perturbation problem. Here its denominator is identified as
\begin{align*}
P(C)=\sum_{r,s,t,u}P(a_{01}y_{rstu})
=P(A=1\mid Z=1)-P(A=1\mid Z=0)>0.
\end{align*}
Consequently, optimizing the weighted numerator jointly and dividing by
$P(C)$ gives sharp bounds.
Replacing the fixed-exclusion equality by an upper bound
$\theta_{\mathrm{ER}}^{\max}$ gives the corresponding sensitivity program.

\subsubsection{Proof of the LATE bounds under monotonicity violations and 
under exclusion violations}
\label{app:late-monotonicity-exclusion}

{
\renewcommand{\repeatedstatementheading}{Theorem~\ref{thm:late_as_func_defiers}}
\begin{repeatedstatement}[Perturbation bounds on the LATE for monotonicity violations]
    \singlespacing
Let $C=\{A(z_1)>A(z_0)\}$ denote the complier event.
Suppose exogeneity and exclusion hold, the observed distribution is
compatible with the model and a fixed specified $\theta_{\mathrm{ND}}$, and
$P(C)  = P(A(z_1)>A(z_0))>0$. Then the perturbation bounds for a particular $\theta_{\text{ND}}$
are {$l(\theta_{\text{ND}})/P(C) \leq LATE \leq u(\theta_{\text{ND}})/P(C)$},
where {$p_{ij}^{k}=P(Y=y_i,A=a_j\mid Z=z_k)$
for $i,j,k\in\{0,1\}$}.
{
The functions $l(\theta_{\mathrm{ND}})$ and $u(\theta_{\mathrm{ND}})$ are,
respectively,
\begin{equation*}
\max\begin{cases}
- p_{10}^{0}, \\
p_{11}^{1} - P(y_1|z_0), \\
p_{11}^{1} + p_{00}^{0} - 1 + \theta_{\text{ND}}, \\
-P(C), \\
p_{10}^{1} -  p_{10}^{0} - \theta_{\text{ND}},\\
p_{01}^{0} -  p_{01}^{1} - \theta_{\text{ND}},\\
P(y_0|z_0) -  P(y_0|z_1) - \theta_{\text{ND}},\\
- p_{01}^{1}, \\
p_{00}^{0} - P(y_0|z_1),
\end{cases},
\quad
\min\begin{cases}
1-p_{10}^{0}-p_{01}^{1}-\theta_{\text{ND}}, \\
p_{11}^{1}, \\
P(y_0|z_0) - p_{01}^{1}, \\
P(y_0|z_0) - P(y_0|z_1) + \theta_{\text{ND}}, \\
p_{00}^{0}, \\
P(C), \\
p_{00}^{0} - p_{00}^{1} + \theta_{\text{ND}}, \\
P(y_1|z_1) - p_{10}^{0}, \\
p_{11}^{1}-p_{11}^{0}+\theta_{\mathrm{ND}}, \\
\end{cases}
\end{equation*}
}
and {$P(C) = P(a_1|z_1) - P(a_1|z_0) + \theta_{\text{ND}}$}.
\end{repeatedstatement}
}

{
\renewcommand{\repeatedstatementheading}{Theorem~\ref{thm:late_as_func_er}}
\begin{repeatedstatement}[Perturbation bounds on the LATE under exclusion violations]
\singlespacing
Here, with exclusion relaxed, write
$\operatorname{LATE}=\E[Y(Z,1)-Y(Z,0)\mid A(z_1)>A(z_0)]$,
which reduces to the usual LATE under exclusion.
Suppose exogeneity and monotonicity hold, the observed distribution is compatible with
the model and the specified $\theta_{\mathrm{ER}}$, and
$P(C)  = P(A(z_1)>A(z_0))>0$. { Then}
$\operatorname{LATE}$ satisfies
$l(\theta_{\mathrm{ER}})/P(C)\leq\operatorname{LATE}
\leq u(\theta_{\mathrm{ER}})/P(C)$,
{using $p_{ij}^{k}$ as above.}
{
\par\noindent
The functions $l(\theta_{\mathrm{ER}})$ and $u(\theta_{\mathrm{ER}})$ are,
respectively,
\medmuskip=2mu
\thickmuskip=3mu
\begin{equation*}
\sum_{k=0}^1 P(z_k)\max\begin{cases}
-P(C), \\
-p_{1-k,k}^{k}, \\
P(y_1|z_1)-P(y_1|z_0)-\theta_{\mathrm{ER}}, \\
p_{01}^{0}-p_{00}^{0}+p_{00}^{1}-p_{01}^{1}-\theta_{\mathrm{ER}}, \\
p_{10}^{1}-p_{11}^{1}-p_{10}^{0}+p_{11}^{0}-\theta_{\mathrm{ER}},
\end{cases},
\;
\sum_{k=0}^1 P(z_k)\min\begin{cases}
P(C), \\
p_{kk}^{k}, \\
P(y_1|z_1)-P(y_1|z_0)+\theta_{\mathrm{ER}}, \\
p_{00}^{0}-p_{01}^{0}-p_{00}^{1}+p_{01}^{1}+\theta_{\mathrm{ER}}, \\
p_{10}^{0}-p_{11}^{0}+p_{11}^{1}-p_{10}^{1}+\theta_{\mathrm{ER}}.
\end{cases}
\end{equation*}
}
with {$P(C) = P(a_1|z_1) - P(a_1|z_0)$}.

{
Let $p_{z1}=P(Z=1)$,
$\min_z=\min\{p_{z1},1-p_{z1}\}$,
and $\max_z=\max\{p_{z1},1-p_{z1}\}$. Define
\begin{align*}
a_{p_{z1}}&=\begin{cases}
p_{10}^{0},&p_{z1}\leq\tfrac12,\\
p_{01}^{1},&p_{z1}>\tfrac12,
\end{cases}
&
b_{p_{z1}}&=\begin{cases}
p_{00}^{0},&p_{z1}\leq\tfrac12,\\
p_{11}^{1},&p_{z1}>\tfrac12.
\end{cases}
\end{align*}
For $\theta=\theta_{\mathrm{ER}}$, the sharp perturbation bounds are
\begin{align*}
L_{\mathrm{ER}}^{\mathrm{pert}}(\theta)
&=\frac{1}{P(C)}\max\Big\{
 l(\theta),\,-\max_z P(C)+\min_z(\theta-1),\\
&\hspace{4em}-\min_z P(C)-|1-2p_{z1}|a_{p_{z1}}
 +\min_z(\theta-1)\Big\},\\
U_{\mathrm{ER}}^{\mathrm{pert}}(\theta)
&=\frac{1}{P(C)}\min\Big\{
 u(\theta),\,\max_z P(C)+\min_z(1-\theta),\\
&\hspace{4em}\min_z P(C)+|1-2p_{z1}|b_{p_{z1}}
 +\min_z(1-\theta)\Big\}.
\end{align*}
}
\end{repeatedstatement}
}

\begin{proof}[Proof of Theorems~\ref{thm:late_as_func_defiers} and~\ref{thm:late_as_func_er}]
First, notice that the denominator is point identifiable
for both problems. Under the monotonicity relaxation case, after
fixing $d = \theta_{\mathrm{ND}}$,
then $P(C)=P(A=1\mid Z=1)-P(A=1\mid Z=0)+ d$. Under the
exclusion restriction relaxation case, $P(C) = P(A=1\mid Z=1)-P(A=1\mid Z=0)$.
So, it is sufficient to calculate bounds for the numerator. The problem
becomes solving both linear programs defined in Appendix~\ref{subsub:ivprograms}.

The linear programs can be solved symbolically using the Python interface
to \texttt{cddlib}, with exact rational arithmetic, which
is employed in the \texttt{SymbolicProblem} class in \texttt{autobounds}.
 Appendix~\ref{app:minimal-iv-code} provides minimal
standalone code for both perturbation calculations that makes use of
\texttt{cddlib} and \texttt{SymPy} but without the \texttt{autobounds}
framework.
As solutions for a linear program, the calculated bounds are sharp.
The final stated bounds are divisions of those bounds by the identified denominator,
assuming $P(C) > 0$.

For the exclusion restriction perturbation function, solve the program for the maximum and minimum, where the objective function is
$N=(1-p_{z1})N_0+p_{z1}N_1$,
over probability distributions
on the 48 principal strata, excluding  the ones that do not satisfy monotonicity. The program must include  the constraints
of observed data and the exact violation proportion
$\theta_{\mathrm{ER}}$.
The solution is obtained by enumerating vertices, using exact rational projection in \texttt{cddlib}.
This
yields the lower and upper numerator envelopes displayed in
Theorem~\ref{thm:late_as_func_er}.
The feasible set is a nonempty compact polytope, so both extrema
are attained, which gives the sharpness of the bounds on the numerator. As $P(C) > 0$ is identified,  then division by it  gives the stated
sharp perturbation bounds.
\end{proof}

\subsubsection{From IV perturbation bounds to sensitivity bounds}
\label{app:iv-sensitivity-envelopes}

\renewcommand{\repeatedstatementheading}{Theorem~\ref{thm:iv-natural-sensitivity}}
\begin{repeatedstatement}[Natural sensitivity bounds for IV assumption violations]
Consider either relaxation separately, under the maintained assumptions
of Theorems~\ref{thm:late_as_func_defiers} and~\ref{thm:late_as_func_er},
respectively, and suppose its sensitivity model is nonempty.
Write $q_a^z=P(A=a\mid Z=z)$ and let $C$ denote the complier event.
\begin{enumerate}[label=(\roman*)]
\item For monotonicity sensitivity, let $\mathcal D$ contain the feasible
values $d\leq\theta_{\mathrm{ND}}^{\max}$ with $P(C)=q_1^1-q_1^0+d>0$.
Let $\mathcal B$ consist of the endpoints of $\mathcal D$ and the values
of $d$ within $\mathcal D$ at which two nonparallel affine terms in
either $l(d)$ or $u(d)$ intersect.
The sharp LATE sensitivity endpoints are
$$
L_{\mathrm{ND}}(\theta_{\mathrm{ND}}^{\max})
=\min_{d\in\mathcal B}\frac{l(d)}{q_1^1-q_1^0+d},\qquad
U_{\mathrm{ND}}(\theta_{\mathrm{ND}}^{\max})
=\max_{d\in\mathcal B}\frac{u(d)}{q_1^1-q_1^0+d},
$$
using one-sided limits at excluded endpoints.
\item For exclusion sensitivity, $P(C)=q_1^1-q_1^0>0$ is fixed and the sensitivity
 envelopes of the displayed bounds for $\operatorname{LATE}$ are
$$
L_{\mathrm{ER}}(\theta_{\mathrm{ER}}^{\max})
=\frac{l(\theta_{\mathrm{ER}}^{\max})}{P(C)},
\qquad
U_{\mathrm{ER}}(\theta_{\mathrm{ER}}^{\max})
=\frac{u(\theta_{\mathrm{ER}}^{\max})}{P(C)}.
$$
{
These sensitivity bounds are sharp.
}
\end{enumerate}
\end{repeatedstatement}

\begin{proof}[Proof of Theorem~\ref{thm:iv-natural-sensitivity}]

We start with the proof for monotonicity sensitivity.

For the proof, consider the lower bound of the 
sensitivity function. The proof for the upper bounds follows 
analogously.

Denote the probability of defiers by $d$. 
The sensitivity lower 
bound is
$$\min_{\substack{d\text{ feasible}\\ d\leq\theta^{\max}}}
L_{\mathrm{pert}}(d).$$
 The proof reduces this potentially infinite set of candidates $d$ to a 
 finite set.

The key idea of this proof is for each expression of each perturbation function, 
one can calculate endpoints and switching points for a feasible interval for $d$.
Then for each candidate $d$, one can evaluate the maximum of the lower bound numerator
 expression and divide them by $P(C)$. Finally, among the many candidates of 
$d$, take the minimum of the resulting lower bound values.

Write $p_{ya}^z=P(Y=y,A=a\mid Z=z)$ and
$q_a^z=P(A=a\mid Z=z)$. With defier proportion $d$, the complier
probability is $P(C)=q_1^1-q_1^0+d$.
Under exogeneity and exclusion, the feasible defier interval is
\begin{align*}
d_-=\max\{0,q_0^1+q_1^0-1,p_{00}^1-p_{00}^0,p_{10}^1-p_{10}^0,p_{01}^0-p_{01}^1,p_{11}^0-p_{11}^1\}, \\
d_+=\min\{q_0^1,q_1^0\}.
\end{align*}

The upper bound can first be calculated by noticing that the observed cells 
 $A=0$ under $Z=1$  and 
$A=1$ under $Z=0$ each contain all the defier 
strata $a_{10}y_{00}, a_{10}y_{01}, a_{10}y_{10}, a_{10}y_{11}$.
The lower bound can be calculated by covering all the other 
12 strata with observed cells and taking 
the smallest sum of their probabilities. 
One minus this upper bound on the probability of all nondefier 
strata will be the lower bound for the defiers. 
Fixing any $d \in [d_-, d_+]$, then the probability of never-takers, always-takers, 
and compliers are respectively identified by $q^1_0 - d$, $q^0_1 - d$, 
$q^1_1 - q^0_1 + d$. Then, for each of those values, the mass can be distributed, 
matching each observed cell, 
across the outcome $y$-strata: $y_{00}, y_{01}, y_{10}, y_{11}$.

Let $\mathcal{D} = [d_-, d_+] \cap [0,\theta_{\mathrm{ND}}^{\max}]$.
 Between intersections of the displayed affine terms,
each LATE endpoint has the form $(a+bd)/(q_1^1-q_1^0+d)$, with
$P(C)=q_1^1-q_1^0+d>0$. Its first derivative with respect to $d$ is
$$
\frac{b(q_1^1-q_1^0)-a}{(q_1^1-q_1^0+d)^2}.
$$
As the sign is constant, 
then the extrema over $\mathcal D$ occur
among the stated endpoints or switching points, so there is no need to 
search for interior ones.

For exclusion sensitivity, impose the constraint
$$
P\{\exists a:Y(0,a)\neq Y(1,a)\}
\leq\theta_{\mathrm{ER}}^{\max}.
$$
The function $l(\theta)$ is nonincreasing and $u(\theta)$ is
nondecreasing, so their values at $\theta_{\mathrm{ER}}^{\max}$ give
valid bounds for the numerator. Then minimizing and maximizing
$(1-p_{z1})N_0+p_{z1}N_1$ over the
48 principal strata not excluded by the monotonicity assumption, under the constraints of observed data
and maximum violation of $\theta_{\mathrm{ER}}^{\max}$, gives $l(\theta_{\mathrm{ER}}^{\max})$ and
$u(\theta_{\mathrm{ER}}^{\max})$ by exact symbolic optimization.
Both extrema are attained on the nonempty compact feasible polytope, which proves sharpness.
Dividing by the identified $P(C)>0$ gives the sharp sensitivity bounds.
\end{proof}

\section{Statistical Inference}
\label{sec:statisticalinference}
\label{app:statistical-inference}

\subsection{Endpoint estimation and confidence limits}

So far, sharp sensitivity curves capture only structural uncertainty,
which remains even if the population distribution of the observed variables is
known. With sample data, the estimated endpoints also contain sampling
uncertainty. Fix an allowed violation vector $\Theta^{\max}$ and suppose that
the data are independent and identically distributed. Let
$L=L(\Theta^{\max})$ and $U=U(\Theta^{\max})$ denote the population
endpoints, and let $[\widehat L_n,\widehat U_n]$ be their plug-in estimates
obtained from the empirical distribution.  Notice that the goal is to
estimate the curves, i.e. the lower and upper bounds
of the partially identified region, not the region itself, which might
contain holes. \red{The confidence intervals here concern the endpoints.
By contrast, \citet{imbens2004confidence} construct confidence intervals
for the true parameter rather than the entire identified region.}

Inference below is pointwise at the fixed sensitivity vector
$\Theta^{\max}$.
Consistency of these estimators requires regularity conditions on the
optimization problem. In particular, suppose that its feasible sets are
nonempty and compact in a neighborhood of the population distribution and
that the endpoint value functions are continuous there. The continuous
mapping theorem then implies
$(\widehat L_n,\widehat U_n)\stackrel{p}{\longrightarrow}(L,U)$.
Proposition~\ref{prop:endpoint-consistency} proves this fact.
A harder problem is if those estimators converge in distribution and 
to which distribution. It is not possible to conclude that 
those estimators converge at root-n rates to Gaussian distributions, 
because maxima and minima are in general only directionally differentiable.
This lack of differentiability means that naive applications of bootstrap 
or standard delta method might fail \citep{andrews2009invalidity},
although specialized versions of both methods can work \citep{bugni2010bootstrap,
fang2019inference}. 

The key result shown here is that confidence intervals for the bounds,
given some $\theta^{\max}$,
can be estimated using recentered
subsampling \citep{romano2010inference,cooper2026learning}.
The validity of this method depends on the conditions that
the estimators of the curves are consistent and
$$\sqrt n
\begin{pmatrix}
\widehat L_n-L\\
\widehat U_n-U
\end{pmatrix}
\rightsquigarrow
\begin{pmatrix}
Z_L\\
Z_U
\end{pmatrix},
$$
where the marginal distribution functions are continuous and strictly
cross the quantile levels used for inference, as specified in
Proposition~\ref{prop:subsampling-validity}.
These limits do not need to be normally distributed.
For bounds expressed as finite maxima or minima of smooth functions,
Proposition~\ref{prop:finite-bound-limit} below establishes consistency
and the limiting distribution directly from the central limit theorem.
Continuity of the limiting distribution and the quantile conditions
still require attention in degenerate cases. This argument does not
establish the same conclusions for arbitrary polynomial programs.
Proposition~\ref{prop:subsampling-validity} proves that subsampling consistently
estimates their marginal distributions when $m\to\infty$ and $m/n\to0$
\citep{politis2001asymptotic}.

This is the procedure.
Set $m=\lfloor n^\gamma\rfloor$, where $0<\gamma<1$.
The empirical applications use $\gamma=0.8$.
Draw $B$ independent subsets of size $m$,
uniformly and without replacement within each subset, recompute
$[\widehat L_m^{(b)},\widehat U_m^{(b)}]$ in each subsample, and form
$T_L^{(b)}=\sqrt m\bigl(\widehat L_m^{(b)}-\widehat L_n\bigr)$ and
$T_U^{(b)}=\sqrt m\bigl(\widehat U_m^{(b)}-\widehat U_n\bigr).$
This choice of $m$ satisfies the two rate conditions. If
$\widehat q_{L,p}$ and $\widehat q_{U,p}$ denote the empirical $p$-quantiles
of these statistics, respectively, the one-sided confidence limits are
\begin{align*}
\ell_n(\theta)=\widehat L_n(\theta)
-\frac{\widehat q_{L,1-\alpha}(\theta)}{\sqrt n},
\qquad
u_n(\theta)=\widehat U_n(\theta)
-\frac{\widehat q_{U,\alpha}(\theta)}{\sqrt n}.
\end{align*}
At each fixed sensitivity value, $\ell_n$ is a lower confidence limit for
$L$, and $u_n$ is an upper confidence limit for $U$, each at asymptotic
level $1-\alpha$. It is important to emphasize one more time, 
those are not a joint $(1 -\alpha)$ confidence interval 
for the identified set, but separate one-sided statements about 
the bounds.

Besides sensitivity curves, 
inference on robustness values has specific challenges.
In the first place, this in population terms represents 
the amount of relaxation a conclusion about an estimand sign 
can withstand. The concern is typically that true
robustness is smaller than estimated. Assume 
the initial estimand is positive. 
As larger robustness does not
undermine this guarantee, one-sided lower confidence intervals 
are sufficient. This is how it is calculated.
For a prespecified
positive effect direction, assume the population bounds are attained,
$L(0)>0$, and $L$ is continuous and nonincreasing. Define
\begin{align*}
\underline\theta_n
=\inf\bigl(\{\theta\in[0,1]:\ell_n(\theta)\leq0\}\cup\{1\}\bigr).
\end{align*}
Then $P(\underline\theta_n\leq\theta^\star)$ has limit inferior at least
$1-\alpha$: coverage at an interior true crossing implies that the lower
confidence limit has reached zero no later. If there is no crossing, however,
$\theta^\star=1$, the inequality holds true automatically, and 
$P(\underline\theta_n\leq\theta^\star) = 1 \geq 1 - \alpha$.
If the estimand is initially negative, 
then it is sufficient to replace $\ell_n$ by $-u_n$.
As no upper confidence bound on robustness is sought, subsamples whose
curves never cross are not discarded. 
If the robustness value is evaluated by grid search, on a fixed grid, 
report the grid point immediately before the first
point where the lower confidence limit is nonpositive. Report zero
if it is nonpositive at the first point, and one if it remains positive
throughout the grid. Rounding down preserves the lower-confidence-bound
guarantee when the true threshold lies between grid points.
The proof is in Appendix~\ref{app:robustness-inversion}.

\subsection{Proofs of consistency and subsampling validity}
\label{app:inference-proofs}

The following results 
concern consistency and pointwise coverage
 of the estimated sensitivity 
bounds given a fixed relaxation parameter
 under the population  
distribution. Therefore, 
they do not target the simultaneous coverage of the whole sensitivity curve.

\begin{proposition}[Consistency of estimated endpoints]
\label{prop:endpoint-consistency}
Let $p$ be the vector of observed cell probabilities in a finite model,
and let $\widehat p_n$ be its empirical estimate from independent,
identically distributed observations. Fix the sensitivity parameter.
Suppose the feasible sets are nonempty and compact on a neighborhood
of $p$ in the probability simplex, the objective is defined and
continuous on those sets, and the endpoint functions $L(p)$ and $U(p)$
are continuous at $p$. Then:
$$
\big(L(\widehat p_n),U(\widehat p_n)\big)
\stackrel{p}{\longrightarrow}\big(L(p),U(p)\big).
$$
\end{proposition}

\begin{proof}
The law of large numbers gives $\widehat p_n\to p$ in probability, 
and in consequence, $\widehat p_n$ is consistent for $p$. The polynomial programs
are constructed over closed subsets of the probability simplex. 
Assume all $p$ probabilities are positive and the bound
 functions are defined near $p$ and continuous at it. 
 By the continuous mapping theorem, therefore,
the bounds are also consistent. For cases where some $p$ cells are 
equal to $0$, then their empirical frequencies are also zero almost surely. 
Then the empirical $\widehat p_n$ stays in subset of the simplex 
where those cells have probability zero, and continuity outside 
that set is irrelevant. If the bound functions are defined 
near $p$ within this subset and continuous at $p$ along it,
then the empirical bounds are again consistent. 
\end{proof}

\begin{proposition}[Limits for finite maxima and minima]
\label{prop:finite-bound-limit}
Fix the sensitivity parameter. Let $p$ be the vector of observed cell
probabilities and suppose
$\sqrt n(\widehat p_n-p)\rightsquigarrow G$, where $G$ is a centered
Gaussian vector. Suppose an endpoint is
$T(p)=\max_{1\leq j\leq J}f_j(p)$, where $J$ is finite and each $f_j$
is continuously differentiable in a neighborhood of $p$. Write
$I(p)=\{j:f_j(p)=T(p)\}$. Then $T(\widehat p_n)$ is consistent and
$$
\sqrt n\{T(\widehat p_n)-T(p)\}
\rightsquigarrow \max_{j\in I(p)}\nabla f_j(p)^\top G.
$$
For a minimum, both maxima in this statement are replaced by minima.
\end{proposition}

\begin{proof}
Consistency of $T(\widehat p_n)$ follows from 
Proposition~\ref{prop:endpoint-consistency}.
Any function $f_j$ that is inactive, i.e. 
is not a maximum,
has a strictly
positive gap below the maximum at $p$. The fact that the number 
of functions $f_j$ is finite and that they are continuous implies
that, with probability tending to one, only the indices in $I(p)$ can
attain the empirical maximum asymptotically. 

For each index in $I(p)$, Taylor expansion gives
\begin{align*}
\sqrt n\{f_j(\widehat p_n)-f_j(p)\}
=\nabla f_j(p)^\top\sqrt n(\widehat p_n-p)+o_p(1).
\end{align*}

The remainders are jointly $o_p(1)$ because there are finitely many
functions. Taking their maximum and applying the continuous mapping
theorem proves the result. 

This proof also works for the minimum by the fact that 
$\min_{j} f_j = - \max_j(- f_j)$. 
\end{proof}

The result in Proposition~\ref{prop:finite-bound-limit} 
holds for multinomial proportions whose Gaussian limit has 
covariance $\operatorname{diag}(p)-pp^\top$ ,  conditional cell probabilities 
 with positive conditional probabilities, and finite sums and nested maxima or minima  
 of smooth expressions. For ratios, such as LATE, the result 
 holds at a fixed population 
 distribution as long as the denominator is strictly positive.

 The limiting distribution need not be continuous. If a bound is
constant in a neighborhood of the population distribution, its
estimate equals that constant with probability tending to one.
In this case, the subsampling confidence limit has coverage tending
to one. Otherwise, the continuity and quantile conditions in the
following proposition still need to be verified.

A separate issue is degeneracy of the limit. For a single finite maximum
or minimum, positive variance of every active Gaussian term is sufficient
for a continuous limiting distribution: an atom in their maximum or
minimum would require an atom in at least one term. A locally constant
bound, for example, instead has a degenerate limit. Thus a positive LATE
denominator alone does not establish all the distributional conditions
of the next proposition. Whenever its continuity and quantile conditions
hold, recentered subsampling estimates the limit derived above, including
at ties between active smooth expressions.

\begin{proposition}[Recentered subsampling and one-sided coverage]
\label{prop:subsampling-validity}
For either endpoint $T\in\{L,U\}$ at a fixed sensitivity value, let
$\widehat T_n$ be a measurable, permutation-invariant statistic of an
i.i.d. sample. Suppose
\begin{align*}
\sqrt n(\widehat T_n-T)\rightsquigarrow Z_T,
\end{align*}
where $Z_T$ has continuous distribution function $F_T$.
Let $m=m_n\to\infty$, $m/n\to0$, and $B=B_n\to\infty$.
Conditional on the data, draw $B$ independent, uniformly selected subsets
of $m$ distinct observations and recompute the same statistic on each.
Define
\begin{align*}
\widehat F_{n,m,B}(x)
=\frac1B\sum_{b=1}^B
\bbone\{\sqrt m(\widehat T_m^{(b)}-\widehat T_n)\leq x\}.
\end{align*}
Then $\sup_x|\widehat F_{n,m,B}(x)-F_T(x)|\to0$ in probability.
For a quantile level $v\in(0,1)$, additionally assume that
$q_{T,v}=\inf\{x:F_T(x)\geq v\}$ satisfies
\begin{align*}
F_T(q_{T,v}-\varepsilon)<v<F_T(q_{T,v}+\varepsilon)
\quad\text{for every }\varepsilon>0.
\end{align*}
Its subsampling estimate $\widehat q_{T,v}$ then converges in probability
to $q_{T,v}$. Applying this condition at $v=1-\alpha$ for $L$ and
$v=\alpha$ for $U$ gives
\begin{align*}
P\left(\widehat L_n-\frac{\widehat q_{L,1-\alpha}}{\sqrt n}\leq L\right)
\longrightarrow1-\alpha,\qquad
P\left(U\leq\widehat U_n-\frac{\widehat q_{U,\alpha}}{\sqrt n}\right)
\longrightarrow1-\alpha.
\end{align*}

Alternatively, suppose the endpoint function 
is constant in a neighborhood of the population distribution.
In the locally constant case, the relevant subsampling quantiles converge to zero, 
and the one-sided confidence limit has coverage tending to one. Then
$$
   \liminf_{n\to\infty}P(\text{confidence limit covers the endpoint})
   \geq1-\alpha.
$$

\end{proposition}

A sufficient condition for incompatibility to vanish is that the true
observed distribution lies in the interior of the compatible region,
relative to its sampling support. At binding compatibility boundaries,
incompatibility may persist. The fallback values assigned to infeasible
samples do not by themselves establish the weak-limit assumptions above,
so the stated coverage result does not automatically extend to these cases.

\begin{proof}
For the distribution formed using all subsets, uniform convergence and
one-sided coverage follow from Theorem~2.1(iii)--(iv) of
\citet{politis2001asymptotic}. In their notation, take
$j_n=k_n=m_n$ and $\tau_n=\sqrt n$. The required conditions follow
from $m\to\infty$ and $m/n\to0$, since
$\tau_m/\tau_n=\sqrt{m/n}\to0$. The assumed weak limit and
continuity of $F_T$ provide their distributional assumptions.
The same conclusions hold for the Monte Carlo 
approximation as \(B\to\infty\).
 The stated strict crossing condition gives quantile 
 consistency and therefore both coverage statements.

For the locally constant case, the plug-in estimate equals the true
constant with probability tending to one, and the fraction of subsample
estimates differing from that constant tends to zero. Then each
relevant subsampling quantile is exactly zero with probability tending
to one, giving coverage tending to one.
\end{proof}

\subsection{Coverage of the one-sided robustness bound}
\label{app:robustness-inversion}

\begin{proposition}[Lower robustness bound]
\label{prop:robustness-inversion}
Suppose the identified sets are nonempty and expanding as $\theta$ increases. Suppose their bounds are
attained and $L$ is continuous on $[0,1]$. 
Assume $L(0)>0$.
Let $\theta^\star=\inf(\{\theta\in[0,1]:L(\theta)\leq0\}\cup\{1\})$
denote the population robustness value.

Let $\ell_n(\theta)$ be the lower confidence limit for the population lower sensitivity bound $L(\theta)$, computed from the sample:
$$
\ell_n(\theta)
=\widehat L_n(\theta)
-\frac{\widehat q_{L,1-\alpha}(\theta)}{\sqrt n},
$$
where \(\widehat q_{L,1-\alpha}(\theta)\) is the \(1-\alpha\) quantile of the recentered subsampling distribution.

Let  $\underline{\theta}_n
=\inf\bigl(\{\theta\in[0,1]:\ell_n(\theta)\leq0\}\cup\{1\}\bigr)$, i.e., 
 the first sensitivity value where the estimated lower confidence limit becomes 
 nonpositive, or $1$ if it never does. 
 
Assume that $\ell_n(\theta)$ has a pointwise asymptotic  coverage of at least $(1-\alpha)$. 
The first nonpositive value of
$\ell_n$, capped at one, is a lower confidence bound
$\underline\theta_n$ satisfying
$$
\liminf_{n\to\infty}P(\underline\theta_n\leq\theta^\star)\geq1-\alpha.
$$

\end{proposition}

\begin{proof}
If $\theta^\star = 1$, then the probability statement is $1 \geq 1 - \alpha$ for any  
$\underline{\theta}_n$.

Suppose, so, $\theta^\star<1$. Then continuity and initial positivity give
$L(\theta^\star)=0$. On the coverage event,
$\ell_n(\theta^\star)\leq L(\theta^\star)=0$,
so $\underline{\theta}_n\leq\theta^\star$.
By the assumed pointwise coverage, this event has probability
with limit inferior at least $1-\alpha$.

Generate a fixed grid $0=t_0<\cdots<t_K=1$. Let $k\geq0$ be the first index with $\ell_n(t_k)\leq0$. Return $t_{k-1}$ if $k>0$, $0$ if $k=0$, and $1$ if no such index exists.

For the grid implementation, let $t_j<\theta^\star\leq t_{j+1}$, as the sensitivity $L$ is nonincreasing at $\theta$, 
$L(t_{j+1})\leq0$. By Proposition~\ref{prop:subsampling-validity}, 
provided its assumptions hold at $t_{j+1}$, the event $\ell_n(t_{j+1})\leq L(t_{j+1})$ 
has probability with limit inferior at least $1-\alpha$. 
On this event, $k\leq j+1$, so the returned value is at most
$t_j<\theta^\star$. If $k=0$, the returned value is zero.
Thus the grid procedure also has asymptotic coverage
of at least $1-\alpha$.

\end{proof}

\clearpage

\section{Minimal code for the IV perturbation bounds}
\label{app:minimal-iv-code}
The following Python script requires libraries SymPy and pycddlib 3 with
its exact rational backend. Run it as \texttt{python minimal\_iv\_bounds.py}.

The script prints lower and upper bounds for the programs that generate
closed-form solutions for the monotonicity and exclusion perturbation
functions, as described in Appendix~\ref{subsub:ivprograms}.
The variable names \texttt{pya\_z} denote
$P(Y=y,A=a\mid Z=z)$, with \texttt{y}, \texttt{a}, and \texttt{z}
replaced by their binary values. The variable \texttt{theta} is the
exact violation proportion, and \texttt{w} is $P(Z=1)$.
The output was simplified by removing terms that are redundant and algebraically
grouping the remaining expressions.

{\footnotesize\singlespacing
\begin{verbatim}
"""Run with Python, SymPy and pycddlib>=3; no autobounds import is needed.

Print exact perturbation bounds with all observed cells, theta and w symbolic.
Only evaluate them on the printed compatibility domain with denominator > 0.
"""
from itertools import product, combinations
from fractions import Fraction
import sympy as s
import cdd.gmp as cdd

p = s.symbols('p00_0 p01_0 p10_0 p00_1 p01_1 p10_1', real=True)
theta, w = s.symbols('theta w', real=True)
x = s.Matrix([1, *p, theta])
delta = p[0] + p[2] - p[3] - p[5]


def hull(points):
    V = cdd.matrix_from_array(
        [[Fraction(str(v)) for v in row] for row in points],
        rep_type=cdd.RepType.GENERATOR)
    H = cdd.copy_inequalities(cdd.polyhedron_from_matrix(V))
    assert not H.lin_set  # These coordinates have full affine dimension.
    return [list(map(s.Rational, row)) for row in H.array]


def bounds(exclusion):
    points = []
    for a0, a1 in product((0, 1), repeat=2):
        if exclusion and a0 > a1:
            continue
        for ys in product((0, 1), repeat=4 if exclusion else 2):
            y = ys if exclusion else ys + ys
            obs = [int((a0, a1)[z] == a and y[2*z+a] == b)
                   for z in (0, 1) for b, a in ((0, 0), (0, 1), (1, 0))]
            violation = int(y[0] != y[2] or y[1] != y[3]) if exclusion else int(a0 > a1)
            N = [int(a0 < a1)*(y[2*z+1]-y[2*z])
                 for z in range(2 if exclusion else 1)]
            points.append([1, *obs, violation, *N])
    d = 2 if exclusion else 1
    H = hull(points)
    domain = [s.Ge((s.Matrix([r])*x)[0], 0)
              for r in hull([v[:-d] for v in points])]
    c = s.Matrix([1-w, w] if exclusion else [1])
    pieces = {1: set(), -1: set()}
    for rows in combinations(H, d):
        A = s.Matrix([r[-d:] for r in rows])
        if not A.det():
            continue
        b = s.Matrix([r[:-d] for r in rows])*x
        for sign in (1, -1):
            lam = A.T.inv()*(sign*c)
            condition = s.reduce_inequalities(
                [v >= 0 for v in lam] + ([w >= 0, w <= 1] if exclusion else []), w)
            if condition == s.false:
                continue
            value = s.expand(-sign*(lam.T*b)[0])
            pieces[sign].add(s.Piecewise((value, condition), (-sign*s.oo, True)))
    denominator = delta if exclusion else delta + theta
    lower = s.Max(*sorted(pieces[1], key=s.default_sort_key), evaluate=False)
    upper = s.Min(*sorted(pieces[-1], key=s.default_sort_key), evaluate=False)
    return lower/denominator, upper/denominator, domain, denominator


if __name__ == '__main__':
    for exclusion in (False, True):
        lower, upper, domain, denominator = bounds(exclusion)
        print('\nEXCLUSION' if exclusion else '\nMONOTONICITY')
        print('Require:', domain, 'and', denominator, '> 0; 0 <= w <= 1')
        print('Lower:', lower)
        print('Upper:', upper)
\end{verbatim}
}

\end{document}